\documentclass[12pt]{article}
\usepackage{amsmath,amssymb,graphics,latexsym,amsfonts}

\begin{document}

\renewcommand{\theequation}{\arabic{equation}}
\renewcommand{\thefigure}{\arabic{figure}}
\newcommand{\rhob}{\mbox{\boldmath $\rho$}}
\newcommand{\omegab}{\mbox{\boldmath $\omega$}}
\newcommand{\Omegab}{\mbox{\boldmath $\Omega$}}

\begin{center}
 Table of contents

\end{center} 

\begin{itemize}
\item
Abstract ---------1

\item 1. Introduction-----------2

\item 2. Observed properties of galactic magnetic fields --4

\item 3. Summary of our present understanding of 
cosmic magnetic field origins------------9

\item 4. Basic equations for magnetic field evolution --12

\item 5.  Cowling's theorem   and Parker's dynamo --15

\item 6.  The alpha-Omega disc dynamo ---18

\item 7. The magnitudes of $ \alpha $ and $ \beta $ in the
interstellar medium ---23

\item 8. Ferri\`{e}re's  dynamo theory based on 
supernova and superbubble explosions---26

\item 9. The validity of the vacuum boundary
 conditions -- 34

\item 10. Arguments against a primordial origin  --37

\item 11. Seed fields ----41

\item 12 A protogalactic theory for magnetic field 
generation --- 45

\item 13. Generation of small scale magnetic fields by 
turbulence ---48

\item  14. The saturation of the small scale magnetic fields 
--------54

\item 15. History of the evolution of a primordial magnetic field
--------58

\item 16. Extragalactic magnetic fields --66

\item 17. Summary and conclusions --67

\item 18.  Publications --- 70

\end{itemize} 

\newpage 
 
\begin{center}
\Large
{
{On the Origin of Cosmic  Magnetic Fields }\\

Russell M. Kulsrud and Ellen G. Zweibel

Princeton University and the University of Wisconsin-Madison}

\end{center}

\abstract

We review the extensive and controversial literature
concerning how the cosmic magnetic fields pervading
nearly all galaxies and clusters of galaxies
actually got started. 
Some  observational evidence supports a  hypothesis
that the field
is already moderately strong at the beginning of the
life of a  galaxy and its disc.  One argument
involves the chemical abundance of the light
elements  Be and B, while a second one is based
on the detection of strong magnetic fields  in
very young high--red--shift galaxies.

Since this problem of initial amplification  of 
cosmic magnetic fields 
involves important plasma problems, it is obvious
that one must know the plasma in which the amplification 
occurs.  Most of this review is devoted to this basic
problem and, for this,  it is necessary to devote ourselves
to reviewing studies that take place in environments
in which the plasma properties are most clearly understood.
For this reason the authors have chosen to restrict themselves
almost completely to studies of dynamo action in our Galaxy.
It is true that one can get a much better idea of the grand
scope of galactic fields in extragalactic systems. 
However, most mature galaxies share the same dilemma
as ours of overcoming important plasma problems.
Since the authors are both trained in plasma physics, 
they  may be biased in pursuing this approach,  but they 
feel this restriction it is justified by the above argument.
In addition, they  feel they  can produce a better review
by staying close to that which they know best. 

In addition they  have chosen not to consider
the saturation problem of the galactic magnetic field
since, if the original dynamo amplification fails,
the saturation question does not enter.

 It is generally accepted
that seed fields, whose strength is of order
$ 10^{ -20} $ gauss,  easily spring up in the era preceding
galaxy formation.  Several mechanisms have been proposed
to  amplify these seed  magnetic fields to a coherent
structure with  the microgauss strengths of the 
currently observed galactic magnetic fields.

The standard and most popular mechanism is the 
alpha-Omega  mean--field dynamo  theory developed
by a number of people in the late sixties.  This
theory and its application to galactic magnetic fields
is discussed in considerable detail in this review.
We point out  certain difficulties with this theory that
make it seem unlikely that this is the whole story.
The main difficulty with this as the only such
amplification mechanism, is rooted in the fact that, on
galactic scales, flux is constant and
is frozen in the interstellar medium.  This implies
that flux must be removed from the galactic discs,
 as is well recognized by the standard theory.

   For our Galaxy this turns out to be a major problem, since
unless the flux and the interstellar mass 
 are somehow separated,  some
interstellar mass must also be removed from the deep
  galactic gravitational well. This is very difficult.
It is 
pointed out that unless the field has a substantial field
strength,  much larger than that of the seed fields,
this separation  can hardly happen.  And of course,
 the alpha--Omega dynamo must 
start from the ultra weak seed field. (It is our
philosophy, expressed in this review, that if an origin
theory is unable to create the magnetic field in our
Galaxy it is essentially incomplete.)
 
Thus, it is more reasonable
 for the first and largest amplification to occur
before the Galaxy forms and the matter embedded
in the field is gravitationally trapped. 
  Two such mechanisms are discussed for such a pregalactic 
origin; 1) the fields  are generated in the turbulence of the protogalaxy
and 2)  the fields  come from giant radio jets.  
Several arguments  against a primordial
origin are also  discussed, as are  ways around them.

Our conclusion as to the most likely origin
of cosmic magnetic fields  is that
they are first produced at moderate field 
strengths by  primordial mechanisms,  and 
then  changed and their strength increased  to
their present value and structure by a galactic disc
dynamo. The 
primordial mechanisms have not yet been seriously
developed as of yet, and this preliminary 
amplification of the magnetic fields is still very  open.
If a convincing case can be made that these 
primordial mechanisms  are necessary,
more effort will of course be devoted to their study.

\newpage 

\section{ Introduction}

It is well established that the universe
is filled with magnetic fields of very large scale
and substantial strength.  The fields exist on
all scales, in planets, stars, galaxies, and clusters
of galaxies (Parker 1979).  But 
with respect to its origin,  the magnetic
field in stars and planets is secondary, while
the field in galaxies is primary.  The situation
for clusters of galaxies is not very clear (Carilli 2002); their
magnitude and structure being rather uncertain.
Therefore,  the best route to understanding cosmic
fields is through discovering their origin  in
galaxies, and in particular in our Galaxy.  
(Parker 1979, Ruzmaikin et al, 1988, 
Beck et al 1996, Zweibel \& Heiles 1997, Kulsrud 1999, 
Carilli \& Taylor 2002,
Widrow 2002, Vallee 2004).

  The idea embraced in this review is: that one has  the clearest 
idea of what happens in our Galaxy.   If one  can not understand
the origin  problem here,  then the cosmic origin  theory 
of magnetic fields has to be considered incomplete.  

 It must be remarked that this choice of reviewing only
the work on dynamos specifically in our Galaxy,  is the choice of the authors
and represents a certain bias on their  part.  Generally,
reviews of galactic magnetic fields 
discuss the magnetic fields in a great variety of extragalactic
systems.
This in general is justified since, by examining the global
shapes and properties of fields in external galaxies one can form a much
better picture of these fields,  than by restricting ourselves
to  the field
in our Galaxy, in which
we see only its  more  local parts.
Moreover, the display of these magnetic fields
has aesthetic beauty which alone should justify this approach.

  However, the authors feel  that 
every one of  these extragalactic fields represent a very difficult
problem from a plasma physics point of view.  If one wants
to understand how all these field in ours and other galaxies
got started from an extremely weak seed field,
 one has to first deal with fields  much weaker
than those  that can be observed.  The problem that needs to
be overcome is the problem of flux conservation, basically
a plasma problem.  Since the authors are trained plasma 
physicists, they need to know the most basic properties 
of the plasma in which this happened, so
these is no better situation to examine than our own 
interstellar medium.
Therefore, their  choice makes  it is possible to critically
examine the basic plasma physics of galactic dynamos in this
weak phase and here at home.

In addition, they do  not seriously consider the problem
of the saturation of the  the interstellar 
magnetic field.  In the opinion of the authors
this problem is really secondary to the origin  of the
field since if the field cannot be amplified by the large 
amount required to reach its present value, the saturation problem does
not enter.

  Although it is widely accepted that
 magnetic fields were not produced in the Big Bang, 
there seems little difficulty with  the creation  of seed
fields in the universe, during the 
period subsequent to recombination,  that is the creation
of fields 
 with  strengths of order $ 10^{ -20} $ gauss.
There are a number of mechanisms that can operate during
the period of galaxy formation  and can  generate such fields.
The Biermann battery (Biermann 1950)
 is a simple such  mechanism\footnote{It is worth noting that magnetogenesis by exotic processes in the
early universe has also been proposed. Because the nature of these fields by the epoch of galaxy formation
is highly uncertain, we have ignored them in this review; for  good discussions see Grasso \& Rubenstein (1995)
and Widrow (2002)}.
On the other hand, the currently observed  field strengths
are  of the order of $ 10^{ -6} $ gauss.  Thus,  there is a long way
to go between fields with  these two  strengths.  Hence, the
main problem with the origin  of cosmic magnetic fields centers on
how the strengths of cosmic magnetic  fields were raised from  
 weak values  of
$ 10^{ -20} $ gauss to the currently  observed 
 microgauss strengths.

In discussing this subject of magnetic field origins,
 we distinguish between (1) amplification 
of  fields that are already somewhat strong 
 so that
the amplification mechanisms can  actually by aided by
the magnetic fields themselves, and (2) amplification of
extremely weak fields,   those  whose strength is so weak that
  they can play no role in the amplification mechanisms.
The latter  are passively amplified by mechanisms that are totally 
unaffected by  their presence.

There is a second division of the problem of amplification
that concerns the nature of the magnetic field itself.
As we will see, it is relatively easy to increase the 
energy of magnetic fields if one allows them to be very tangled,
changing their direction on very small scales.  It is much
more difficult to produce coherent magnetic fields that 
 change their
direction only on very large scales, as is the case
for the magnetic field in our Galaxy. 

Since amplification of the magnetic field energy
is relatively easy to understand, whether the field is
very weak or whether it is  strong
(Batchelor 1950, Kazantsev 1968, Kraichnan \& Nagarajan 1967,
Kulsrud \& Anderson 1992, Boldyrev \& Cattaneo 2004), 
 the real
problem of concern for a  theory of magnetic field origin  
is this coherency problem, especially for very weak fields.  

Why are we interested in the generation and amplification 
of cosmic fields?  There are several reasons.
First, until we can be sure we understand this problem,
we cannot be sure that we really understand the cosmological
evolution  of  the  universe.   Second, the
actual structure of the observed magnetic fields is not very
well determined because   most
of the measurements of the magnetic field use techniques,
such as Faraday rotation,  
that  average magnetic fields over large  distances,
(Zweibel \& Heiles 1997, Heiles, 1998).
If we knew theoretically how the magnetic fields were generated,
this would give  extra leverage to determining their local 
structure.  Finally,  many of 
the very mechanisms that produce
magnetic  fields are astronomically interesting, and 
 important in themselves.

Why are galactic magnetic fields astrophysically important?
Without their universal presence in the interstellar medium
its  astrophysical properties
would be very different. At the present time,
 magnetic fields play a crucial
role in the way stars  form (Spitzer 1978). They
also control the origin  and confinement 
of cosmic rays,   which in turn play important astrophysical
roles. Further, magnetic
 fields  are  an important ingredient in the equilibrium 
balance of the galactic disc. 
 
Why is the origin  of  magnetic fields so difficult to understand?
First,  they are difficult to observe because they 
are so weak and far away. Second, to understand the physics of
their origin  one needs to understand  astrophysical plasma physics
(Kulsrud 2005,  Cowling 1953),
 fluid dynamics, and many other fields of  astrophysics. 
Plasma physics on galactic scales is still not a well
developed subject and the details of how it works
are  hard to
observe.  More importantly, since the origin  of cosmic
fields occurs either  over the entire life of the Galaxy,
(Ruzmaikin et al 1988) or possibly in a pregalactic age
before galactic discs formed, it is very difficult
 to gain observational  knowledge of the  early generation
mechanisms.  The early 
stages in  other galaxies are  observed at large red 
shift where large  distances  make these observations difficult
to make, (Kronberg, 1994,
Welter, Perry \& Kronberg  1984, Watson \& Perry 1991, 
Wolfe, Lanzetta, \& Oren 1992, Oren \& Wolfe 1995.) 
For these reasons {\it  definitive}  progress  in uncovering
magnetic field origins in the universe  has been slow.

  A main goal of this review is to arrive at some sort
of  conclusion as to whether at the stage when the galactic
disc formed the magnetic field was still extremely weak 
and the amplification occurred during the entire age of the disc,
or whether the the fields already were substantial before the
galactic disc started to form.  If the former is the case,
we will call the origin galactic and the dynamo generating
it the galactic dynamo.  In the latter case we call the
origin  pregalactic.  (We avoid using primordial which suggest
a much earlier origin  then occurs,  say,  during the protogalactic
era.)

In the next two sections 
 we briefly review the salient 
observations and present  a historical
introduction to galactic dynamo theory.
 The remainder of the review  discusses current theories of magnetic
field origin and concludes with a critical summary.

\section{Observed properties of galactic magnetic fields}

  Our knowledge of galactic magnetic fields rests on four observational
 pillars. Measurements of Faraday rotation
and Zeeman splitting give the magnetic field perpendicular 
to the plane of the sky, integrated along the line of
sight. These effects are direction sensitive, and contributions
 from oppositely directed fields tend to cancel each
other. Observations of the polarized synchrotron continuum, and
 polarized emission and absorption by magnetically
aligned dust grains, give the line--of--sight integrated magnetic
 field components in the plane of the sky. These
diagnostics are sensitive to orientation, not direction,
 but 90$^{\circ}$ swings in orientation also cancel.
 In addition to
line--of--sight averaging,  finite angular
resolution of the telescope causes plane--of--sky averaging.

  According to these observations, the mean orientation of the 
magnetic field is parallel to the Galactic plane
and nearly azimuthal; the deviation is consistent with assignment
 along the spiral arms (Heiles 1996). This orientation
is consistent with the effects of induction in a system with
 strong differential rotation and a spiral density wave pattern
(Roberts \& Yuan 1970). Assuming that the Galactic
halo rotates more slowly than the disk, induction would act on 
a vertical field so as to produce a reversal in the
azimuthal field across the Galactic plane (so-called dipole symmetry).
 The traditional view has been  that the field does not reverse
across the plane (Beck et al 1996), although some authors favor 
asymmetry (Han 2003, Han et al 1997, Han 2001). This
important question is still open.

The Galactic field within several kiloparsecs (kpc) of the Sun has
 both mean and random components (Rand \& Kulkarni
1989, Han et al. 2006, Beck 2007), with the mean component being of
 order 1.4-2$\mu$G and the rms field about 5-6$\mu$G.   The rms value 
for the random 
field is derived from the assumption that the random  field is isotropic
and from a measurement of its  line--of--sight component. 
 There {\it  is}  some 
evidence that the fluctuations are anisotropic, with 
more power parallel to the mean field than perpendicular to it (Zweibel 1996,
Brown \& Taylor 2001). There is also evidence that the
 mean field reverses with Galactocentric radius (so-called
bisymmetric spiral structure), but the locations and
  frequency of reversal are quite uncertain 
(see Han \& Wielebinski 2002 for a review, 
and Weisberg et al. 2004, Han et al. 2006,
Brown et al 2007 for more recent studies). The
discrepant results of this important and difficult 
measurement reflect the high level of noise
 (fluctuations greater than the mean),
uncertain distances to the background pulsars against
 which Faraday rotation is measured, uncertainty in the
 location of Galactic
spiral arms (Vallee 2005), and a possible systematic spatial
 variation in the field structure which is unaccounted for in the models.

There are  some complications in the interpretation introduced  by 
the spiral arms perturbing the direction of the magnetic field.  
Even if the unperturbed field is toroidal, {\it  these spiral arm}
 perturbations 
give the impression that the global field is aligned along the 
spiral arms and its lines of force are a spiral 
[Lin, Yuan and Shu, 1969,  Manchester, 1974 
(section iv page 642)].

How far back in time are galactic magnetic fields detectable?
 Young galaxies and their environment have been probed
 through
the absorption lines found in quasar spectra whose
redshift is different than that of the quasar. 
These absorption lines are impressed on the quasar light
as it passes through clouds of gas, and these clouds
 are interpreted
as young galaxies.  Most of these systems are rather
thin and  are referred to
as part of a Lyman alpha forest.  However, some of these
systems are much thicker and display both metallic lines
(particularly lines of MgII) and very broad damped  Lyman alpha
lines.  The latter are referred to as damped Lyman alpha
systems.  More important for our purpose, if the quasar
emits polarized radio waves,  the plane of polarization
of these  waves would be Faraday rotated by a significant
amount if these systems had coherent magnetic fields
(Perry 1994).

To determine this possibility  Welter, Kronberg and Perry
(1984) searched for Faraday rotation in a number of
radio emitting quasars,  and found  a definite
correlation between those which  had a rotation measure
and those which have  metallic absorption lines.
The major difficulty with these observations is
the correct subtraction 
 of the Faraday rotation produced by the magnetic field
of {\it our} Galaxy, which does not vary smoothly with
 angular position.
Welter et. al. found 
 five  unambiguous cases.  These results were
reaffirmed by Watson \& Perry  (1991).  Since the metallic
absorption line systems had red shifts that were fairly
large (of order two), these systems probably represent
young galaxies in an early state of formation.
  This data was reanalyzed
 by Wolfe and his collaborators (Oren and Wolfe, 1995),
 but in a  different
manner.  They restricted their analysis to 61 quasars
with MgII absorption lines and separated out, as a subclass,
11 of these that also had damped Lyman alpha lines.  They
found that the incidence of Faraday rotation in the 11
damped systems was higher than that in the remaining
50 undamped systems with a 99.8 per cent confidence
level.

In this analysis  they concluded  that the errors
introduced by the Faraday rotation in
 our  Galaxy  were larger than those
assumed by 
Welter et al by a factor of three. Thus,  they
only found two cases in which they were certain
that there  was Faraday rotation.

The detection of only two cases with
definite intrinsic rotation measures seems to make
a weak case for extragalactic fields in these damped systems.
But these cases were those systems with the lowest red
shift.
For   the other systems of larger redshifts $z$,
any intrinsic Faraday rotation produced by them is diluted
by a factor of $ (1+z)^2 $.  This is  because the frequency
of the radio waves passing through them is higher
by the factor $ (1+z) $ and the amount of Faraday rotation
decreases with the frequency squared.  Thus,
the other members of the damped class could very
well have magnetic fields of the same strength as
the lower red shift members without producing
a detectable signal.
  This bolsters their correct subtraction of  the galactic 
component of the Faraday rotation.
Taking this into account,  Oren and Wolfe conclude
with 98 per cent  confidence that
all such systems have Faraday
rotation and substantial magnetic fields.

Another important window on the history of the Galactic magnetic fields over cosmic time 
is provided by analyzing the chemical composition of the
atmospheres of the oldest stars in the Galactic halo (Zweibel 2003).
As a result of observations from the Hubble space
telescope the chemical abundances of these very
early stars have now been analyzed.   The light elements
lithium, beryllium, and boron
have been found in even the oldest stars, e.g.   those
with $ 10^{ -3 }  $ times solar abundances.  In addition,
the amount of beryllium and boron in them increases
with the iron abundance and in fact is directly 
proportional to it.  Since the early stars are produced
from the interstellar medium their composition should
reflect that of the interstellar medium. (Primas et al. 1999,
Duncan et al. 1998, Garcia-Lopez et al. 1998, 1999)

Now, it is known that no beryllium 
was produced in Big Bang nucleosynthesis, and that 
it is very difficult to make it  in stars, since it
quickly burns up.  The leading theory is that it was
made by cosmic ray nucleosynthesis,
(Meneguzzi, Audouze \& Reeves, 1971, Reeves,  1994,  2007). The situation for boron is ambiguous, because this element
can also be produced by neutrinos during Type II supernova explosions; see Ramaty et al. 1997. We will still include
boron in the arguments for cosmic rays and magnetic fields early in the history of the Galaxy, but this caveat should
be kept in mind.

The linearity of the Be and B abundances with iron
is explained by assuming their creation is by
spallation of the low energy (tens of MeV) carbon
and oxygen cosmic rays.
If it were due to the complementary process,
spallation breakup of interstellar carbon and
oxygen nuclei, one must take into account that the
latter themselves are produced by stellar
nucleosynthesis and supernovae. and their
abundance  should increase proportionally
 to the amount of iron
produced in supernovae.  Thus, their abundance
should increase with that of iron in the interstellar medium,
and the  abundance  of these light elements
should be quadratic with iron abundance or in time. 

However, to get linearity in time with spallation
of carbon and oxygen cosmic rays, one needs to assume
that the composition of cosmic rays must not change
with time.  This would seem to be a stumbling block
since cosmic rays are assumed to be accelerated
by  shocks from  interstellar--medium nuclei.
Therefore, they would be    expected to also reflect the changing abundances
in the interstellar medium and would be
expected to  increase their abundance in
carbon and oxygen with time.

 Ramaty and others (Ramaty, Lingenfelter, Kozlovsky (2000),
Ramaty,  Scully,  Lingenfelter, 2000a,
 Lingenfelter, Higdon, Ramaty, 2000b, Ramaty, Tatischeff et al. 2000)
  argue that the acceleration of cosmic rays
occurs mainly inside  superbubbles, and that the material
inside these superbubbles is the material that has
just emerged from the supernova generating
 the superbubble.  This freshly produced matter has
not been diluted with the rest of interstellar medium
in the superbubble region where  cosmic ray acceleration occurs.
Thus,
the relative  abundance of different cosmic ray nuclei
should be  constant in
time and determined by the undiluted material emerging from
supernovae.   On the other hand, iron in
 the interstellar material from which stars emerge
{\it  has}  been diluted, since it was produced in supernovae.
 Therefore, its 
abundance relative its solar abundance
increases with time at a constant rate determined
by the rate of supernova explosions.

This constancy of the chemical abundance of   cosmic rays
has been supported
by detailed numerical simulations of Ramaty and others,
and does lead to an explanation
of the observations  that the abundance
of beryllium and boron is directly proportional to
the abundance of iron in old stars (Garcia-Lopez et al. 1998, 1999)

Now to explain the numerical value of this ratio,
 the cosmic ray intensity at tens to hundreds of MeV per nucleon
in the early Galaxy had to be  the same then as now. Zweibel (2003)
 has shown that magnetic fields several orders of magnitude weaker
than now,  suffice for cosmic ray acceleration and diffusive propagation 
at these energies.    On the other hand, if the field
is very weak such a cosmic ray intensity might  not be
confined by a very weak field. This is because, if most of the mass in the
interstellar medium was in the form of discrete, cold clouds, as 
appears to be the case today, then a high cosmic ray pressure
between the clouds, which must be  confined by magnetic tension,
 would
inflate the field lines to infinity (Parker 1979).

 This can be quantified by a simple two dimensional model.
Let  the cloud distribution be two dimensional and have a  scale height $H$ 
and the clouds be on planes separated
by a mean distance $\ell $.  Let the cosmic ray pressure be proportional
to the magnetic field strength squared by a factor $ \beta/\alpha $.
Then the model shows that the lines bow out above their
average height in
the clouds by a factor 
\[
 \frac{1}{\cos {\sqrt{(\beta/\alpha + 1)}\ell/H } }
\] 
Thus, if the field is very weak,  $ \beta / \alpha $ is very large
and there is no solution,  implying the field lines would bow out to infinity.
 
Although the amount of cold interstellar material in the early 
Galaxy was doubtless lower than today
because of the lower metallicity,
thereby somewhat easing the cosmic ray confinement problem, we are
 faced with a situation in which primeval galaxies already 
must have had substantial 
magnetic fields, at a stage 
 too early to have been produced by a conventional dynamo.

 Finally,  the problem
of the abundance  of the Li$^6$ isotope
which is not at all  linear with that of  iron
is entirely up in the air.
 (Reeves 2007, Ramaty, Tatischeff et al. 2000).
   Li$^6$ also is not  produced by 
Big Bang nucleosynthesis and, because it depletes so easily,
 can only be produced  with great difficulty in stars.
Thus, as long as we do not understand
the process that produces lithium 6,   we cannot
be sure that this process does not in some
way produce beryllium and boron as well, without the aid
of cosmic rays.  Until the Lithium problem  is resolved, we cannot
be certain that the beryllium boron argument
 proves that the origin  of   Galactic field is
pregalactic. 

It is interesting that boron has been observed in
other galaxies at large red shifts (Prochaska et al 2003).
Thus,  if the ideas concerning  the origin  of boron hold
up, this provides  even stronger evidence that 
magnetic fields are already  present at the
formation time of galactic discs and their origin 
is pregalactic.

\section{Summary of our present understanding of
cosmic magnetic field origins}

 Origin  theories divide into two further parts: (1) the origin
in small bodies such as planets and (2) the origin  in larger bodies
such as galaxies.  This division occurs because,
for a large body such as a galaxy, 
the  resistive decay lifetime of its magnetic fields is much longer
than the age of the universe.    For these bodies a so-called fast dynamo 
is required.  On the other hand in 
 small bodies the decay lifetime is much shorter and the required
dynamo is called a slow dynamo.  
For the Earth the problem of  where  the field 
came from,  and  how it is sustained against resistive decay
is easier than the same problem for fields in large bodies
(Parker 1979, Spitzer 1978).
If a magnetic field starts to decay, the inductance of the body
in which it resides produces a backvoltage (or E.M.F.) that
keeps its  currents flowing against its  resistance.  
Thus, one can roughly say that the lifetime of the magnetic
field against decay (if no other mechanisms are present)
is its total inductance divided by its total resistance.  
 Inductance is proportional to the size of a
 body, $ L $ while 
 resistance is inversely proportional
to $ L $ so that the lifetime is proportional to $ L^2 $.
Thus, the Earth has a relatively short magnetic decay
time of order a few tens of thousands of years, while
that of the Galactic disc is many orders of 
magnitude longer than  the age of the universe.
Thus, dynamo  mechanisms aside, the Earth's field would decay
away in a time very short compared to its age, and
therefore, there must be a mechanism to sustain it, 
just as   to sustain a current in a laboratory
circuit, one needs a battery or dynamo (Parker 1955, 
Elsasser 1946, 1950.)  

On the other hand, while one need not worry about the
galactic field decaying because of its enormous inductance
(Fermi 1949),
one has to worry about how to get the currents started 
in the interstellar medium to produce  the galactic field: 
as the currents rise the back voltage is so large
that a very strong generator is needed to balance the
back voltage.  This turns out to be the essence of
the problem behind the origin  of galactic fields,
(Hoyle, 1958).

This  discussion  of magnetic field generation  and decay
is not quite correct.  It treats the bodies as being static
and at rest, while in both cases the bodies are either
fluid or gaseous with motions generated by their dynamics.
When a fluid moves across a magnetic field, an electric
field 
$ {\bf E} = - {\bf  v} \times {\bf B}/c $ 
exists in the frame   in which the velocity is measured, and this
electric field results in the dynamo action 
that is needed to balance 
the magnetic field against resistive decay in case of the Earth,
and balance  inductance in the case of the Galaxy.
The problem is to find a reasonable fluid velocity 
that would properly balance the inductive and resistive
effects that must occur during the
evolution of the field (its decay or  growth).

In 1955 Parker  was able to find  such velocities and  to propose a model
containing, non axisymmetric motions, that explained 
how the Earth's field could be sustained against decay (Parker 1955). 
To do this,  he built  on the work of many others 
(Elsasser, 1946, 1950).
Parker  found that a  dynamo should exist in the liquid
core of the Earth and the fluid motions
producing this dynamo action are 
 a combination of differential rotation of the
core, and a multitude of rising and falling  convective cells,
that are twisted by the Coriolis force of the Earth's
rotation.

His solution was gradually  improved  in the next decade
(Backus, 1958),  until finally, in 1966,
  Steenbeck, Krause, and R\"{a}dler (1966) developed  a refined theory
for dynamos, the  {\it mean--field theory}, 
which  consist of such turbulent motions

Once this theory was accepted as correct, 
it was  applied to the problem of the origin  of 
the Galactic magnetic field. 
Parker in the United States (Parker, 1971a)
and Vainshtein and
Ruzmaikin in Russia (Vainshtein \& Ruzmaikin, 1972)
 in the early seventies  showed
that motions, similar to the terrestrial motions,
exist in the Galaxy, and that  they could overcome
the inductance problem and generate the Galactic 
field.

Although the dynamo mechanism appears   viable, it
still does  not resolve the primary origin question:
was  the galactic field actually produced, from an initially
weak field,   during the galactic  lifetime, 
or was  it generated earlier, before the
galaxy formed?  That is, for  the latter case,  was a fairly strong field 
already present  at the onset of formation
of the galactic disc, and then, its final form determined by this
dynamo action? (See Kulsrud, 1999).

In the last analysis, this
 question can only be answered  by observations. 
 As pointed out above,  direct observations of the early magnetic fields 
are very difficult. At the present moment there are 
 two  ways to
make progress on  the origin  problem:  First,  a careful theoretical analysis
 of the proposed dynamo model and  its assumed velocities; 
second, an analysis of the effects of an early 
magnetic field on independent  astrophysical problems such
as  star formation and cosmic ray properties, discussed
in \S 2.
 If the field changed
drastically from being very weak to its present value,
these properties, which {\it are} subject to
direct observations, should also change.  
 The
stellar luminosity function would be expected to
change from its early form  to its present form.
Also, the properties of cosmic rays should drastically
change.  Their  energy density should  either increase from
an initial small  value  to its present value, or should remain
constant.  Its  variation  can be
inferred  from  the measurement  of  the abundances
 of the isotopes of   elements that can only be produced by cosmic ray 
spallation, as discussed in the previous section.

In the seventies, when the galactic dynamo origin
of magnetic fields was first proposed,
 the velocities needed for the  dynamo 
 were not  well known 
 and the  estimates made of their magnitudes led to too
weak a dynamo to be able to amplify the strength of seed fields to 
 the current strength  magnetic field, 
over the finite age of the Galaxy.   More precise estimates
were made in  a series of papers by
 Ferri\`{e}re in the nineties (Ferri\`{e}re, 1992a,b 1993a,b,
1996, 1998, Ferri\`{e}re  and Schmitt, 2000).
In these papers  she   pointed out the importance
of the expansional  motions of random superbubbles; [giant
bubbles produced by multiple supernovae that arise
in stellar clusters (McCray and Snow 1979)].

  These motions last long enough, and  are sufficiently fast
for the necessary amplification of the fields. However,
they work only if significant magnetic flux 
can be  expelled from the Galaxy.  This is possible  only
 if the magnetic field lines can separate from
the heavy plasma and evaporate away leaving the plasma
in the superbubble to fall back into the Galactic disc.
But arguments that this
is possible assume that   the magnetic field 
is  already fairly strong (Brandenburg et al, 1995, Moss et al 1998).
Thus, the field strengths they require   would be inadequate 
to make the dynamo work in the early stages of
amplification if  the initial  field at the beginning of the Galaxy
were a weak  seed field. 

    To be sure the dynamo could work during early galactic 
times, one needs to consider
the possibility that a field of considerable strength 
and of  pregalactic origin,  existed
prior to the  galactic disc.
If the pregalactic  field  is strong enough to solve the  problems
associated with the galactic dynamo theory,  then one
may have a double origin. First,  a pregalactic magnetic field 
of moderate strength must develop, 
   and then  a subsequent amplification of it
to its present strength  and configuration 
be produced by the conventional galactic dynamo.

 This pregalactic field could arise as a natural part of the galactic 
formation process.
For example, it is likely that
 the formation of the first massive stars preceded the formation 
of galactic disks,
 occurring instead in the $\sim 10^6 M_{\odot}$ halos that appear 
to be the first bound structures formed (Tegmark et al 1997,
Gilmore et al. 2006).
Any magnetic field present in the ambient gas or in the supernova
 ejecta would have been amplified by turbulence in the halo gas.
Thus, the disks that formed as the halos merged, could have had 
moderately strong magnetic fields from the outset.
Turbulent amplification can  also occur during the 
gravitational formation of larger structures where
very strong large scale turbulent velocities are present
(Ryu et al 1993, Kang et al. 1994).

 These velocities are sufficiently large
that sufficiently strong  magnetic fields can be generated to provide
a pregalactic  origin (Kulsrud \& Anderson 1992).
 If these fields are formed on the 
scale of the largest turbulent eddies which drive them, this would
provide the coherence needed for a 
pregalactic  origin  of the
galactic field.  
However, the theory of these random turbulent fields is still under
development and a number of tricky plasma problems
have arisen involving them.
(See \S 14.) 
Simulations of turbulent amplification show a pileup
 of magnetic power at the resistive scale
in situations where the resistive scale is much smaller
 than the viscous scale (Schekochihin \& Cowley,  2004).
Even smaller scale fluctuations may arise in the collisionless
 case, because of anisotropic plasma pressure
(Schekochihin et al 2005b).
 It appears that the  fate of a primordial
theory based on the protogalactic physics may hinge on the resolutions
of these plasma problems.  
The question of generation of significant 
fields at earlier epochs in the universe than that
of galactic disc  formation are still being debated.

 It has also been proposed that
the pregalactic fields could arise from the fields observed to
be present in giant radio jets, (Daly \& Loeb, 1990,
  Colgate \& Li 2000, Kronberg et al. 2001,
Furlanetto \& Loeb, 2001).
 So far, none of these proposals has been 
 developed to a sufficient extent to judge whether
they lead to a viable solution.

\section{ Basic equations for magnetic field evolution }

Nearly all theories of magnetic field origin  rely
on the magnetohydrodynamics (MHD) description of plasma physics.
It turns out that the larger the system the better these
equations describe magnetic field evolution. 
 These equations  essentially parallel the 
ordinary electromagnetic equations.  One
equation describes how  the motion of a  
plasma affects the magnetic field, and the other
describes how the magnetic field affects the motion
of the plasma.  Although these equations
are familiar,  we sketch their derivation, since this
derivation demonstrates the accuracy of these equations
for our purpose.

We start first with the effect  the plasma motion 
 has on the magnetic field 
(Cowling 1953). This
effect results solely from the fact that in the frame
of the moving plasma the electric field $ {\bf E'} $,
  to all  intents and purposes, must balance the resistive
term,  so that we
have  
\begin{equation}\label{eq:1}  
{\bf E'} = {\bf E}  + \frac{{\bf  v} \times {\bf B} }{c}=
\eta {\bf  j} ,
\end{equation} 
where  $ {\bf  v} $ is the plasma motion, the left hand side 
is the electric field in the frame of the plasma, $ \eta $
is the resistivity  and $ {\bf  j} $ is the plasma current
density.
The right hand side is essentially zero
since  from Ampere's law the current density is inversely
proportional to $ 1/L $, and on galactic scales is usually
very small,  while the resistivity  is the same as that in
a laboratory plasma.  Combining this with Faraday's law   
$ \partial {\bf B}/\partial t = - c \nabla \times {\bf E} $,
and Ampere's law $ \nabla \times {\bf B} = 4 \pi {\bf  j}/c $,
we get 
\begin{equation}\label{eq:2} 
\frac{\partial {\bf B} }{\partial t}=
\nabla \times ({\bf  v} \times {\bf B})  +
\frac{\eta c^2}{4 \pi} \nabla^2 {\bf B} .
\end{equation}
We  drop the displacement term from Ampere's law 
and, for simplicity take $ \eta $ to be a constant,
to get the last term.  

The neglect of the resistivity  term in galactic physics
becomes obvious if we approximate $ \nabla^2 {\bf B} $ as
$ {\bf B}/L^2 $ and set the last term equal to
$ {\bf B}/T_{decay} $.  For a plasma at a temperature
of $ 10^{ 4} $ degrees Kelvin, the resistive term
$ \eta c^2/4 \pi \approx 10^{ 7} \mbox{cm}^2/sec $ so
\begin{equation}\label{eq:3} 
T_{decay} \approx 10^{ 26} L_{ps}^2 \mbox{  years }  ,
\end{equation}
where $ L_{ps} $ is the scale size in parsecs.
Since the galactic scale is many parsecs, the incredible
smallness of the resistive  term in equation (2) is
evident.  Even if the magnetic field is extremely tangled,
so that its length scale is reduced to a fraction of
a parsec, the resistivity  term is still very small.
(For $ L > 10^{ 12} $ cm the effective decay time is
still longer than the current Hubble time.)

Thus, in almost all cases, it is permissible to drop
this term, and   this is valid to an extraordinary
degree.  There are exceptions to this in cases
where extremely thin current layers form.  Also
there are additional, very weak terms in Ohm's law that we  come
to later, which  can play a role in producing seed fields.
An example of these terms is  the so called Biermann battery.

For the bulk of the theories of magnetic field origin
the {\it ideal} magnetic differential equation
\begin{equation} \label{eq:4} 
\frac{\partial {\bf B} }{\partial t}=
\nabla \times ({\bf  v} \times {\bf B})  
\end{equation}
is sufficiently correct.

The ideal equation  has an important 
implication for cosmic magnetic field origin theories.
It implies that magnetic field  lines have
a physical reality beyond their mathematical description
for a magnetic field.  (Their mathematical definition
is: that their signed direction gives the signed direction
of the magnetic field and their density can be assigned
so as to give the magnetic 
field strength.)

Their reality consists of the fact
 that the magnetic field lines
cannot be created or destroyed once they are embedded
in the large scale plasma.  They are bodily carried
by the motion of the plasma in which they are embedded,
and  continue to properly represent the
magnetic field.  The argument that flux lines have a physical reality is
standard,  and is called flux freezing.  It is 
discussed  in many textbook on plasma physics
(see,  for example,  Newcomb 1958,
 Kulsrud 1964, Alfv\'{e}n \& Falth\"{a}mmer, 1963, Moffatt, 1978,
 Parker 1979, Kulsrud 2005).

The  mechanism for the
creation and  destruction of field lines is the resistivity,   which is
negligibly small.  But, by its very definition,
an origin theory starts off with a very weak field and
produces a strong field.  Because of flux freezing, this
 has to be done without
 changing the number of lines of force.  

It is a fact that stretching a tube of force by 
motions that preserve its volume,  
does increase the local field strength.
This stretching  occurs in a finite volume  by doubling the
line back and forth as in Figure 3 of \S 14, so that the mean 
value of the field strength, taken over the volume, does not change 
and, thus, does not lead to an increase in the coherent field.

It turns out that  standard dynamo theories actually work this way.
They fold lines of force back and forth increasing the 
pointwise field strength. The dynamo works
in such a way that
the field lines pointing in 
one direction are  near the center of the plasma volume
and the field pointing in the reverse  direction are
concentrated on the  outside.  Thus, 
the magnetic  field becomes coherent over roughly the inner half
of the volume with flux of one sign, and coherent over
the outer  half with  flux of the other sign. 
   In \S  6, we  show  how this remarkable process
happens when we discuss
the conventional $ \alpha - \Omega $ theory for the 
exponential growth of a magnetic field in the galactic disc.
However,   to
 actually
produce a strong magnetic field of constant sign over 
an  entire volume, the magnetic flux 
 in the outer region must be expelled.

A leading question then becomes whether  the flux can  actually
be expelled to complete the galactic dynamo action.
If one were content with a field of reversing sign, constant
over one half the volume, but with zero  total net flux,
one could avoid this expulsion problem.  However, the observed magnetic
field in our Galaxy  appears to have net flux,
so that  such a solution to  the origin problem for our
Galaxy is not available.  The
 same difficulty occurs in  other galaxies as well.

Moreover, there is an issue of scale. The typical 
 strength of seed fields at the onset of the disc
is  12 - 14 orders
of magnitude below present day galactic fields. 
In order to bring them to their present strength, the field lines
have to be stretched by 12 - 14 orders of magnitude as well.
  This would seem to imply that the field lines
 fold  down to length scales that are also smaller by 
 12 - 14 orders
of magnitude or, without any flux expulsion, down to the  resistive scale.
But by invoking flux expulsion at every doubling
of the flux  one can avoid this problem.  In fact, only tiny pieces,
of field lines $ \sim 10^{ -12}-10^{ -14} $ of the total initial line length 
are  stretched by this  large 
order of magnitude while  the rest of the pieces of the
line are  removed 
from the disc.

The other half of the MHD picture of how magnetic
fields affect plasma motions is of less interest in the
origin  problem since the magnetic fields we consider are too weak
to affect the plasma motions.  Their effect is the introduction
of a magnetic force term 
\begin{equation} \label{eq:5} 
 \frac{{\bf  j} \times {\bf B}}{c} 
\end{equation}
into the hydrodynamic equation. As remarked in \S 1, 
  we do not consider the
saturation problem in this review since it clearly does
not arise unless a successful build up of the field to
large strength is not achieved.

\section{ Cowling's theorem and    Parker's dynamo}

We have remarked that the motion
of a plasma acts like a dynamo in that, as a plasma
moves perpendicular to a magnetic field, a
substantial  electric
field, $ -{\bf  v} \times {\bf B}/c $,  arises in the laboratory frame.
This was first  pointed out  by Larmor (1919).  In fact, this
is just how a dynamo in a power station generates electricity.
Larmor expressed the hope  that velocities could be found
which  could generate
astrophysical magnetic fields.  The first attempts to
create magnetic fields  in this way in spheroidal bodies naturally assumed
that the velocities and fields were axisymmetric.

Axisymmetric fields are generally broken into 
poloidal components,  whose lines of force lie in
meridional planes,  and toroidal components, whose
lines of force form circles about the axis.  Imagine starting
with a pure poloidal field with a zero toroidal component,
and imagine that the plasma is differentially rotating
with a rotation rate that varies along a poloidal
field line.  It is clear that this rotation will drag
different parts of the line in the toroidal direction at
different rates, so the line will not stay poloidal but
will develop a toroidal component.  If this continues,
the toroidal component will grow stronger and stronger
at a linear rate in time.  

From this it is easy to see that a toroidal magnetic 
field is easy to generate from a poloidal field by
differential rotation, (see Kulsrud 2005). 
 On the other hand,
if one starts with a pure toroidal field, it is impossible
to produce a poloidal component.  Rotation will
have no effect on the field and any symmetric poloidal 
motion will just move the toroidal field lines around
 leaving them as circles, although the circles will change
their position. 

 Cowling (1934)  showed that this is the general
situation:  if one starts with a magnetic field that
has both axisymmetric toroidal and poloidal components,
then axisymmetric motions can increase the toroidal
flux, but they will just leave the total poloidal flux unchanged.
This result is known as Cowling's theorem and basically
defeats any attempt to exponentially amplify a very weak field 
to a strong one by axisymmetric motions.  (The original
purpose of the theorem was show it was impossible sustain
the Earth's field against its relatively rapid 
resistive decay. However,  it applies equally to any attempt
to produce a galactic dynamo that starts with a weak axisymmetric
  coherent field,
 and tries to increase
it  to a coherent field of order $ 10^{ -6} $ gauss
by axisymmetric motions.)
To increase the toroidal magnetic field linearly by purely
differential rotation would
take $ 10^{ 14 } $ differential turns, while the
Galaxy has only rotated about fifty times.  
(It should be remarked that it is possible to
amplify a non axisymmetric magnetic field by symmetric
velocities, but this does not seem to have been attempted for
a galactic dynamo.)

Thus, to exponentially increase the flux of an initial
small seed field to the present Galactic value, or to sustain
the Earth's field against decay,
  it is necessary to deal with nonaxisymmetric 
motions.
This would seem to require complicated numerical simulations,
which  were beyond the power of computers of the 1950's.  

However, as previously mentioned, 
 Parker (1955) did come up with 
a physically correct model  containing non axisymmetric motions
that  could  successfully, either 
amplify  the Earth's field, or  
sustain it against resistive decay.  
He considered random small convection cells of rising and falling
fluid in the Earth's liquid core.  Imagine, as
above, that there is  a purely toroidal magnetic field inside the Earth's
core.  Then it is true that by itself a rising convection cell
would distort this field, but only in a  plane
not producing any net poloidal component.  However,
due to the Coriolis force of the Earth's rotation
the distortion of the toroidal field would be twisted into
the poloidal plane and a poloidal component would emerge.
  If one only had this convection, the poloidal 
field would not grow significantly.  However,  when combined
with the  differential rotation of the Earth's core,
the poloidal distortion would produce a change in
the initial toroidal component of the field.
(This differential rotation in  the Earth's core
{\it also} results from Coriolis forces.) 

 It turns out
that the change in the toroidal field is actually
an increase.
  Thus, one has the possibility of exponential 
growth.  In the lifetime of a convection cell
the poloidal component  increases by an amount
proportional to the toroidal component,   and the toroidal 
component increases by an amount 
proportional to the poloidal component.
Since these two increases are coherent, both 
components grow exponentially.  Parker showed that all
the signs are consistent with growth.   The
 downward motions produce
changes with the same sign as the upward motions, and
the same changes apply to both the northern and
southern hemispheres.  Thus,  all of the random convection
cells act to increase the mean magnetic  field of the earth
in the same direction.

In detail, a rising   convective cell rotates in the opposite
direction to the Earth  due to the Coriolis force.
This is because it laterally expands increasing its moment of inertia.
This leads to a twist in a distorted toroidal field 
line with the upper part of the line gaining a northward
component and the lower part a southward component.
The Earth  rotates slower at larger distance from its axis, so
the differential rotation acting on  this new poloidal 
field loop produces an additional toroidal field in
the same direction as the original field reinforcing it.
A sinking convection cell contracts and rotates oppositely,
yielding a loop with the same sense as the rising cell,
and also reinforcing the toroidal component. 

The cells in the southern hemisphere act the same,
reinforcing the toroidal field in the southern hemisphere
which is in the opposite direction to that in the northern
hemisphere.  Thus, because the sign of the Coriolis
force is properly correlated with the differential rotation
of the Earth's core, the convective cells all lead to
a growth in the toroidal field,  or  compensate
for the resistive decay and lead to a non decaying terrestrial
magnetic field.

This theory was intuitively satisfactory, and was also 
subsequently supported by the  more rigorous analysis
of Backus (1958),  and by later numerical simulations
(Glatzmaier \& Roberts 1995, Ogden et al 2006).
It can most easily be illustrated, in the galactic case,
by the example of
a supernova exploding in a rotating galaxy.  See figure 1,
(Kulsrud 2000).
The convection cell is illustrated by the exploding
remnant that is forced to rotate backward by a Coriolis
force induced by Galactic rotation.
  The role of differential rotation is also supplied
by  Galactic rotation.  In this Galactic  case, the lines
  need to be 
expelled to infinity, in contrast to the terrestrial
case where they  need merely   be 
expelled into the nonconducting mantle and effectively
 disappear.

However, to  successfully apply Parker's theory
to the Galactic magnetic field problem a more systematic
treatment of turbulent amplification of magnetic fields
is required. 
This was provided by the mean field dynamo theory 
of Steenbeck, Krause and R\"{a}dler (1966).  
They treat the turbulent motions in a more general 
way than  Parker treated the Earth, replacing 
the convective cells by a quantity they called
kinetic helicity, $  {\bf  v} \cdot \nabla \times {\bf  v} 
$.  They denoted its effect on magnetic fields by $ \alpha $ where 
\begin{equation}\label{eq:6} 
\alpha = -\frac{\tau }{3} <{\bf  v} \cdot \nabla \times {\bf  v} > ,
\end{equation}
where the angular brackets denote turbulent ensemble averages and
 $ \tau $ is the decorrelation time of the turbulent motions.
(To obtain  a scalar $ \alpha $,  isotropic turbulence must be  assumed.)
Note that helicity has the same sign of twist relative to the
motion $ {\bf  v} $ as do Parker's convective cells,
and serves the same function in amplifying the
field on which they act. ($ \tau  \nabla \times {\bf  v} $) 
corresponds to  the angular twist of the convective cell
during a single decorrelation time.  

Because the normal resistivity  is usually very
small, they also introduce a turbulent resistivity  
which they called $ \beta $ where
\begin{equation} \label{eq:7} 
\beta =\frac{\tau }{2} <{\bf  v}^2> .
 \end{equation}
This turbulent resistivity  does not act the same
as normal resistivity  in that it produces no
dissipation or change in field line connectivity. 
It is an effective mixing term 
produced by 
random turbulent motions,  and smooths out the
fields on a larger scale 
than the turbulent motions. It can be justified as follows:
$ \tau $ {\bf  v} is the displacement $ \Delta {\bf  r}  $,
so that $ \beta \approx (\Delta {\bf  r} )^2 /\tau 
$ is the expression for a random walk.

\section{The alpha-Omega disc dynamo}

The mean field dynamo theory introduced by  Steenbeck et al (1966)
was a very important step towards applying dynamo action
in many contexts: the Earth, the Sun, stars, galaxies,
and clusters of galaxies.  We  present it in
the context of our  Galactic disc, 
which is  very thin.  This allows us to treat the dynamo  locally
in radius and angle and makes it 
 one dimensional with the significant  space coordinate 
 perpendicular to the disc.  This is consistent
with the general Galactic field, and
 is the case most relevant to the Galactic dynamo 
 (Ruzmaikin, Shukurov, \& Sokoloff 1988).

We start with a very weak field such that the motions
can be considered to be independent of the magnetic field,
a limit  called the {\it kinematic limit}. Thus, one 
is only concerned with the  magnetic differential equation (2).
Because resistivity  is so small, we can start with the
ideal equation (4).  In this equation we split the 
velocity $ {\bf  v} $ into two parts,  a part associated
with the turbulent random motions $ \delta {\bf  v} $, and
a part $ {\bf U} $ which is smooth and coherent.  In the Galactic case
$ {\bf  U} $  is the differential rotation of the Galaxy.  Thus,
\begin{equation}\label{eq:8} 
{\bf  v} = {\bf U} + \delta {\bf  v} .
\end{equation}

 The magnetic field is also broken up into its mean
part $ \bar{{\bf B} } $ and its random part
$ \delta {\bf B} $, so 
\begin{equation}\label{eq:9 }
{\bf B}=\bar{{\bf B} } +\delta {\bf B} .
\end{equation}  
Substituting these into  equation (4)
and ensemble averaging over the turbulence we get,
\begin{equation}\label{eq:10}
\frac{\partial {\bf \bar{B}}}{\partial t }=
\nabla \times ({\bf U} \times \bar{{\bf B} }) +
\nabla \times \left( \left<(\delta {\bf  v} \times \delta {\bf B} \right>
\right) .
\end{equation}

Thus, an extra term associated with
the random velocities has been added to the magnetic differential 
equation for the smooth field.  This term can break
the ideal flux constraint for the mean field $ \bar{{\bf B} }$,
but not for the true field $ {\bf B} $.

To complete the process of obtaining a dynamo equation,
one has to solve a magnetic differential equation for
$ \delta {\bf B} $ in terms of $ \delta {\bf  v} $.
This is carried out  by what is known as a quasilinear expansion
(Sagdeev and Galeev 1969),
 which is presented  in many places and will not be given
here [see Ruzmaikin, Shukurov, \& Sokoloff, 1988;
Parker, 1979; Krause \& R\"{a}dler, 1980;
 Moffatt 1978; and  Kulsrud 2005).

 The result for the mean induction generated by the 
fluctuating fields can be written
\begin{equation}\label{eq:11}
\left<\delta {\bf  v}  \times \delta {\bf B} \right> =
\alpha \bar{{\bf   B}} - \beta \nabla\times \bar{{\bf B} }.
\end{equation}
[In general, $\alpha$ and $\beta$ are tensors, but here we
 understand them to be scalars and defined through equations (6) and
(7)].
Substituting this result into the above equation gives the 
famous mean field dynamo equation
\begin{equation}\label{eq:12}
\frac{\partial \bar{{\bf B} } }{\partial t}=
\nabla \times ({\bf U} \times \bar{ {\bf B}} ) +
\nabla  \times (\alpha \bar{{\bf B} }   )+
\beta \nabla^2 \bar{{\bf B} }. 
\end{equation}

Applying this equation to a local part of 
 the Galactic disc, in which we introduce
cylindrical coordinates $ r, \theta, z, $ and keeping only
derivatives in the thin vertical direction, $ z $, we get
\begin{eqnarray} \label{eq:13} 
\frac{\partial B_r}{\partial t} & = &  
-\frac{\partial }{\partial z} (\alpha B_{\theta} )
+ \beta \frac{\partial^2 B_r }{\partial z^2}   ,
\nonumber\\  
\frac{\partial B_{\theta} }{\partial t} & = & - \Omega B_r 
+ \beta \frac{\partial^2 B_{\theta}  }{\partial z^2} .   
\end{eqnarray} 
 (We have dropped the bars; the magnetic field $ {\bf B} $
is understood to be only the mean field.)

 We have  substituted the Galactic rotation  
$r  \Omega \hat{\theta } $ for
 ${\bf U}  $,  and used the fact that
$ r \Omega $ is a constant to find its derivative $ d \Omega /dr
= -\Omega/r $. 
We have also dropped the term $\partial_z(\alpha B_r)$ in the
 equation for $B_{\theta}$, which is small 
relative to the differential rotation term.

The usual procedure is to look for a growing mode, proportional
to $ e^{\gamma t} $ and to solve equations (13) as an eigenvalue
problem.  The boundary conditions that are customarily invoked
are that the disc is confined to a region $ -h <z <h $. 
On the presumption
that the diffusion coefficient, $ \beta $,  is very large
outside of the disc, the magnetic field is taken to
be zero for $ |z|>h $,  and therefore the boundary conditions
are:  $ B_r  $ and $ B_{\theta} $
 are zero at $ |z| = h $.  These conditions 
are termed the {\it vacuum boundary conditions}. 
As we will    show later (in \S 9),  
   the assumption of these conditions is
  the  critically important  point in
the  disc dynamo origin for the Galactic field.  As discussed
 in \S 3, in order to preserve flux conservation,
 flux of one sign 
 must be removed from the disc.  This is
exactly what these boundary conditions accomplish.  However,
the weak point in the Galactic dynamo theory 
is the assumption that $ \beta $ is large outside
the disc.  This too is discussed in \S 9.

Although the assumption that $ \beta $ is constant is non
objectionable,  $ \alpha $ cannot be constant
 since it has to be of
opposite sign above and below the disc. This can be seen
physically if one interprets $ \alpha $ as the effect
of twisting convection cells.  

In fact,  consider
a cell rising above  the midplane of the disc,
that is $ z $ and $ v_z >0$.  As it rises it
expands,  and the Coriolis force of galactic rotation
rotates it backward relative to normal Galactic rotation. 
(This follow easily  if one considers that the moment
of inertia of the convective cell is increasing so its
angular rotation relative to a fixed frame must decrease
to preserve its angular momentum.)  The same backward
rotation applies for a descending cell below the
galactic midplane.  However, for this cell $ v_z< 0 $,
and  $\alpha \sim -{\bf  v} \cdot (\nabla \times {\bf  v}) $, the helicity,
 reverses sign.  Inspection of equation (6)
of the last section  shows that,
since  Galactic rotation is positive around $ \hat{z}$, 
 helicity is negative above the midplane and $ \alpha $
is positive.  

Given the uncertainty in the properties of interstellar 
turbulence, it is difficult to give an explicit form 
 for $ \alpha $, so various forms have been  tried in order to
explicitly solve  the eigenvalue equations. 
The simplest form $ \alpha = \alpha_0 z/h $, is probably
as good as any.  

To  solve the eigenvalue equations (13), they are cast into
 dimensionless form.  
\begin{eqnarray} \label{eq:14} 
z' & = &  z/h, \quad t'= \beta t/h^2, \quad \gamma' = \gamma h^2/\beta ,
\nonumber\\ 
B_{\theta} & = & B'(\beta/h\alpha_0), \quad  B_r  =  B'_r(\beta/h\alpha_0) . 
\end{eqnarray}
so that the dimensionless eigenvalue equations become
\begin{eqnarray}\label{eq:15} 
\gamma' B_r = -\frac{\partial }{\partial z'}(z' B') ,
+ \frac{\partial^2 B_r}{\partial z^{'2}} ,
\nonumber\\ 
\gamma'  B' = D B_r + \frac{\partial^2 B'}{\partial z^{'2}} 
\end{eqnarray} 
where the dimensionless number 
\begin{equation}\label{eq:16} 
D= -\frac{\Omega \alpha_0 h^3}{\beta^2} ,
\end{equation}
determines the eigenvalue $ \gamma' $.  (Since $ \Omega, 
h , \alpha_0$ and $ \beta $ are positive, 
$ D $  is negative.)

$ D $ is known as the {\it  dynamo number}. 
It represents competition between the growth terms
 (shear and the $\alpha$ effect), and diffusion (the $\beta$ effect) and
determines whether
$ \gamma $ is real and positive.  There is a negative
critical value $ D_c $ such that if $ D<D_c $, then 
$ \gamma $ is positive and the magnetic field will grow
 exponentially.  The dimensional growth time, if $ \gamma'
= 1 $, is $ h^2/\beta  $. 

Following Parker or Ruzmaikin, Shukurov and Sokoloff,
  one obtains an approximate idea
of the size of $ \beta $ in the interstellar medium,
by taking the random turbulent 
velocity $ \delta v \approx 10 \mbox{km/sec } $, 
and the correlation length $ \delta v \tau \approx 100 $ pc.
This  gives $ \beta \approx 1.5 \times 10^{ 26} \mbox{ cm}^2$/sec.
A possible value for $ h $ is 300 pc. This gives for the
dimensionless growth time $ 1.5 \times 10^{ 16} \mbox{sec }   
= 5 \times  10^{ 8} $ years. That is to say, if the eigenvalue
problem leads to $ \gamma' = 1  $,  an initial seed field 
will exponentiate every 500 million years, and there could
be roughly twenty  efoldings during the age of the Galactic disc.
This would raise an initial field by a factor of $ 10^{ 8 } $ 
from say a field of $ 10^{ -14} $ gauss to $ 10^{ -6} $  
gauss. 

It should be noted that in the Galaxy today, while 300 pc 
is probably a reasonable scale height for $\alpha$, the
observed magnetic scale height appears to be  1500 pc.
  However, since the field is anchored in the thinner 
interstellar mass, 300 pc is the better value. 
 The field at 1500 pc. is confined by magnetic tension 
which has not been taken into account
in the simplified dynamo theories.
 Using
 a larger value of $h$ would decrease the rate 
of diffusion, increase $\vert D\vert$, and increase the
 growth rate of dynamo waves.

Since the $ \alpha-\Omega $ dynamo leads to exponential growth of
 the mean field, it must eventually saturate.  
This is expected to happen by nonlinear quenching of $ \alpha $
 and $ \beta $.  
Many papers have been devoted to exploring this quenching. 
(For example, see the excellent papers of Blackman \& Field, 
1999, 2000 or Kleeorin et al. 200, 2002, 2003). 
The crux of the issue is whether the dynamo is quenched
 when the mean field $\bar B$ comes to equipartition with
the turbulence (which is approximately the situation in our Galaxy),
 or whether small scale fields suppress the
dynamo long before $\bar B$ reaches equipartition. We return 
to the issue of the small scale fields in \S 14.
For an extensive review the saturation problem 
consult the Physics Report of Brandenburg and Subramanian (2005), 
and also Shukurov (2004).

Proceeding with the mean field analysis and examining
equations (13),  we see that they are symmetric
in $ z $ so that there  are solutions with two different  parities.
In one both magnetic field components are even
in $ z $,  and in the other they are  both  odd.   The first solution
is called the {\it quadrupole} solution. It  has a much smaller
critical dynamo number than the odd one which is called
the {\it dipole}  solution.  The quadrupole solution
is the one whose parity 
agrees most closely with the present  Galactic field, which
appears to be  symmetric with respect  to the midplane.
However, this is the solution in which the net flux in the
disc changes and, therefore,  requires the flux expulsion
provided by the vacuum boundary conditions.  Its critical dynamo 
number is about $ -13 $ for the form we have chosen for $ \alpha(z) $.

If one had  closed boundary conditions with 
\newline $ \partial B_r/\partial z = \partial B_{\theta} /\partial z=0 $,
no such flux problem arises, and one would not need flux
expulsion.  For this case, the critical number for 
the dipole solution is $ -4 $, and a growing quadrupole solution does
not exist.  Thus, knowing nothing about the observations
one would naturally expect the galactic disc dynamo to generate 
this dipole solution, which both avoids the  flux expulsion problem
and has a lower critical dynamo number. 
The fact that  the Galaxy  possess the more difficult
quadrupole solution with its larger critical dynamo
number seems to imply that the field did not originate
from a seed field with random symmetry,  but had an already
established quadrupole symmetry and a respectable field strength 
from  the beginning.  
We  discuss the flux expulsion problem in \S 9.

We conclude this section with an estimate of the dynamo
number.  The estimates of the value of $ \alpha_0 $ and $  \beta $
given by Parker, and RSS,
will be presented in the next section.  
It will be shown that earlier estimates converged on a
value of $ \alpha $ of about 0.1 km/sec.  Taking the above values
of $ \beta $ and $  h $, and  $ \Omega \approx 10^{ -15} $/sec,
we find that $ D $ is about unity, a value an order of magnitude
too small for dynamo amplification of the quadrupole field.
However, as seen, 
in \S 7  the properties inferred for the interstellar
turbulent motions driving the dynamo are  too pessimistic. 
[See Field (1995)]. During the  
nineties Ferri\`{e}re (1992a,b, 1993a,b)  recognized that the observed
turbulent motions were mostly the intrinsic 
 motions of expanding supernovae remnants and  
expanding superbubble shells.  These motions have much 
longer correlation in scale and  time than
that  previously assumed for the  turbulence, 
 and thus lead to larger values
for the dynamo number and for the growth rates.  We discuss the
dynamos based on this turbulence in \S 8.

\section{The magnitudes of $\alpha$  and  $\beta$  in the interstellar medium}

As mentioned in the previous section,
the most popular theory for the origin  of the
Galactic field is based on the mean field   dynamo theory
first developed by Steenbeck, Krause, \& R\"{a}dler (1966). 
For a systematic presentation see Krause and R\"{a}dler
(1980).   This theory
was applied to the Galactic disc dynamo by 
Vainshtein \& Ruzmaikin (1972) in the early seventies.
At the same time and completely independently of
the work of  Steenbeck et al.,  and of Vainshtein \& Ruzmaikin,
  Parker developed his
own dynamo theory (Parker 1970a,b,c, 1971a,b,c,d,e, 1973a, 1979),
 and his own application to  the galactic dynamo problem.

The work of Vainshtein and  Ruzmaikin is best summarized
in the book ``Magnetic fields in galaxies `` written by
Ruzmaikin, Shukurov, and Sokoloff (1988). which we will
refer to in the remainder of this review as RSS.
 Parker's work is expanded and detailed 
 in his book
``Cosmical Magnetic Fields '', (1979).  The equations
that Parker derived are completely equivalent to those
of Steenbeck, Krause and R\"{a}dler, and in his application
to the Galactic field, are also equivalent to those of RSS.

The success of the dynamo theory for an  origin  of
the Galactic field that is based on the  amplification of
weak fields during the life of the galactic disc,
 hinges on the values of the 
coefficients $ \alpha , \beta $ 
and the effective Galactic half thickness  $ h $.  
These  are encapsulated into 
 the dynamo number $ D $, in  equation (16).

Let us assume that when the Galactic disc is first
formed, it possessed  a very weak field with field 
strength ranging from $ 10^{ -16 } $, (which can arise
by compression from an even weaker  field strength 
$ 10^{ -20 } $) generated  earlier by a Biermann battery,
or by  other means.

Then, to increase this field to its present strength of
a few microgauss and   quadrupole symmetry,
certain conditions must be satisfied.
The dynamo number $ D $ defined in the previous
section must be negative and exceed 13 in magnitude
by a large enough amount that the growth  rate
is fast enough to increase the Galactic field from
its initial very small value to its present value
of a few microgauss.  For example, if the growth time were
five hundred million years and the age of the
Galactic disc ten billion years,  then the field must   exponentiate
twenty times an  increase by a factor of $ 10^{ 8}  $.
This  would amplify an initial field of order $ 10^{ -14} $
gauss to the present observed value.  A somewhat faster 
growth than this could amplify the field from 
$ 10^{ -16 } $ gauss.  The question reduces to 
whether the interstellar values of the coefficients
can achieve so  fast a growth rate.

 Both Parker (1979) and RSS give rough estimates 
of the dynamo number and when reduced
to the same definitions the values based
on rather different estimates come out to 
be remarkably close.

In this discussion we defer  the problem
 of flux escape and the justification for
 vacuum boundary conditions to \S 9.
 
The  critical dynamo number for any growth at all is $D = -13 $.
The dimensional  growth time,  $  h^2/\beta $, was shown in the
last section to be five hundred million  years.
If we take such a growth time as adequate for dynamo generation
of the field, then we need the dimensionless growth rate
$ \gamma' \approx 1 $.  

Then from the figure VII.2 in RSS we see that the required
dynamo number is $ D =-25 $. If the absolute value
of $ D $ is smaller, then 
the growth time  is longer.

Parker estimates $ D $ in his book (Parker 1979),
making the best estimates he can of $ \alpha $ and $ \beta $.
He bases his estimate of $ \alpha $ ($\Gamma$ in his notation)
on a model he constructs for a rising
convection cell in the Galactic disc.  He  arrives at the estimate
\begin{equation}\label{eq:17} 
\alpha \approx \frac{1}{8} \pi \epsilon L \Omega
 \approx 0.04 L \Omega , 
\end{equation}
where for his model $ \epsilon \approx 0.1 $, and $ L $
is the radius or spacing of convection cells. For this 
he takes $ L \approx  100 \mbox{pc} $.  This leads to 
\begin{equation}\label{eq:18} 
\alpha \approx 0.1 \mbox{km/sec }   .
\end{equation}

He estimates $ \beta $ (which he calls $ \eta_T $)
 from another model for diffusion.  He finds that
\begin{equation}\label{eq:19} 
\beta \approx 0.2 v L 
\end{equation} 
where $ v  $  is the observed velocity of the interstellar clouds
which  he assumes is typical of the observed interstellar turbulent motions.

Thus, his value for  $ D $ (which he calls $ (Kh)^3 $ )
is 
\begin{equation}\label{eq:20} 
(Kh)^3 = D \approx - (\frac{3}{2}) (h/L)^3 (\Omega \tau )^2
\end{equation}
where $ \tau = L/v $ is his estimate of the correlation
time.   He takes $ h=400 \mbox{ps } $.
 His factor of $ 3/2 $ is based on assuming  Keplerian
rotation  of the  Galactic disc, and should actually be replaced
by unity.  This equation (after this replacement) leads
to his estimate of the dynamo number $ D \approx -4 $.

This estimate is based on very  rough assumptions, so
the fact that it is less than the critical number for
excitation of the $ \alpha-\Omega $ dynamo in the 
the quadrupole mode is not of  too great a concern.  It
is interesting that in spite of the very large and small
numbers going into the definition of the dynamo number
it turns out to be of the correct order of magnitude  for
exciting the dipole mode.

RSS  also made an estimate of the dynamo number.
Their  estimate is based on a less careful estimate of the
$ \alpha-\Omega $ dynamo coefficients, but leads to a value
rather close to Parker's estimate.  Their estimate
for $ \alpha $ given on page 181 of their book  is 
$ \alpha = L^2/\Omega/h \approx 1 \mbox{km/s } $,
but apparently they left out the factor of 
one third in the definition of $ \alpha $.  Taking the same
number $ L = 100 pc, h = 400 pc $ we see that their estimate
for $ \alpha $
is larger than Parker's estimate of $ \Gamma = .04 L \Omega $
by a factor of about six.  (With the factor of one third th is
would be twice Parker's value.)  Their estimate for $ \beta $
is $ \beta \approx \frac{1}{3}  L v $ larger than Parker's
estimate by a factor of five thirds.  These values lead to
their estimate for $ D \approx 10 $.  
If we restore the missing factor of $ 1/3 $ in RSS's  estimate
for  $ \alpha $, their  estimate reduces to $ D \approx 3 $,
  not far from Parker's estimate of $ D \approx  4 $.

Both of the two estimates for $ D $ are  substantially smaller than
that  necessary to
produce the present value of the Galactic field starting 
from a reasonable seed value in the lifetime of the Galaxy.

We have given these estimates in some detail because for a while
these were the principal estimates for the action
of the  $ \alpha-\Omega $ Galactic dynamo and the astrophysical
basis of these estimates can be easily understood.
If taken seriously, they result in a rather discouraging numerical
situation for the Galactic dynamo.  Clearly, to make the
dynamo work some way to increase the effective value of $ \alpha $
is needed.  

It is interesting that for closed boundary conditions 
(no flux escape) the critical dynamo number is about 
-4 for the dipole mode. These boundary conditions correspond
to no net flux in the local region of the Sun.
Since the observations seem to require the nonzero
flux quadrupole mode,    one has a feeling that
if the 
Galactic dynamo is the origin  of its field, the Galaxy would
choose the more difficult quadrupole mode for its
magnetic field structure.

It is intriguing and
provocative that more recent Faraday rotation
 observations seem to indicate
that there actually is some asymmetry about the Galactic midplane
(Han, 2001, 2002, 2006, Han Manchester, Berkhuijsen and Beck 1997)
These results are based on Faraday rotation measurements of
pulsars at Galactic latitudes $ b>10\deg $ in the second and
third quadrants.  The other 
measurements still are quite symmetric.   

As mentioned in the last section,
 the remaining  option is, of course,  that
the magnetic field is pregalactic,  and initially had  the
quadrupole symmetry.

This was  the dynamo situation in 1991.  But the
whole picture was changed substantially by a  careful
 analytic series   of  excellent papers published by Katia Ferri\`{e}re
(Ferri\`{e}re 1992, 1993a,b, 1996, 1998, Ferri\`{e}re \& Schmidt 2000).
These papers  turned the whole situation around leading 
to a much more definitive analysis of the astrophysical
 situation,  a more precise evaluation of the dynamo coefficients,
and a much clearer picture of
the helicity of
 interstellar turbulence.
It turns out that the turbulence of interest is not that
resulting from stirring of the interstellar medium by
stellar winds and supernova.  The motions
of importance are those actually present in the  supernovae
and superbubble  explosions
themselves.    We discuss Ferri\`{e}re's analysis in the next section.

\section{ Ferri\`{e}re's dynamo theory based on supernova
and superbubble explosions}

The estimates of dynamo action in the Galactic disc
by Parker and RSS were derived from what little was known
in the seventies and
eighties  about interstellar turbulence, as well
as an educated guess as  to the effect 
galactic rotation would have in  producing  kinetic helicity.
It has always been correctly   assumed 
that the main source of the observed turbulent motions 
 is due to  supernova explosions and stellar winds.
However, it was earlier supposed that this turbulence arose due to
the injection of supernova energy into the interstellar medium,
stirring it up into turbulent motions. This did not appear
to make the turbulence strong enough to amplify the Galactic
field.

 Ferri\`{e}re took  a more direct approach to the dynamo  problem,
using the actual velocities of the expanding supernova 
shells during the explosions themselves as the
mechanism for producing $ \alpha $ and $ \beta $.  That is,
she asserted that the magnetic field is  produced 
during the actual explosions, and much of the random motion we see in the 
interstellar clouds and interpret as fluid turbulence, is
the actual expansion  velocities of  the explosions themselves.

To analyze this, she made use of  a blast wave model for  supernovae
in which the interstellar magnetic field 
and the ambient interstellar medium at a  particular time are
swept up into  a thin rapidly expanding
sheet.
She employs 
  a definite expression  for  $ r $, the  radius of this sheet,
as a function of time 
\begin{equation} \label{eq:21} 
\frac{r}{14\mbox{pc} }= \frac{E_{51}^{0.22}}{n_0^{0.26}}
\left( \frac{t}{10^{ 4} \mbox{  years}}\right)^{0.3}   .
\end{equation}
(Ferri\`{e}re 1992a, Weaver et al, 1977). 
She then derives  the behavior  of
the plasma motions  and the magnetic field evolution 
 in this sheet,  under
the influence of  Galactic rotation.  As the shell expands,
its moment of inertia increases by a large factor and the
shell tries to come to rest in the laboratory frame,  thus
appearing to rotate  backwards in the frame of the rotating 
Galaxy.
Since the initial field and the matter in which it is embedded
were initially rotating with the Galaxy,
the effect of the Coriolis forces  produce helical motions
that  twist the  magnetic field out of its
original plane.  If the field is initially toroidal,
then   this twist  partially rotates it into the radial,
or poloidal, direction and produces the $\alpha$ effect needed
to complete the dynamo.  

The actual angle through which the field is twisted is 
roughly the angle  that the
Galaxy rotates  at the  radius of the supernova,
i.e. the angle it would rotate during the lifetime of the
supernova, a  few hundred thousand
 years. This twist 
is actually quite 
 small,  so that a perturbation calculation is adequate for the twist
of the field.
  However,  under the influence of the  many supernovae 
that a given piece of the interstellar medium suffers,
the twist accumulates.  Since it is always in the
same direction, it accumulates  to a finite 
value in  few hundred million years.  

Ferri\`{e}re calculates this effect with considerable precision
taking into account that,  relative to a fixed point, $ P $, 
the many  different supernova remnants which overlap this point
occur with  different directions  and  distances
from $ P $.  Some supernovae occur quite close to
$ P $, and produce  a stronger effect than those that
explode farther away.  
 It is known that the frequency of supernovae at any 
given place is roughly comparable to  the reciprocal
lifetime of a typical supernova remnant, so that the angular rate
 of the accumulated  twisting of  the piece of the
field line at point $ P $ is  a finite 
 fraction
of the rotation of the Galaxy.  Thus,  it is  no longer 
necessary to make a very rough guess as to  the value of $\alpha$.
By using her technique 
the full $\alpha$ tensor
can  be calculated with some 
 precision.  Its value  depends
mainly  on the frequency and spatial distribution of 
the supernovae.  These are reasonably well known,
as  are the details of  the expansion.

Further, it  was   possible to  calculate  the
 other parameter of the dynamo theory, $ \beta $. This is
obtained by evaluating the mixing of the plasma and magnetic field
while they are in the expanding shell.

  There is 
another  important parameter she introduces into  the dynamo theory, 
the escape velocity, $ V_{esc} $ of the field lines
and plasma.  Because  supernovae occur 
 closer to the midplane of the Galaxy than the bulk of
the interstellar medium and its  magnetic field, the
result of any explosion on the average is to expel
the plasma and field lines  away from the midplane.  Ferri\`{e}re
points out that
this average displacement
per unit time is equivalent to $V_{esc}$,  but
she  keeps it in the turbulent term in the dynamo equation.

A  difficulty with her calculation is that she only includes
the effect of the supernova remnant on the plasma and 
field when the remnant is  expanding,
and does not consider  the effect  on the field when its plasma 
falls back  to its original
position, after the supernova expansion concludes.
  Her assumption is that the magnetic field would
not be acted on coherently during the fall back,  so that
the twisting that occurs during the explosion would persist. 
This again involves the physical process of 
 the magnetic flux separating  from the
plasma.  This remains a serious problem in
 her calculations which  we also discuss 
in \S 9.

\begin{figure}
\rotatebox{0}{ \scalebox{0.55}{ \includegraphics{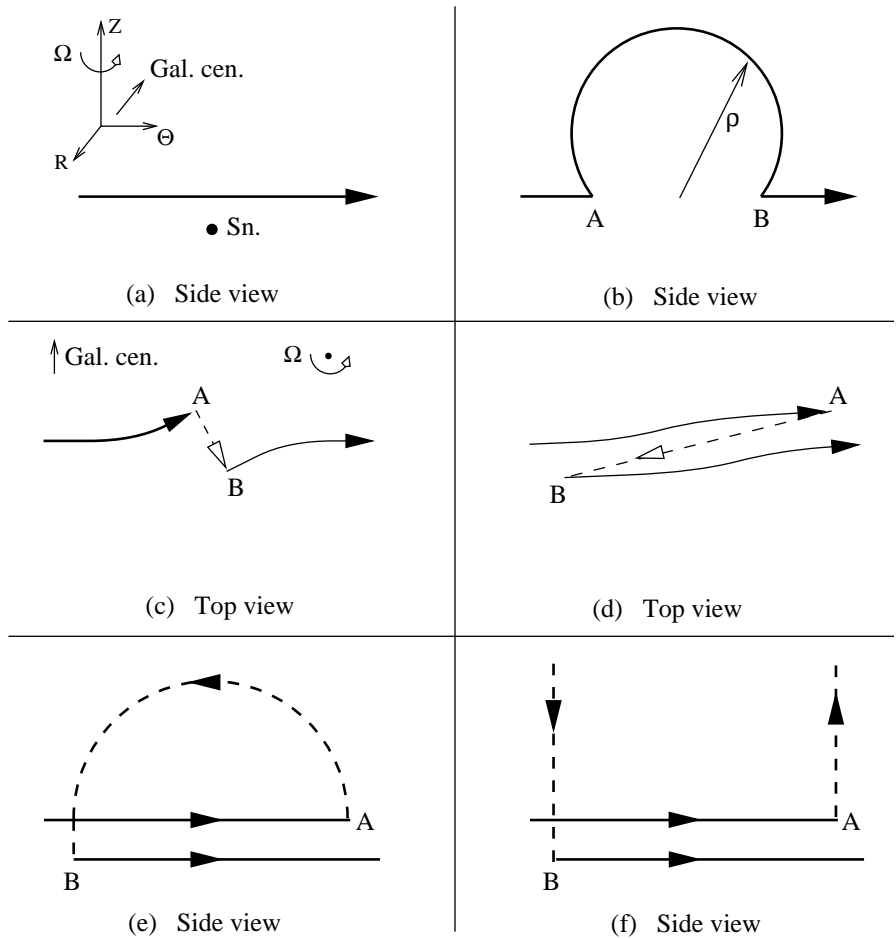} } }
\caption{The operation of the $ \alpha-\Omega $ dynamo.
In (a) and (b) a supernova blows the magnetic
loop. It is twisted into the poloidal plane by Coriolis forces as
is seen in the top view of (c). The dashed line represents the upper 
part of the loop.  In (d) the lower part of the line is stretched
by differential rotation.  Then the upper part of the line is
removed to infinity as in (e) and (f).}
\label{Fig1}
\end{figure}
\vspace*{1in}

A pictorial view of how the supernova explosion actually 
 amplifies 
the magnetic field without violating flux conservation 
is shown in Figure 1, (Kulsrud, 2000)  The various panels are alternate
views from the side towards the Galactic center and from above.
In panel (a), the supernova is shown just before it goes
off and an undisturbed magnetic line in the toroidal direction 
is also shown. After the   supernova remnant  has  expanded to a 
radius $ \rho $  and the radiative phase of the explosion is reached, 
nearly  all of the swept up 
matter is contained in a thin shell at its boundary 
and the magnetic field lines are  embedded in this shell.
This is shown in panel (b).    Outside of the bubble 
the field and plasma are undisturbed.  Because the matter has been
shifted from the place it occupied before the explosion 
where it uniformly filled the sphere to the shell,
the matter in the shell has  increased its
moment of inertia by a large factor.  By
conservation of angular momentum the supernova bubble
 decreases its rotational velocity in the laboratory frame
and rotates backward in the Galactic frame, 
as shown  from a top view in panel (c).  The piece of the original
line in its original  plane is shown by the solid line,  and the part
overlapping the supernova bubble is shown by the dotted line.

The rotation has displaced the footpoints of the
line in the radial direction in
this plane.  But this displacement plus the differential rotation
of the Galaxy displaces the lines in the toroidal  direction, as
shown in another top view in  panel (d).  Thus, in the original
horizontal plane the field line has doubled over itself, increasing
its field strength and the magnetic flux in this plane.  On the
other hand, the field overlapping the supernova has developed
negative toroidal flux compensating for this increase and still 
conserving total toroidal flux.  If the supernova continues
to expand indefinitely, or at least the flux line continues
to rise indefinitely,  as shown by the side views in panels (e), 
and (f),  the part of the field line  on top
of the supernova bubble would be  removed
from the Galactic disc and the toroidal flux
in the disc  will have increased.  Since negative flux is 
 removed  far away, it is   no
longer  counted when one describes the Galactic field.
No Faraday rotation measure passes through this removed flux and
the rotation measure in the disc is now doubled.  
In this manner the problem of flux conservation is solved.
  [Note that if the two vertical parts of the line in panel
(f) should magnetically reconnect then the lower part of the
reconnected line would return to the disc and cancel any 
net gain in magnetic flux there.]

The last of the panels in this picture  represents how the vacuum
boundary conditions, assumed in the conventional 
 $ \alpha-\Omega $ dynamo,
are imagined to act.  One can see the difficulty with which
this dynamo is faced.   The question is reduced to whether
the flux in 
the expanding plasmas can actually escape from the Galaxy.
Generally, except for supernovae that occur in the halo, the supernova 
bubble is slowed down below the Galactic escape velocity
by gravity and ram pressure on the interstellar medium.
It  stops long before it reaches
more than a hundred parsecs or so above the plane.  
Then the danger to the dynamo field amplification is that the 
matter in the supernova shell falls back down and any  
unwanted  flux still embedded in it cancels out the amplified
field.  The result in this case  is that there is  little or 
no amplification at all.  

The separation of flux from plasmas during the expansion is crucial
to the amount of amplification.  If the field is very weak, there
is no way for the plasma to know it has flux embedded in it,  and
there will be no separation at all.

\begin{figure}
\rotatebox{0}{ \scalebox{0.55}{ \includegraphics{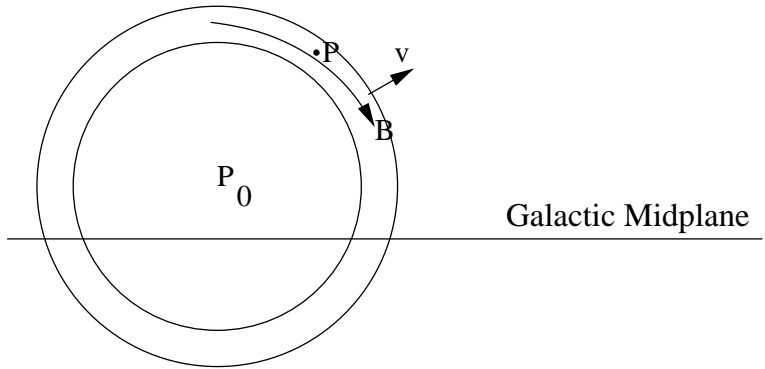} } }
\caption{The supernova shell passing point $ P $}
\label{Fig2}
\end{figure}
\vspace*{1in}

To calculate the dynamo action of a supernova 
Ferri\`{e}re   selects some
point $ P $ at which she calculates the
dynamo action.  (See Figure 2.)
  She  then  calculates the electric
field $ {\bf E} $  at this point during the time in which  the  shell
passes over it.  This is  given by 
\begin{equation}\label{eq:22} 
{\bf E} = - \frac{\delta {\bf  v} \times {\bf B} }{c}
\end{equation}
 at this point while  the thin supernova  shell
passes over this point.  ($ \delta {\bf  v} $ is the expansion
velocity which though large she denotes by  a $ \delta $
since she regards it as random when summed over many supernovae.)
She treats the Coriolis forces and the magnetic
field twisting as  first order in a perturbation expansion.
To first order 
she takes   $ {\bf B} $ from flux freezing.  That is she 
calculates the field as though it were simply swept up
by the snowplow effect of the supernova shock using 
the blast solution for the shock motion given in equation (21).

With this model the electric field at the point $ P $ is nonzero
only when  the shell passes over it.  There is no velocity before
the shell reaches $ P $,  and after it passes, the magnetic 
field is reduced to zero. 
 The time integration of 
$ {\bf E} $ at $ P $  is just such as to produce
the field which cancels the original field leaving $ B $
zero as one expects.  
Although the shell may be very thin,  and  the time
over which the $ {\bf E} $ is nonzero very  
short,   the magnetic field
in the shell is very  large and the actual magnitude of $ {\bf E} $
multiplied by the time when it is nonzero is finite 
and independent of this thinness.

To properly include the twisted  part of the changed magnetic 
field Ferri\`{e}re evaluates the first order velocity 
$ {\delta \bf  v} $ in the shell,
taking into account the effect of the
Coriolis force on the  zero order velocity, and
$ \delta {\bf B} $,  the first order change
of the magnetic  field  produced by the
$ \delta {\bf  v} $.

Substituting her result for $ \delta {\bf  v} $ and
$ \delta {\bf B} $ into the expression for $ {\bf E} $ given by equation
(22), integrating it over time at the fixed  $ P $,
 summing this result over  all  supernovae 
that interact with the point $ P $ during some long
time $ t $,  and  then dividing by $ t $,
she finds the long time average of 
$ {\bf E} $ at $ P $.  She assumes supernovae occur
randomly with a distribution that depends only
on $ z $,  the height of above the plane.  The resulting
value she finds for $ {\bf E}   $ gives her the mean value
for $ \alpha $ and $ \beta $ to use in the dynamo 
equation to calculate the  evolution of the mean 
field $ \bar{{\bf B} } $.  See equation (11) of \S 6.

  The result is a tensor for
$ \alpha $ which depends on the original 
orientation of $ {\bf B} $ before she started her averaging
over supernovae.  If $ {\bf B} $ is in the toroidal
direction, i.e. the same direction as the Galactic rotation,
then $ \alpha $ is a scalar, the same scalar originally
considered by Parker, RSS,  and others.  However, in
general, ${\bf B} $ will have other components and the
tensor character of $ \alpha $ becomes important.  
(Ferri\`{e}re, 1998). 

In addition to $ \alpha $ she also obtains expressions
for the $ \beta $ tensor and for $ {\bf  V}_{esc} $,
which is a vector in the $ \hat{z} $ direction.

Using the same technique, Ferri\`{e}re also considered
the contribution  of superbubbles to the Galactic value of
$ \alpha $.  A superbubble is  the result of many supernovae
going off effectively simultaneously. The simultaneous
explosion of many supernovae at the same time and place
arises from the formation of stars close
together in clusters at  nearly the same time.   
The more massive stars exhaust their nuclear fuel 
at nearly the same time and  one finds that, during the
lifetime of a single supernova, many others explode.
The gaseous remnants overlap and  form a low density bubble
of high pressure gas.  This  drives a shock that arises 
from  the energy input
of all the supernovae, rather than from  the kinetic
energy of the individual explosions.  

As far as driving the dynamo effects the superbubble 
acts just as a single supernova does. In the thin shell,
driven outward  at high velocity, the Coriolis
forces act to twist the lines.  However, a superbubble is
much bigger than a single supernova shock and lasts
much longer.  As a result,  the effect on the dynamo is
much greater than that of individual supernovae,
 even when one takes account of the smaller number of 
superbubbles than supernovae.
The radial time dependence  of the superbubble shock
which Ferri\`{e}re employs is (Ferri\`{e}re 1992a,  McCray \& Kafatos, 1987)  
\begin{equation}\label{eq:23} 
\frac{r_s}{ 267 \mbox{pc}} = \left( \frac{L_{38}}{n_0} \right)^{0.2} 
\left( \frac{t}{10^{ 7}\mbox{yr }} \right)^{0.6} ,
\end{equation}
where 
$ L_{38} $ is the luminosity in units of $ 10^{ 38} $ ergs/s
and $r_s$ is the cylindrical radius.

 Comparing the times and distances
 in this equation with  those for the single
supernova, one can see that the superbubbles extend to
  much greater  distances and last a great deal  longer.
  The larger distances and times allow the galactic
Coriolis forces to act longer, and  produce a larger effect 
in the  superbubble shells than in
 single supernova shells.  This is compensated somewhat 
by the fact that there are typically thirty  times more supernovae
than superbubbles.  However, even taking this into account Ferri\`{e}re found
that the effect of all the superbubbles was larger than 
that of all the supernovae by a factor of order
seven,  even though the total number of supernovae in 
superbubbles is  comparable with their  total number of supernova.

Making use of these calculations,
 Ferri\`{e}re developed the Galactic dynamo 
to a  new level of precision.  
She established
several new results  in these papers.  One result  was
that the   $ \alpha $,   in the alpha
effect is a tensor unless the initial
Galactic magnetic  field is 
purely  in the toroidal direction.   She 
calculated  the $ \beta $ tensor
using the same procedures. In addition,
she found  a new systematic velocity 
$ V_{esc} $ that represents the rate at which the magnetic
field lines rise above the midplane of the Galaxy (during
the expansion phase of the bubbles).  
Ignoring any fallback of the magnetic flux, she took this velocity 
 to represent the   rate of flux loss to the disc.

There is one  correction that  should  be made
to her calculations.  She  calculates the mean $\alpha$
effect at a fixed point. That is to say, she
assumes that the $\alpha $ tensor 
which  she 
calculates  gives the rate of evolution
of the poloidal field at this fixed point.  But
there is a term missing in her calculation that when
introduced implies that her value of $\alpha$ actually
gives the evolution following the expansion of the shell.
This seems  intuitively
clear when one notes that, after the shell passes point
$ P $,  there are  no field  lines left at $ P $. 
 All the field lines have been  assumed to be swept up by the
shell.  This extra term disappears if one carries out
the calculation in the moving frame and expresses $\alpha$ 
as a function of the Lagrangian coordinate, the
fluid element.  Her method of calculation working with
the averaged time integral of the electric field,
which she calls the $ E.M.F. $, tends to hide this
extra term.  

Correcting for this by
 making  this simple transformation from an
Eulerian independent variable to the Lagrangian variable 
should make her calculation more correct.

It is interesting that she calculates the actual change
of the field in Lagrangian coordinates, and then uses
this changed field to calculate the dynamo coefficients
$ \alpha $ and $ \beta $ which are subsequently 
used to calculate the mean change in $ \bar{{\bf B} } $.
This roundabout procedure is an attempt to relate
her calculation to the standard $ \alpha-\Omega $
mean field dynamo theory.  However, the approach
from supernova and superbubble to the dynamo does not
quite fit the assumptions of the dynamo theory which
assume that the turbulence scales are small compared
to the large scales on which the mean field is to be
calculated.   The supernova and superbubble remnants	
are on a scale of the total thickness of the Galactic disk. 
 It would be neater
if she just stopped with the changed fields in
any given Lagrangian element of plasma and directly averaged
the changes occurring in a given plasma element rather
than referring them to the dynamo coefficients.

\section{The validity of the vacuum boundary conditions}

As emphasized in \S 6 and \S 7,    boundary conditions 
at $ z = \pm h $, are needed to compute the rate of growth of the dynamo 
mode.  To see that these conditions are  directly related to the 
change of flux in the disc for the quadrupole eigenfunction,
integrate the first  equation of (13) 
 from $ -h $ to $ + h $.  This yields
\begin{equation}\label{eq:24} 
\frac{d}{dt} \int_{-h}^{+h} B_r =
\left. \beta \frac{\partial B_r}{\partial z} 
\right|_{-h}^{+} -\left. \alpha B_{\theta}\right|_{-h}^{+h} ,
\end{equation}
where $ B_r $, $B_{\theta}$  are assumed symmetric.
 We see that the first term $ -\beta \partial B_r/\partial z $
is the rate of escape of the $ B_r $ flux through
 the boundary while from  the zero boundary conditions
on $ B_r $ and $ B_{\theta} $, the  second term  vanishes.
   Although the flux lines
of the mean field are not frozen in the plasma its global
flux inside the disc still satisfies the conservation relation
(24).  
The $ B_{\theta} $ flux is not {\it per se} conserved,
since it grows as the $ B_r $ flux is wound up, but this does
not increase the number of lines of force.

As discussed in \S 6, the boundary conditions
customarily assumed at $ \pm h $  are that the fields
vanish, and this is the natural condition if the flux which
reaches the boundary,  escapes instantaneously.  Physically
this is difficult to imagine because,  flux freezing
implies that 
any  flux that escapes is embedded in
interstellar matter.  Thus, for the dynamo to work,
a large portion of the interstellar medium must be removed
at every efolding of the  magnetic field. [However,
see Brandenburg, Moss \& Shukurov, (1995),
 Dobler, Poezd, \& Shukurov, (1996) and   
Shukurov, (2004), where this problem is discussed.]
Let a fraction $ f $ of the
entire interstellar medium be removed for every
exponential  increase of the magnetic field.  Then,
if the field strength increases from $B_0$ to $B_1$, the mass
 of the interstellar medium decreases from $M_0$ to 
$M_1=M_0(B_1/B_0)^{-f}$.  To raise the field from
$B_0 =10^{ -16 } $ gauss to $B_1= 10^{ -6} $ gauss with $f=1/3$ 
would require $M_0\sim 2000 M_1$, which seems untenable, and
inconsistent with the progressive chemical enrichment of the 
interstellar medium over time observed from stellar composition
measurements.

These numbers give one
pause when considering these  assumed boundary conditions.  

  Further, the energy to remove this amount of mass to infinity 
requires $ 1/2 v_{esc}^2 $ per unit mass,  where $ v_{esc} \sim 
500 \mbox{km/s} $ is the escape velocity (Smith et al. 2007).  This gives  of order
$ 10^{  15} $ ergs /g.  Taking $ M_1 \sim 10^{ 9} $ solar masses
we find an energy requirement of about $ 10^{ 60} $ ergs.
Thus, an amount
of order the total luminosity emitted over the disc life, is needed to
amplify the magnetic 
 field by the galactic dynamo under the above assumptions.
Clearly,  a better way is needed to remove flux.

A more analytic treatment of the boundary conditions is
discussed in the review
article of Beck et al 1996.  They 
state that the turbulent resistivity  $ \beta $ is
much larger by a factor of say $ A^2 $ 
in the halo, (the region outside the disc
$ |z|>h $) larger than in the disc, and that    $ \alpha $ 
vanishes in the halo.  In this case, the
components of the halo  magnetic field, in the dimensionless
units of equation (14) of \S 6,  are proportional to
$ e^{-A z'} $ and the boundary conditions 
at $ z' = \pm 1 $ should be that the field there
is smaller than the average field by about  $ (\mp 1/A) $.  
For large $ A $
the boundary conditions become essentially the
vacuum condition $ {\bf B} =0 $.

To justify this large increase in the halo value of
$ \beta $, Beck et al. quote results from Poezd,  Shukurov \&
Sokoloff (1993).  These authors 
estimate  $ \beta $ from measurements  of the random
velocities 
in the halo made by  Kulkarni \& Fich (1985). The results were
 $\tilde{v} \approx  10 $ km/sec,  and the correlation
length $ \ell \approx $ 100 pc. This increases $ \beta $
by a factor of one hundred above its disc value. 
 But these measurements are made on the bare halo
at a time when there is no escaping flux or mass.  Surely
if there was a large amount of mass diffusing outwards,
the velocity fluctuations would be much smaller due  to
the large amount of energy involved in lifting the 
mass-loaded flux outward.  

Beck et. al. (1996)  also appeal to  arguments given by
Brandenburg et al (1995), who attempt to treat the
actual outward  escape of flux  by invoking reconnection 
of it with the halo field.  Again, the reconnection 
for this merging is based on
replacing the normal resistivity $ \eta $   by 
the very large turbulent resistivity  $ \beta $. 

Now it 
is not clear that there is enough magnetic flux in the
halo field to match the expelled 
flux and  accommodate this reconnection. Further, $ \beta  $ 
is not a true resistivity  and can only mix the field lines
of different sign.  The field lines rising up from the
disc would still be loaded with the heavy interstellar mass while
the other lines would be much lighter.  Then even after the lines are mixed,
the gravitational field could separate them 
allowing the original field lines to sink back and cancel
any gain from the dynamo.  

It must be borne in mind that, when trying to arrive at the {\it origin }
of the Galactic   magnetic field,  we are primarily interested
in building up the field from an extremely weak initial 
value.  When the field is weak, such mixing is unlikely to
occur.  Also any true reconnection due to real resistivity  
would be too slow to be of interest. 

A third possibility to justify the boundary conditions, 
at least physically,  is to imagine that the escaping
flux is lifted in huge arcs such as occur in
superbubbles, and to have the mass slide down the flux
tubes.  This would leave  the field lines at the tops of
the arches unloaded and lighter so that this piece
of them might escape by 
buoyancy or cosmic ray pressure (Parker 1992,  Hanasz et al 2004).
However, again this can  not work in the early dynamo stages
when the field is weak because the magnetic field  can have no dynamic
consequence;  the fluid flows in a direction 
independent of the magnetic field and probably just 
falls straight down.  

It also does not seem to work when the field 
is stronger, at least for the superbubble case, because
the mass at the top is on a nearly  horizontal line, and the
rate of  flow down the field lines is too  slow 
to release much flux.  This is  demonstrated in
the paper of Rafikov \& Kulsrud (2000).  Further, 
the presence of cosmic rays 
on the field lines in this latter case also
inhibits the downflow along the lines.
This is 
 because  an Alfv\'{e}n wave
instability is produced by any relative motion
of the cosmic rays and the matter exceeding  the 
Alfv\'{e}n speed.  (Rafikov \& Kulsrud  2000,
Kulsrud \& Pearce 1969.)
This instability couples the parallel motion
of the plasma and the cosmic rays and has 
 the result that the cosmic ray pressure stops 
the downsliding.

The most reasonable way for the flux to escape and properly
complete the $ \alpha \Omega $ dynamo theory would be to have
the superbubbles blow the mass entirely out of the Galaxy.
But, in our Galaxy, except for the rare case where a superbubble
 occurs high in the
halo, the superbubble shell is slowed down by the ram pressure of
the swept up material.  Then  the  gravitational downpull is enough  to
stop it before it has gone far enough to  escape 
from the Galaxy.  This is found to be the case in the
 many numerical simulations
(MacLow \& McCray 1988).  

  However, as emphasized  by Heiles (1990)
the number of supernovae in a superbubble 
assumed in these simulations is taken as the  average number (thirty)
and further the superbubble is taken to start at the midplane.
Superbubbles which contain more than the average number of  supernova and
which  start at a position that is 
high enough  above the central plane of the disk that lies 
above  the  bulk of the interstellar medium will 
lead to a blowout that  expels their   mass entirely out of the galaxy,
(See in figure 1f.) These events  
which are rare, should  occur independently
of the magnetic  field strength and  happen at any time
during the 
life of the disc, even when the magnetic field is weak.   

As long as the vertical field lines do not reconnect
some net flux expulsion will occur and make
the galactic dynamo viable although if these events are rare
the dynamo will be weak.  One cannot ignore this possibility.
However, if this is the case,  the  dynamo would 
 be extremely inhomogeneous. 
The resulting  field lines would   be discontinuous
as viewed inside the disc (but actually connected through
 intergalactic space). Such a configuration has not been considered
in interpretations of the galactic field.  Whether
 it could be consistent with observations of the 
interstellar field is unclear.  However, it would 
certainly change  the general picture of the galactic field.

For smaller galaxies in which the 
gravitational field is weaker or in starburst galaxies
where the superbubbles are  stronger the flux expulsion
 problem is eased and
 the $ \alpha--\Omega $ dynamo actually  could 
more easily produce magnetic fields.

The $ \alpha--\Omega $  dynamo   is a  beautiful
mechanism.  It  only appears to  be limited by these
difficult boundary conditions issues.  However, they 
arev very hard to get around when the magnetic field is very weak.
In fact, it does work quite well for the dipole symmetry
where the total flux is zero and does not need to change.
If the $ \alpha-\Omega $ dynamo is the main agent for growing
the magnetic field strength, it seems very surprising that
the Galactic symmetry is  not odd.

It is entirely possible that a reasonable way to
separate out the unwanted flux may emerge and 
resolve the Galactic dynamo problems.  This would go
a long way toward establishing an  origin  
for  cosmic magnetic fields. Any  indirect evidence
as to the state of  the Galactic field  at the 
inception of the disc  would 
also go far toward  resolving whether the $ \alpha--\Omega $ dynamo 
is the primary source of the field.
Indeed,  there is actually  some evidence which, while still  uncertain,
does  bear on the problem.  This evidence, 
discussed in \S 2, comes from
the abundance of the light elements Li, Be, and B
in very old stars.

\section{Arguments against a primordial origin  }

  We have raised some difficulties with
the galactic disc dynamo as the sole origin
of the Galactic magnetic field.   There are also
a number of arguments that have been 
 raised against a pregalactic origin.
These have been expressed  by Woltjer (1969), Parker (1973b,
1979 p. 519 and p. 525), and  Rosnesr \&  Ducca, (1988).
[But see Kulsrud (1990), Howard and Kulsrud (1997) and 
Kulsrud (2006).]

The mechanisms for the production
of a pregalactic field have been developed with much less 
precision than  those for the galactic dynamo.
  The plasma physics of the very  early
universe is only partially known.  
In addition,  events that occur very early on
have a coherence scale smaller than the Hubble
radius at the corresponding redshift $ z $.
  In fact, the corresponding
comoving scale (the present scale of the cosmos
that had the Hubble radius at red shift $ z $) 
is equal to the present Hubble radius
times $ (1+z)^{-1/2} $.
The observed  scale of the Galactic field
is at least $ 1 $ kpc which corresponds to a region
of the universe of about $ 10 $ kpc before
collapse to the Galaxy.   Thus, any magnetic  field
generated at a red shift  $ z> 10^{ 13} $, corresponding
to a temperature $ 10^{ 9 } $ 
electron volts,
would lead to a  field whose present
 scale is smaller
than $ 1 $ kpc in the Galaxy.  After the time 
corresponding to this temperature,
which is the electron -  neutrino decoupling  temperature,
the plasma physics seems well understood and 
there are few grounds for believing that anything
mysterious could generate a magnetic field.

A  very weak field with strength of order $ 10^{-20}  $
gauss could  reasonably be generated by action
of the Biermann battery mechanism discussed in the next section,
(Biermann 1950), or by interaction   of the electrons with the 
CMB photons (Mishustin \& Ruzmaikin 1972).  
But such a field could only serve
as a seed field, and this
gis generally accepted  as a reasonable 
seed field for   galactic dynamos.

One  possibility that has been taken seriously
for a pregalactic  origin, is a field generated
at  moderate redshift by  turbulence 
that occurs during structure
formation, say in the protogalaxy,
(Pudritz \& Silk, 1989, Kulsrud, Cen,  Ostriker \& Ryu, 1997).
  This possibility
will be discussed in \S 12.  Another 
possibility that is seriously considered  is that
the magnetic field is somehow generated when 
a large radio jet is formed,  say by dynamo generation
by a  
black hole at its center.  Magnetic fields  of  substantial strength 
are  presently 
observed in such giant radio jets.  There are even some
Faraday rotation measurements of  such fields that
indicate the magnetic flux is large enough that if  
spread out throughout the universe it would provide
the total  flux currently observed in all the galaxies.  
If this should happen before galaxies form, then the field
is already  present in the plasma out of which the galaxies 
form,   and would thus have  a pregalactic  origin
(Rees \& Seti, 1968, Daly \& Loeb, 1990, Furlanetto \& Loeb, 2001).
So far, the details of this filling of the universe 
with flux have not been worked out with any precision,
but the presence of this amount of flux, if supported
by the Faraday rotation measurements,  is very suggestive
as an origin  for  pregalactic 
 fields, (Colgate \& Li,2000, Li at al. 2006a,b).

Before taking up the discussion of the first of these proposals
for a pregalactic  origin  let us discuss the
arguments that have been made against any primordial origin 
as opposed to a galactic dynamo origin, (see 
Kulsrud, 1990.)

There are essentially two   arguments 
against a pregalactic magnetic field:  (1) the winding up
argument and (2) the escape of any magnetic field
from the Galactic disc in a short time by turbulent diffusion.

One can explain  the winding up argument as follows:
 Suppose one starts out with a uniform
unidirectional field $ {\bf B}_0 $ in the $  \hat{{\bf  x}} $ 
direction  that   extends  over the entire Galactic disc.
 Then in cylindrical coordinate the radial
component of the field is $ B_r = B_0 \cos \theta $.  Now, at
each radial circle $ r $ of the Galaxy this  radial field
is rotated at a rate $ \Omega(r) $, so after a time $ t $
the field is rotated by an angle $ \Omega t $,  and
\begin{equation}\label{eq:25}  
 B_r(r,\theta,t ) = B_0 \cos (\theta - \Omega(r) t) .
\end{equation} 
By flux freezing, this field develops a toroidal component
\begin{equation}\label{eq:26}
B_{\theta} = B_r(r,\theta,t) \times r  \frac{d \Omega }{d r}t .
\end{equation} 
This   is   a slowly varying function of 
$  r $  times  $ B_r $. But, if we look at the variation
of $ B_r $ with  $ r $ at fixed $ \theta $ and $ t $,  we see 
 that $ B_r $ oscillates in $ r $ with a wavelength
equal to $ \Delta r = 2 \pi/(t |d \Omega /d r|)$.
Now,   near the sun, $ \Omega r $ is a constant, 
so $ d \Omega /d r = - \Omega /r $ and  $ \Delta r/r =
2 \pi /\Omega t \approx 1/50 $.  Thus, the toroidal field 
would change sign every 100 pc. Parker states that this
is contrary to observations, and clearly such a field would
produce a zero average Faraday rotation measure. 

This winding up argument is one of the basic arguments against a primordial
field origin.  However, we are  not sure  that the correlation
distance of the Galactic field is 1 kpc.  We can only be sure
that  if it does reverse rapidly,  it does  not average out over
such  distances, since there are  rotation measurements 
 of pulsars much further away than this, (Rand \& Kulkarni 1989).
(The one-kiloparsec
correlation distance is derived  from models that take
into account the possible reversal of the field 
between the spiral arms that have  this 
spacing, but ignore the models with short reversals.)  

On the other hand, the resolution of pulsar 
measurements is not as fine  as 100 pc so that
it is entirely possible that the field does reverse
over 100 pc but does not average out.  In fact,
if the initial $ {\bf B}_0 $ field were not uniform,
but varied by a factor of two over the diameter
of the Galaxy,  then an inspection of the above argument
shows that the toroidal component of the field 
$ B_{\theta} $ would still reverse with a scale of
100, 
 but the toroidal field would one sign would be twice
as large as that of the other sign and the
Faraday rotation measurements would not average out. 
(See Howard \& Kulsrud, 1997 and \S 15 for a similar result
that would occur with ambipolar diffusion).
It would be an interesting challenge
to find out whether  this is the case, or
whether  the field is actually
of constant sign over such small distances. The field of our 
example could
account for the puzzling fact that the fluctuations in synchrotron
polarization direction are of  different amplitude 
 than the fluctuations associated
with Faraday rotation measurements
 (Zweibel \& Heiles 1997, Brown \& Taylor 2001).
 This discussion is in no way able to establish
that the Galactic field does indeed reverse over
a small length.  But the possibility of it does weaken the
argument against a primordial origin.

The second argument, the escape argument,
 described by  Parker  on page 522
of his book,   concerns the turbulent resistive decay of a
pregalactic   magnetic field.
The decay rate by ordinary resistivity  is very slow and
can be disregarded.  However, if one replaces
the ordinary resistivity  by the turbulent resistivity,   
$ \beta \approx 10^{ 26} \mbox{cm}^2 $ /sec,  then the decay
time is 300 million years and Parker argues that a primordial field 
would not survive for
the lifetime of the Galaxy.  However, we see
in \S 9  that although this diffusion can
mix the fields in the disc, they cannot remove it
from the disc unless the field is  strong enough that
it is able to separate itself physically from the
interstellar material.  (Without the separation, the
diffusion mechanism would have to lift the interstellar medium
entirely out of the Galaxy which we showed was energetically
very difficult.)  

Thus,  neither of these  arguments
can definitively  rule out the possibility that 
there was an initial large scale magnetic 
field   of pregalactic origin.

In \S 15  we describe  a modification of such 
a model in which ambipolar diffusion plays an important
role and an initial field stronger than $ 10^{ -8 }
$  gauss could be brought up to full strength 
without any dynamo action at all.  However, total
neglect of the $ \alpha-\Omega $ dynamo is, of  course,
 not realistic (Kulsrud 1990, Howard \& Kulsrud 1997,
Kulsrud 2006).

There is,  an additional consideration
that depends more  on our current understanding of galaxy evolution. Our
Galaxy currently has of order 10$^9$ M$_{\odot}$ of interstellar
 material. There are many sources and sinks of gas: stellar winds,
planetary nebula ejection, supernova explosions, infall of 
intergalactic clouds and small galaxies,  star formation, and
a possible galactic wind. Estimating the rates for all these
 processes leads to the conclusion that the interstellar gas is 
replaced    mainly through cycling in and out of stars,
on 10$^9$ year timescales, about 1/10 the age of the 
oldest stars in the Galaxy. If the magnetic field is in a steady state,
dynamo processes must operate over that time to act on the new 
material. Of course, the new material may already by magnetized at a
level far above the estimated pregalactic  field strength, although 
it is unlikely to have the correct topology.  Another alternative is that the
already existing magnetic field somehow mixes into the new material,
 possibly by ambipolar diffusion (Heitsch et al 2004).

\section{Seed fields }

It is reasonable to assume that there was   a time before which
 there were absolutely no magnetic fields in the universe.
If flux freezing were an exact constraint, there
would  be no fields after this time either. This also
follows directly from the magnetic differential 
equation (4) of \S 4.    How
{\it do} magnetic fields get started in the first place?

The most popular mechanism  is  the Biermann
battery (Biermann, 1950).  A magnetic field 
can  arise from extra terms in Ohm's law 
which are not included in   equation (1).

These terms arise  as follows:

First, we rederive the magnetic differential equation 
keeping the additional terms in Ohm's law.  In fact,
Ohm's law can be considered to be simply the equation of
motion of the electron fluid (Spitzer 1962),
\begin{equation} \label{eq:27}
n_e m \frac{d {\bf  v}_e }{d t} = 
- n_e e \left( {\bf E} + {\bf  v}_e \times {\bf B}  \right)
- \nabla p_e + n_e m {\bf  g} +{\bf F}_{ei} ,
\end{equation}
where $ {\bf F}_{ei} $ is the electron-ion frictional
force,  which is related to the $ \eta {\bf j} $ 
term.  We drop this term.  We also drop the
inertial term $ n_e m (d {\bf  v}_e/d t) $ and the gravitational term 
$ n_e m {\bf  g} $ because of the smallness of the electron 
mass density.  Dividing the resulting  equation by $ n_e e $ 
we get 
\begin{equation} \label{eq:28} 
{\bf E} + \frac{{\bf  v}_e \times {\bf B} }{c}
= - \frac{\nabla p_e}{n_e e} .
\end{equation}

To see physically how the Biermann battery works
it must be realized that (in the absence of a magnetic
field)  any accumulation of the electron pressure
at any point naturally  leads to a tendency
for the electrons to leave this point.  But this automatically
leads to a charge imbalance producing an electrostatic
field,  and this field is just strong
enough to resist this tendency.  (The charge
imbalance to produce such a field is generally minute.)
 This field is
\begin{equation}\label{eq:29a}
{\bf E} = \frac{\nabla p_e}{-n_e e }  .
\end{equation} 
$ {\bf E} $  depends on  $ p_e $ and $ n_e $ the electron pressure
and density.   If $ n_e $ is a constant in space,
the electric field is curl free and  purely electrostatic.
Thus, it will drive no current just as a battery with
its terminals unattached to wires produces no current.
However,  if $ n_e $ is not constant, $ {\bf E} $ may
not be curl free and there can be a path in the plasma
around which there would be a potential drop.
 Thus, a current would
flow along this path and produce a changing magnetic field 
which would in turn induce an electric field  to
balance this potential drop.  The key to the importance
of this term is its dependence on $ n_e $ in addition
to its proportionality to  $ \nabla p_e $ which enables
it to act as a battery.

Now, the resulting current is known to be
 tiny,  so $ {\bf  v}_e \approx 
{\bf  v}_i $
to a very high degree of approximation.  Further,  even
if the plasma is  partially  ionized,
the electron temperature is kept  close to the
neutral  temperature by collisions  so that $ 
p_e/n_e = p/n(1+\chi) =Mp/\rho (1+\chi) $, 
where $ \chi $ is the degree of ionization, 
which for the moment we take to be constant.
Thus, 
\begin{equation} \label{eq:29}  
{\bf E} + \frac{{\bf  v} \times {\bf B} }{c}=
-\frac{M }{e (1 + \chi ) }  \frac{\nabla p}{\rho } .
\end{equation}

Finally, taking the curl of this equation and
combining it with the induction equation, one gets the
modified magnetic differential equation,
\begin{equation}\label{eq:30} 
\frac{\partial {\bf B} }{\partial t} =
\nabla \times ({\bf  v} \times {\bf B} ) 
+\frac{\nabla p \times \nabla \rho }{\rho^2} 
\frac{M c}{e(1+\chi)}  .
\end{equation}

 It is well known that the  vorticity 
$ {\bf  \omegab} = \nabla \times {\bf  v}  $ satisfies
a very similar equation  
\begin{equation} \label{eq:31} 
\frac{\partial {\bf  \omegab} }{\partial t}=
\nabla \times ({\bf  v} \times {\bf  \omegab} )
-\frac{\nabla p \times \nabla \rho }{\rho^2}  
+ \nu \nabla^2 {\bf  \omegab} .
\end{equation}
Thus, as long as viscosity is small, these equations 
are identical up to the factor $ -e/M c (1+\chi) $.

Therefore,   $  e {\bf B}/M c =  {\bf  \Omegab} $ 
and $ - (1+\chi) {\bf  \omegab} $ satisfy the same equations.
  It is reasonable to  assume that up
until a certain time vorticity is also zero.
Thus, we have that   the magnetic field and the
vorticity  satisfy the same 
zero initial conditions. 
($ {\bf  \Omegab = \omegab } = 0 $ ).  Because 
the  differential equations have a unique solution we find that
\begin{equation}\label{eq:32}
{\bf  \Omegab } = - \frac{{\bf  \omegab} }{(1 + \chi) } .
\end{equation}
is valid at least until the viscous dissipation term
(or the resistivity  term)  becomes important.

Thus, when vorticity finally starts to grow, the magnetic
field will grow with it.  However,  the 
resulting magnetic field strength due to this term is always tiny.
To get an idea of its strength in a gravitational forming
structure  assume that its rotational energy 
is of the  order of its  gravitational energy, and thus the
rate of rotation should be comparable to the free fall time,
$ 1/\sqrt{ 4 \pi G \rho } $.   $ G $ is the gravitational 
constant and $ \rho $  the density.  Now,  $ e/Mc =
10^{ 4} $ sec/gauss, so we have
\begin{equation}\label{eq:33} 
B \approx 10^{ -4} \sqrt{ 4 \pi G \rho}
\approx 10^{ -19} \sqrt{ n} . 
\end{equation} 
where $ n $ is the  hydrogen number density $ \mbox{cm}^{-3} $.
Thus, for typical cosmic densities, the    
resulting magnetic field  is well below $ 10^{ -21}$ 
gauss.  

Actually, equation (31)  for the evolution of the  magnetic field 
 only holds as long as the viscosity term plays
no role in the evolution of $   \omegab $;
that is, as long as the vorticity itself does not saturate.
But when the vorticity becomes strong there is 
Kolmogorov nonlinear  coupling to small scales, establishing a cascade 
which is terminated by viscosity acting on small scales.
Effectively, vorticity on all scales
reaches saturation, even though the effective viscosity  on
large scales is small.  Thus,  the Biermann battery continues to operate,
increasing $ B $ above the value given 
equation (33).  

If one examines equation (31),  which gives the  evolution
of $ {\bf B}  $, we see that there are two limits.   When 
the magnetic field is extremely small, the first term on the right,
the dynamo term,  is negligible and the field grows only by the second
term,  the Biermann battery term.  One can estimate
its rate of increase  by assuming there is a finite angle between
$ \nabla \rho $ and $ \nabla p $, and taking the scale size of
 both $ p $ and $ \rho $ to be $ L $.  Then the Biermann
battery term  is of  order of magnitude $  p/\rho L^2 \approx 
v_s^2/L^2 $, where $ v_s $ is the speed of sound.  Thus, 
 $ {\bf \Omegab}  $  increases roughly as $ (v_s /L)^2 t $.   

However,  when $ B $ becomes larger,  
the dynamo term becomes important, i.e. when $ v \Omega /L
\approx v_s^2/L^2 $.  If, for simplicity,  we take
 $ v \approx v_s $,
then this happens when $ v_s/\Omega  \approx L $. But $ v_s/\Omega $ is
 the ion gyration radius $r_i$,  so the dynamo 
term becomes significant when $r_i$ decreases to
become comparable to $ L $.  (Initially, of course,  the gyration radius
is infinite.)  After this time the Biermann battery continues
to operate, assuming that there is still a finite angle between
$ \nabla \rho $ and $ \nabla p $.   But it is quickly outstripped
by the dynamo term, which  increases the magnetic field 
strength  $ B $  exponentially.  This is   because 
the dynamo  term is proportional to $ B $.

Recall that we derived equation  (31) assuming the ionization
 fraction $\chi$ is constant.
Following the birth of the first massive stars and quasars, the
 Universe was pervaded by a network of ionization fronts that
spread out from sources of ionizing radiation. 
During this so-called
epoch of reionization ($z\sim 10$; Spergel et al 2007), the low
 density gas was ionized first, leaving behind pockets of denser gas
that was ionized later. Large scale temperature and density
 gradients in the dense gas interacted with the large electron pressure
gradient in the front itself, allowing the
 Biermann battery to operate more efficiently. 
Gnedin et al. (2000) showed that the magnetic fields generated
in this fashion are somewhat larger than the fields generated 
in vortex sheets ($\sim 10^{-17}-10^{-18}$G), but are still far too weak
to explain galactic fields without
further  substantial  amplification.

The rough assumptions going into these relations  cannot
be taken too seriously,  but they do give a simple picture
of what goes on in the initial build up  of seed fields.
To gain more precision concerning the dynamo  action,
we need to specify the velocity
field in more detail.  This is done in the next section,
in which the assumption is made that the velocity field
has a  turbulent spectrum.
 This appears to be the
case in numerical simulations of the formation
of gravitational structures,  primarily in protogalaxies.
(Ryu et al. 1994, Kang et al 1994, Kulsrud et al, 1997)

We have devoted considerable  space to the Biermann battery
because it can be  treated rather definitively.
There are other proposed origins for seed fields.
The earliest is due to Harrison who invoked rotating
structures in the early universe (Harrison, 1970). 
 He assumes that  a blob of plasma
rotates in the presence of the cosmic background radiation,
(and assumes the radiation 
is  non rotating because  for sufficiently large $ z $ its  mass
is larger than the plasma mass.)  Then he points out that 
 the rotation of the electrons in the blob will be 
slowed down by Thompson  scattering off the 
non rotating radiation 
field,  while the ions   rotate without slowing down.
  The resulting
current will produce a weak  magnetic seed field.
The seed field will increase with time inducing a back
electric field  which will tend to keep the electrons rotating.
The balance between these two forces determines the actual
rate of increase of the field.  Under most
reasonable assumptions,  the resulting field strength
is a little smaller, but  of the same
order of magnitude  as that of 
the Biermann battery,  $ < 10^{ -20} $ gauss.

Since the Biermann battery mechanism produces
an increasing magnetic field, there must also be a
back E.M.F. resisting this growth.  But in the
case of the Biermann battery, this back E.M.F.
must be there to balance the Biermann term in
Ohm's law (equation (28)) in the first place.
One could consider the process in the reverse
order, for  it is clear that the magnetic field must grow to induce
an electric field to balance this term.

Another  origin  for seed fields has been proposed by
Martin Rees,  (Rees, 1987, 1994, 2005, 2006).
 He suggests that supernova remnants
similar to the Crab Nebula have fields of order $ 10^{ -4} $
gauss due to the wind up of the central pulsar.
The fields are coherent, but contain fluxes of
opposite signs on each side of the remnant.
He estimates that there could be $ N \approx 10^{ 6}
$ such remnants in the early stage of the galaxy.
Of course, the fields from these remnants, when they
expand enough will overlap and cancel so that
the resulting fields will be proportional to
$ N^{x} $.   But,  because they may interact, $ x $
may not be $ 1/2 $, but could be as small as
$ 1/3 $.  Even so,  the resulting fields would
be in the range of $ 10^{ -9} $ gauss, a rather
substantial seed field!  However, one might
worry that the rms field would be larger than
this, perhaps larger by a factor of $ N^{1-x} $
unless magnetic reconnection or some other
process reduces it.  The details of this
interesting suggestion have not been carried
out since they depend on plasma processes
such as magnetic reconnection which are very
little understood.  
One should compare this suggestion with 
the suggestion mentioned in \S 16,   that primordial
field can arise from radio jets with large magnetic 
fields of alternate signs and these  can seed the universe
with magnetic fields prior to galactic formation. This 
seeding  involves
very similar problems of cancellation.

In any event, there seems to be little difficulty
with producing an initial field whose strength
is of order $ 10^{ -20} $ gauss, and   that can serve as
a seed field for dynamo amplification to magnetic fields 
of larger  strength.

\section{A protogalactic theory for magnetic field generation}

Since there are a number of difficulties with
the disc--dynamo origin  for galactic fields,
it is of interest to consider  the possibility
that the field could be generated during  the time
in which  the galaxy itself is formed.  Such a theory
was first proposed by Pudritz and Silk(1989).  It was
later developed in some detail  
by Kulsrud and coworkers (Kulsrud et al, 1997).  This latter theory
is based on the observation that there must be
considerable turbulence generated during the 
initial collapse of the cosmic plasma to form the
protogalaxy.   This turbulence is 
generated by the   shocks that result from the
steepening of the cosmic fluid as 
the gravitational instabilities become  nonlinear. 
These shocks heat the plasma to a Jeans temperature
that stops the collapse  and leads to a temporary
virialized state.  At the same time the shocks, which
are generally of finite extent, produce shear 
velocities that evolve into Kolmogorov turbulence.  
This has been established in  numerical simulations 
of structure formation (Ryu et al, 1993, Kang 
et al 1994).  The turbulence in these simulations
has  a Kolmogorov spectrum. Its largest
eddy is  comparable to the size of the protogalaxy, 
and its total energy is comparable to that of 
the total thermal and gravitational energies of the protogalaxy,
(Chandrasekhar, 1949).

Under these conditions a very weak  magnetic field will  grow 
because the pressure and density gradients  
produced by the shocks are  not parallel.  This
development was verified
by integrating the Biermann magnetic differential  equation (31)
of the last  section,   taking the velocities directly from the
numerical simulations (Kulsrud et al 1997).  A magnetic field of order
$ 10^{ -21} $ gauss develops  in this simulation
and is found to satisfy  equation (33)  at nearly every 
pixel of the numerical grid.  However, the expected
dynamo amplification is  not found  because this amplification 
is primarily produced by the  smallest turbulence eddies 
whose scale is  much smaller than the grid size
of the simulation. 
These smaller eddies dominate 
 because their  rate of turn over in 
Kolmogorov turbulence is faster than the larger eddies.

The rate of dynamo growth of the magnetic field 
is approximately  their  turnover rate. (See the next section.)
 However,  Kulsrud and his coworkers 
used  Kolmogorov  theory together with the theory of magnetic
field generation developed  by Kulsrud \& Anderson (1992) to 
estimate the  rate of growth produced by the smallest eddy.
(Kulsrud et al.  1997). They find that  the  growth is 
much faster than the rate of  collapse of the
protogalaxy.  In fact, the magnetic energy   doubles roughly 
one hundred times in this collapse time. (The
Kulsrud-Anderson theory is discussed in the next section.)

 During this time the magnetic energy  saturates
and reaches  equipartition with the energy of the smallest
Kolmogorov eddy.  Further, the Kulsrud---Anderson
theory shows that  the magnetic field has 
a scale much  smaller than that of the smallest
eddy.  If this were the end of the turbulent dynamo process,
the resulting field would not be of much interest since the
size of the smallest eddy is a thousand times smaller than the size
of the protogalaxy (estimated to be approximately 100 kpc).
 After the collapse of the protogalaxy, 
the field would still be coherent
on a scale more than  a thousand times smaller than the radius
of the Galactic disk.  However, even though the 
magnetic field has  such a small scale its  energy 
will be that of the smallest eddy.  This  is of the  order of
one  per cent
of the total binding energy of the protogalaxy.
Such an energy is not negligible, and is  
strong enough to affect the early formation of stars.

Due to the limited spatial resolution of the simulation, it is impossible 
to numerically  study the evolution of the field produced by the 
entire turbulent spectrum. In order to form a large scale 
field by this process, there has  to be an inverse
 cascade of magnetic
energy  to the larger scales. Whether this can in fact occur 
is still an open question in dynamo theory.

 However, an  effective inverse cascade can 
occur by direct amplification
of the seed fields by the larger eddies, (see Kulsrud 2005, 2006).

To predict how the field  develops one needs to 
 gain  a  deeper understanding of the behavior 
of magnetic fields in the presence of  strong turbulence,
a difficult problem of plasma physics. 
Schekochihin, Cowley and their  colleagues have  published  a number of
theoretical  and numerical papers discussing 
problems concerning the 
structure and saturation of these small scale fields.
They consider this problem 
both in the context
of protogalaxies,  and in  the intracluster medium in
 clusters of galaxies. (Schekochihin \& Cowley  2004, 2006
Schekochihin et al 2004, 2005a,b).
Their results are  discussed in \S 14. 
In his book, Kulsrud makes an  estimate of this behavior
(see Kulsrud 2005, 2006). 

When one considers the small scale turbulent eddies
even in a rotating body, one finds they have very
little helicity. 
The behavior of magnetic fields in non helical   turbulence 
is to be compared with their behavior in the
$ \alpha - \Omega $ dynamo theory which is strongly dependent
on  helical motions.
The latter theory, combined with proper boundary conditions, 
actually generates net magnetic flux
and a mean magnetic field in the entire
region of interest, while the former theory 
 produces fields that are coherent on the
scale of the relevant eddy.  However, the
$ \alpha-\Omega $ theory is usually developed
in a thin disc and the  horizontal 
coherence length of the magnetic field
 is usually assumed  to be a few times  the disc
thickness. 

 The radial coherence length of the present 
Galactic field inferred  from modeling pulsar rotation measures
is at least a kiloparsec,
but not necessarily   larger.  
Indeed, it  seems to reverse in  different spiral
arms.  If we compare the distance between the 
spiral arms  with the size of turbulent eddies,
inferred to exist from the numerical simulations
of protogalaxies, 
we see that, if  eddies as large as one tenth
of the protogalaxy generate the main fields, then
the resulting fields, after collapse, will have
a sufficiently large  coherence length to
provide a reasonable origin  for the pregalactic  field.

Further, even  if the protogalaxy 
provides a smaller scale field it could  be large enough
for the mean field disc dynamo to work properly
and develop larger coherence lengths.
 As emphasized before, the field might  be large enough to provide 
the escape of flux needed to validate  the 
required  boundary conditions.

This conclusion  for definiteness is based on
the assumption of top down formation of the protogalaxy.
It is not  drastically modified even if  our Galaxy 
were  assembled through a hierarchical process, with small
structures ($M\sim 10^6M_{\odot}$;
about 10$^{-4}$-10$^{-5}$ of the present galactic mass) forming
 first. Such structures are estimated to be a
few hundred pc in size, and to have virial velocities of order 
10 km/s. If magnetic fields grow on the scales of the largest
eddies in these systems, the result would not be very different
 from eddy scale fields in a massive galactic halo. But the problem
of eliminating the small scale fields and growing large scale 
fields remains.

  In summary, amplification of the magnetic field in
the protogalaxy could address problems with the disk dynamo 
related to its operation at very low field  strength. The flux
expulsion problem, however, still remains even at the present field strength.

\section{Generation of small scale magnetic fields by turbulence}

Earlier, in \S 6, we discussed how, based
on turbulence with helicity flows, the $ \alpha-\Omega $
dynamo can generate magnetic fields with net flux,
inside a disc. 
In the last section we considered how magnetic fields
could be produced by dynamo action in
turbulence with no helical flows. An  astrophysical
application of such a dynamo raises  the possibility
of producing substantial 
pregalactic fields in the protogalaxy.
  Another application is the production of fields
in the intracluster plasma in clusters of galaxies. 
(We do not consider the latter, as the main point
of this review is to uncover the processes involved
in generating cosmic magnetic fields, and in this review
we have reduced the problem to the challenge of producing
a magnetic field in our Galaxy.)

In the case of the protogalaxy, we have seen 
that  magnetic fields produced  by nonhelical turbulence
 have no mean flux overall, but 
could have mean
flux in large subregions.  Other subregions may  have
a mean flux with a different direction.
However, such a field can be the source of the 
Galactic field which itself has  no overall net flux;
only  mean flux over large regions.
Magnetic fields in protogalaxies are generated
by random, nonhelical, small-scale turbulence.

Such a
theory of field generation 
has been developed by
a number of people (Kraichnan and Nagarajan 1967,
Kazantsev, 1968, Kulsrud \& Anderson 1992, Subramanian \&
Barrow, 2002, Boldyrev \& Cattaneo 2004).
Here  we review the Kulsrud--Anderson theory
in some detail.

Assume that the statistics of the turbulence is 
given by a turbulent spectrum such as that
of Kolmogorov.  For simplicity, assume that
the turbulence is homogeneous,  isotropic,  and
steady in time.
 Assume that initially there is  some very
weak magnetic field,   whose statistics  are  not necessarily
isotropic, but are  homogeneous. 

Then one can evolve the magnetic field by integrating the
magnetic differential equation (2), with the resistive term set to zero,
(In almost all cosmic situations it is permissible to
neglect the resistivity  except at extremely  small   
scales).  The velocity in this equation can be assumed to be 
given at least statistically 
and unaffected by the magnetic forces.
 This is the {\it kinematic} assumption.

 Consider the system to be enclosed in a
large  periodic
box so that both the velocity and the magnetic field can
be described by Fourier transforms
\begin{eqnarray}\label{eq:36} 
{\bf  v} & = &  \int d^3 {\bf k}  {\bf  v}_{{\bf k}}(t) 
e^{i{\bf k} \cdot {\bf  r} } ,
\nonumber\\  
{\bf  B} & = &  \int d^3 {\bf k} {\bf  B}_{{\bf k}}(t)  
e^{i{\bf k} \cdot {\bf  r} } .
\end{eqnarray} 

Since the turbulent velocities are clearly random,
we describe them statistically by an ensemble 
average, and make use of the 
{\it random phase approximation} (Sagdeev and Galeev, 1969,
Kulsrud 2005)
\begin{equation}\label{eq:37} 
<{\bf  v}^*_{{\bf k'}}(t) {\bf  v}_{{\bf k}}(t') > =
J(k)({\bf  I} - {\bf \hat{k}} {\bf \hat{k}}) 
\delta({\bf k}'-{\bf k} ) \delta_{\tau}(t-t')    .
\end{equation}
Here the angular brackets denote  the ensemble average,
of the velocities at  different wave numbers and times,
 $ {\bf  I} $ is the unit dyadic, and the hat denotes 
a unit vector.  The delta function  in $ {\bf k}' -{\bf k} $
results from the random phase approximation.  Its significance
is that there are a large number of independently positioned
velocity waves in the large box and taking the Fourier transform
of these waves yields a delta function  factor between different 
wave numbers.  $ \delta_{\tau} (t-t') $ is  the correlation function
 of the velocities at a fixed position at different times.  
If we normalize this function to unity when $ t=t' $,
then $ J(k) $ is the energy   spectrum of the turbulence
\begin{equation}\label{eq:38} 
\frac{1}{2} <{\bf  v}^2> = \int d^3 {\bf k}    J(k) .
\end{equation}

As a result of the turbulent velocities and the magnetic
differential equation  the 
one dimensional magnetic energy spectrum $ M(k) $,  given by 
\begin{equation}\label{eq:39} 
\frac{<B^2>}{8 \pi } = \int M( k ) dk
\end{equation}
evolves in time according to the one dimensional mode-coupling
equation 
\begin{equation} \label{eq:40} 
\frac{\partial M(k)}{\partial t} = 
\int K(k,k_0) M(k_0) d k_0 - 2 \beta k^2 M(k) .
\end{equation} 
 The kernel $ K(k,k_0)  $ is 
\begin{equation}\label{eq:41} 
K(k,k_0) = 2 \pi \tau k^4 \int \frac{\sin^3 \theta d \theta }{k_1^2}
(k^2 + k^2_0 - k k_0 \cos \theta ) J(k_1) ,
\end{equation}
where $ k_1 $ is defined by 
\begin{equation}\label{eq:42} 
k^2_1 = k^2 + k_0^2 - 2 k k_0 \cos \theta  .
\end{equation}

The quantity $ \beta $ is the same turbulent diffusion coefficient 
as in the $ \alpha-\Omega $ theory and is 
\begin{equation}\label{eq:43} 
\beta  =   \frac{2}{3} \tau  \int J(k^{''} ) d^3 {\bf k}^{''} .
\end{equation}
$ \tau $ is the decorrelation time given by 
\begin{equation}\label{eq:44} 
\tau = \int^\infty_0  \delta_{\tau} (t) dt  .
\end{equation} 
(The mode coupling equation is strictly valid
only if $ \tau $ is sufficiently small compared
to the rate of evolution of the magnetic spectrum.
This is called  the short correlation--time approximation.)

The derivation of the appropriate form of the mode coupling
 equation is given in a number of places 
(Kraichnan \& Nagarajan 1967,
Kazantsev, 1968, Kulsrud \& Anderson 1992).  The form given
here was derived in  the paper of Kulsrud \& Anderson.
In this derivation, no  assumption as to the isotropy
of the magnetic field statistics is made.  $ M(k) $
refers to the three dimensional magnetic spectrum averaged over
all solid angles.  

This equation has a number of important implications.
The most important one relates to the growth of the
total  magnetic  energy
\begin{equation}\label{eq:45} 
{\cal E} = \int M(k) d k  .
\end{equation}

If one integrates  equation (39) over $ k $ one finds  the evolution
equation for $ {\cal E} $, 
\begin{equation}\label{eq:46} 
\frac{d {\cal E} }{d t}  = 2 \gamma {\cal E}  ,
\end{equation}
where
\begin{equation} \label{eq:47}  
\gamma = \frac{\tau }{3} \int k^2 J(k) d^3 {\bf k} =
\frac{4 \pi \tau }{3} \int k^4 J(k) dk  .
\end{equation}   
A very similar equation was discovered by Batchelor (1950).

This equation has an  easy interpretation for
Kolmogorov turbulence.  Namely, eddy velocities
with wave number $ k $ have a turn over rate
of $ k \tilde{v}  $ where $ \tilde{v}^2 = J(k) d^3 {\bf k}  $
 is the mean square turbulent  velocity in the range
$ d^3 {\bf k} $.  The decorrelation time $ \tau $ is 
$\sim  1/k \tilde{v} $.   The change $ \delta B/B $
during the decorrelation time, is 
 plus or minus the  rotation angle of the eddy
 $ \Delta \theta = k \tilde{v}  \tau $. Thus, 
$ <(\delta B/B)^2/\tau > = k^2 \tilde{v}^2 \tau =k^2 J(k)\tau
d^3 {\bf k} $, the contribution of the integral in 
$ d^3 {\bf k}$  to $ \gamma $ in equation (46). In short,
every turnover of any eddy produces a relative
change in   the magnetic energy  by a factor of order two.
Note also, that each eddy acts independently of the others
in increasing the magnetic energy.

How does this change in magnetic energy produced
by a single eddy manifest itself
in the change of the actual field?  Surprisingly,
the magnetic field change produced by any 
eddy   occurs on scales
much  smaller than the eddy itself.  It had been  previously
 assumed that the magnetic energy 
is produced on the same scale as the eddy.  (See for example
the treatment in RSS, 1988).  This is because, 
in the presence of a uniform or large scale eddy,
the main interaction was previously  assumed to be between the eddy 
 and this  large scale field.  Actually the stronger
interaction is between the eddy and the {\it small}
scale magnetic field,
and this produces energy at twice the magnetic wave number.
As  a result, magnetic  energy rapidly propagates to 
larger and larger wave numbers. 
The wave number, at which the amplified energy 
peaks,  increases exponentially, and the field develops
a very much finer structure than that of the eddy.
This result might be anticipated since by
flux freezing  the magnetic  field follows the displacement
of the plasma rather than its  velocity.  Since the
displacement is the integral of the velocity one expects
the displacement to also exponentially develop
fine structure.  

These results are demonstrated analytically in 
Kulsrud \& Anderson (1992) as follows:
Consider that the turbulent---eddy scales are larger than
 $ 1/k_1 $.  Then expand the mode coupling equation 
(39)   for larger values of the magnetic wave numbers,
$ k $ and $ k_0 $,  where we see
from the notation in the mode coupling equation (39)
that magnetic energy at the large wave number 
$ k_0 $ is  transferred to the also large wave number $ k $.
The result is the simple partial  differential equation 
\begin{equation}\label{eq:48} 
\frac{\partial M}{\partial t} =
\frac{\gamma }{5} \left( 
k^2 \frac{\partial^2 M}{\partial k^2} 
-2 k \frac{\partial M}{\partial k} 
+ 6 M \right) 
- 2 k^2 \lambda_S M
\end{equation}
where $ \lambda_S = \eta_S c^2/4 \pi  $ is the magnetic diffusivity
and $ \eta_S $ is the Spitzer resistivity. 
(The resistivity  term is  added  by hand.)

Without the resistivity   term, 
this  equation has a Green's function solution
\begin{equation}\label{eq:49}
M(k,t) = \int_{- \infty}^t  M(\frac{k}{k_{ref}},t' )
G \left( \frac{k}{k_{ref}},t-t') \right)  d t'  ,
\end{equation} 
where $ G $ is the Green's function
  \begin{equation}\label{eq:50}
G(k,\tau ) = \sqrt{ \frac{5}{4 \pi }}
\frac{k^{3/2} e^{(3/4) \gamma \tau }\ln k}{\gamma^{1/2} \tau^{3/2}}
 e^{-(5/4) \ln^2(k)/\gamma \tau }  .
\end{equation}
This solution expresses $ M(k,t) $ for general $ k $
in terms of its time variation  at some reference wave
number $ k_{ref}  $.

Thus,  if $ M(k/k_{ref} ,t') $ is a delta function 
in time then as $ t $ increases
 the magnetic energy increases as  $ k^{3/2}  $ up to
a peak   near $ e^{5 \gamma t/3} $, beyond which it decreases.
At the same time it increases at a fixed 
$ k $ by the exponential factor $ e^{3\gamma t/4} $.
Thus, the energy grows exponentially  at fixed k and spreads out 
over a range that also  increases  exponentially.
The combination of these
two effects, local growth and growing range, leads to
the growth in the total energy at the rate $ 2 \gamma $.

Since the scale to which the magnetic energy extends
gets small so fast, one expects it to reach a resistive 
scale,  i.e.  a scale at which the resistivity  becomes
important. Restoring the resistivity  term in
equation (47),  one finds a purely growing solution
\begin{equation}\label{eq:51}
M(k,t) =\mbox{ const  }  e^{(3/4) \gamma t } 
k^{3/2} K_0 \left( \sqrt{ \frac{10 \lambda_S }{ \gamma }} k \right) 
\end{equation}
where $ K_0 $ is the Bessel function of the second kind.
This shows that the wave number corresponding to the resistive  
scale is $ k_{\eta } \approx \sqrt{ \gamma/10 \lambda_S } $.
  At larger wavelengths
$ M $ drops off exponentially as the magnetic energy is destroyed
by resistivity.  However, even with  resistive  damping
the total magnetic energy still  increases exponentially,
but  at the slower rate $ 3 \gamma /4 $.  

One can fix the constant so that the energy at some $ k_{ref} $ is 
\newline $ M(k_{ref},t)~= e^{3 \gamma t/4} $.  If one does that, 
then one finds that
at the time the small scale energy first reaches $ k_{\eta} $, 
the total magnetic  energy is larger than that  at $ k_{ref} $,
$ k_{ref} M(k_{ref} ), $ by the factor $ \tilde{v} /k_{ref} \lambda_S  $, 
which is the effective magnetic Reynold's number.

For application to protogalactic magnetic fields, 
it is of interest to estimate the order of magnitude of 
the  growth rate of the small scale turbulent energy 
while  still in the kinematic limit.

In this case, the growth rate is the rate of turnover
of the smallest eddy,
which we take to have wave number $k_{\nu}$.
The eddy turnover time is 
$( k \tilde{v}_k)^{-1} $ where $ \tilde{v}_k $ is the velocity at
the eddy with wave number $ k $.  For Kolmogorov turbulence
$ \tilde{v}_k  \sim k^{-1/3} $ so the turnover rate is
proportional to  $ k^{2/3} $.  The ratio of the turnover
rate of the smallest eddy to that of the largest eddy
is $ (k/k_0)^{2/3} $ where $ k_0 $  is the wave number of the
largest eddy.  Now, if the turbulent energy  is in equipartition 
with the gravitational energy, it  is easy to see that the
turnover rate of the largest eddy is equal to the free fall
time of the protogalaxy.  (This is also equal to the
decay time of the turbulence.)  But $ k_{\nu}/k_0 = R^{3/4} $
where $ R=2 \pi  \tilde{v}_0/k_0 \nu $ is the fluid  Reynold's number.
$ \nu \approx v_{th} \ell   $ is the kinematic viscosity for an
ion thermal speed
$ v_{th} $ and mean free path $ \ell = 1/n \sigma $,
where $ n $ is the ion density and
$ \sigma $ is the Coulomb cross section. 

Combining these results and setting the free fall time
equal to $ \tau_D $ we get 
\begin{equation}  \label{eq:52}
\gamma \tau_D \approx \frac{k \tilde{ v}_k}{k_0 \tilde{v}}
 \approx (R^{3/4})^{2/3} = R^{1/2}  ,
\end{equation} 
or 
\begin{equation}\label{eq:53}
  \gamma \tau_D =
\left(\frac{2 \pi v_0}{k_0 \nu }\right) =
 \frac{v_0 L_0}{v_{th} \ell } \nonumber\\  
  \approx  \frac{L}{\ell }
=\sqrt{ (L n \sigma)}  ,
\end{equation} 
where $ L_0=2 \pi/k_0 $ and we  take $ v_0 \approx v_{th} $
from equipartition. 

Now more specifically we take the baryonic mass of the protogalaxy 
to be $ M_B = 10^{ 11} $ solar masses and the protogalactic radius
to be one hundred kiloparsecs.  The dark mass may be ten times this.
Then the Jeans temperature $ T_J \approx G M m_H/L $ is $
10^{ 7} $ degrees Kelvin or $ T_{ev} \approx 10^{ 3 } $ eV. The coulomb
cross section $ \sigma \approx 10^{ -12}/T^2 \approx 10^{ -18} 
\mbox{cm}^2 $.   
 
 The baryonic density
is given by $ 4 \pi L^3/3 n = M_B $ or $ n \approx 10^{ -3 }
\mbox{ cm}^{-3} $.  Therefore from our equation we get
\begin{equation}\label{eq:54} 
2 \gamma \tau_D\approx 55  .
\end{equation}
(  A more precise estimate and  for a  general galaxy 
yields  $  2 \gamma \tau_D \approx 70/\sqrt{M_{11}} $ and depends
only on the baryonic mass.  $ M_{11} $ is the baryonic mass
of the protogalaxy in units of $ 10^{ 11} $ solar masses.)

Thus,  we see that the smallest eddy turns over seventy  times
during the  free fall time and
in the kinematic limit  the small scale magnetic energy
 grows by a factor  $2 \times  10^{ 30} $  while the rms
field strength would increase by $ 1.7  \times 10^{ 15 } $.  
By equation (33),
 the Biermann battery  generates a field 
of roughly  $ 10^{ -18} $ gauss on this scale,
which leads to a field of  $ 3 \times 10^{ -3 } $  gauss.
However, the energy density of the smallest eddy is only 
$ 1/2 (n m_H v_0^2 / R^{1/2})  \approx 10^{ -14} 
\mbox{ergs }/\mbox{cm}^3 $ so the kinematic assumption
breaks down and saturation of both the smallest eddy
and the magnetic field must occur.

   This amplification  proceeds even  faster in the
 smaller, cooler, less viscous structure such as the
 $\sim 10^6 M_{\odot}$ halos, with
which galaxy formation is generally believed to begin.
However, the estimate of the $ \gamma $ is very dependent
on the turbulence theory which is highly uncertain.
The  estimate given in Kulsrud et al (1997)
gives a value of $ \gamma \tau_D$ which is larger by
$ 300 $ 
and this would produce an increase of the  field energy
by $ 6 \times 10^{ 32 } $,  and a  resulting field
strength of about one  gauss,  and a much larger
magnetic energy density than the smallest eddy.
Again,  saturation of the magnetic field must
occur. Saturation is discussed in the next
section.

\section{The saturation of the small scale magnetic fields}

In the previous  section we discussed  magnetic
fields generated by turbulence and  found   that the small scale
magnetic energy, that   at scales below the smallest
turbulent  eddy,  increases very rapidly, doubling
essentially every turnover time of this smallest eddy.
In the kinematic limit,  
the energy propagates down to the resistive 
scale after which time it is larger than the magnetic energy at
the eddy scale by a factor of the magnetic Reynolds number.

But such a large energy is clearly beyond the kinematic
limit.  The magnetic field  must produce nonnegligible
velocities which saturate the magnetic energy at a much smaller
value.  For large magnetic Reynolds 
numbers the kinematic limit breaks down before the resistive 
scale is reached.  This is because
 the small scale magnetic energy saturates.

This  effect was      pointed out  in a different way by
Cattaneo and Vainshtein (1992), and  Diamond \& Gruzinov (1997).
(Also see Gruzinov \& Diamond (1994, 1995).
  The statement that these authors
make is that the $ \alpha $ coefficients
of the $ \alpha-\Omega $ theory are  suppressed when
the large scale magnetic energy is larger than
the turbulent  kinetic energy  divided by  magnetic Reynolds number.
This puts a severe restriction on any dynamo theory
since the magnetic Reynolds number for galactic 
 turbulence can be as large as $ 10^{ 19} $.
However, if we view their statement
from the point of view of small scale theory
and replace $ M(k_{ref} ) $ by the large scale field taken
at the eddy scale, we see that if their condition is
satisfied and $ M(k_{ref} ) $ is smaller than the
eddy energy divided by the magnetic Reynold's number,
then the  small scale fields reach the resistivity  
scale before saturation.  Otherwise, they are not  able
to  reach the resistive scale. Instead, energy equipartition
of the total magnetic energy of the 
small scale fields is reached with the smallest eddy, 
 and its dynamo effect
is modified.  But  after equipartition is
reached, the small scale energy  stops growing, and
 never gets much larger  than the eddy energy.
Thus,  the suppression of the dynamo coefficients
is not  limited by the  large
factor of  the magnetic Reynold's number.

To understand the saturation process it is necessary to understand
the structure of the small scale fields.  This structure has
been found numerically by Cowley et al. (2004),
motivated by a theoretical
hypothesis of Cowley.  They found  that the magnetic fields on
small scales are far from isotropic, and indeed consist of
folded magnetic field loops  whose length is that of the
eddy size while the thickness of the folds is very small.
(See Figure 3  where such folds are displayed.)  The folded
nature does not emerge from the mode coupling equation 
(39)  of the last section, since this equation 
only refers to $ M(k)$, the magnetic energy
at a given $ k $ averaged over all solid angles.  This averaging
hides the structure of the fields.  As can be inferred from the
figure, the Fourier transform of a folded field assigns a
wave number equal to the reciprocal thickness of the fold
and the long length drops out.

\begin{figure}
\rotatebox{0}{ \scalebox{0.55}{ \includegraphics{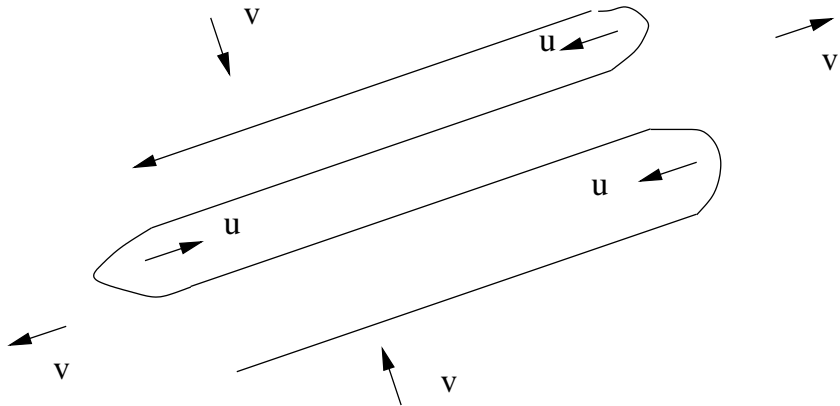} } }
\caption{A folded field.  It is amplified by the stretching
velocities $ {\bf  v}'s $, due to the smallest turbulent eddy.
At equipartition   it can
be unfolded by the velocities $ {\bf  u}'s $ at the Alfv\'{e}n 
speed if Braginski viscosity holds.}
\label{Fig3}
\end{figure}
\vspace*{1in}

On the basis of these folded fields it easy  to understand how
magnetic fields on scales much smaller than the
inner turbulent  scale can be amplified. 
To the
folded fields on such small scales the smallest turbulent
eddy appears quite large and the magnetic folds only  feel 
the effect of  a part of the eddy, the shearing part.
The shear of this part  appears constant in space.  This shear
can also be envisioned as an  incompressible flow
that is compressing in one direction and expanding
in a perpendicular direction.  If the compression direction
is perpendicular to the fold, the fold will be squeezed and
stretched, and its  magnetic field strength will be increased.
Of course, if the compression is parallel to the fold
and the stretching is perpendicular,  then the field strength 
will be decreased.  But for random shifts in direction
of the shearing flow, 
the net effect will be an increase in the mean square of 
the field strength of the folds as well as a decrease
in their thickness.  This mechanism exactly mirrors the
more complex mode-coupling calculations of the last section.
The  validity of   this modeling of the amplification 
 is borne out by numerical simulations.
(Maron et al 2004, Schekochihin \& Cowley 2004). 

From inspection of Figure 3 one can also  see how  the folded
magnetic  fields should tend to saturate. Indeed,
neglecting any viscosity, we see that the folded fields
have tension at their ends and this tension
 should produce a shortening
of the fields at a velocity $ u $  equal to the effective Alfv\'{e}n 
speed based on the strength $ \tilde{  B} $ of the folded fields, divided by
their length.    
 To see this, note that the current density $ j $ perpendicular 
to the fold 
is of order  $  \tilde{ B} /4 \pi \delta $ where $ \delta $ is
the thickness of a single fold.  The field perpendicular
to the fold is also not zero as appears  in the figure,
but is of order $ (\delta /L ) \tilde{  B} $ where $ L $ is
the length of the fold.  Thus, the force on the plasma 
along the fold  is $ F =(\tilde{  B})^2/4 \pi L $, and this leads
to a velocity $ u \approx  (F/\rho ) \times t $ where $ t \approx 
L/u $ is the time to unfold.  Thus,  $ u^2 \approx  (\tilde{  B})^2/
4 \pi \rho $ and so  $ u $  is approximately  equal to
the Alfv\'{e}n speed.
This implies that at saturation, when the small scale
magnetic energy  is the same as the kinetic energy of
the eddy, the unfolding time is comparable to the
eddy turnover time,  and to the exponentiation time
of the field; the folds no longer grow or decay.
In other words,  the small scale fields saturate 
by unfolding as fast as they grow when their energy is in
equipartition with the kinetic energy of the 
smallest eddy.

The same level of saturation is found in the numerical simulations
but the saturation seems to have a different origin since
the folded fields continue to elongate in these simulations.
The different behavior is attributable to the fact that
the numerical simulations include  fluid viscosity  
and this prevents the unwinding motions since the unwinding
velocity on one fold is in the opposite direction to that
on the next fold.  Because of the very great thinness
of the folds, the viscous forces between these flows is
enormous.  (Recall that viscosity terminates the turbulent 
spectrum at the scale of the smallest kinetic eddy whose size
is much larger than the thickness of the folds.)  However,
the viscosity used in the simulations is ordinary scalar
hydrodynamic viscosity.  Because of the magnetic field, the
folds can not transmit shear forces across the magnetic field
for distances larger than the ion gyration radius, a distance
generally much thinner than the thickness of the folds.
The proper viscosity to use in this case  is the Braginski tensor   
viscosity, which does not allow such a perpendicular transfer 
of shear forces.  (Braginski 1965, Malyshkin \& Kulsrud 2002.)

These pictures for the evolution of the folded fields, 
the one based on ordinary viscosity and the other based on
Braginski viscosity, lead to the same equipartition
of the fields generated by the smallest eddy.  However,
they should lead to differences when the next larger
eddy starts to play a role.  For ordinary viscosity 
the next larger eddy should stretch the length of the folds  to 
equal itself.  For Braginski viscosity this
tendency should be suppressed and the magnetic field
should develop a more isotropic structure on the scale 
of the next eddy.   

  In this case, the resulting field 
should have a much larger scale, possibly large  enough
to appear very coherent.  In the case of the protogalaxy, 
the resulting field supplied to the galactic disc
should have a scale comparable with observations 
of the present galactic fields.  On the other hand,
for normal viscosity the magnetic field should end
up very thin and with folds of the length of the largest
eddy.  Its energy   should be in equipartition with
the turbulent energy up to the  eddy for which nonlinear
growth can arise.  This energy can be a fair fraction of
the turbulent energy and thus of the binding energy 
of the entire protogalaxy, so it is possible that
the magnetic field will alter the collapse to the disc.

At  the moment  numerical simulations of 
magnetic field evolution in the presence of
turbulence employ  ordinary
viscosity because Braginski viscosity is numerically difficult
to handle.  As yet, the results from Braginski viscosity are all
based on theoretical estimations, and we must wait for
more sophisticated numerical programs,
 yet to be developed, before the saturation is properly
understood. However,  it is generally
agreed that Braginski viscosity is the correct viscosity 
when  the ion gyration radius is very small compared to 
all the other scales.

An additional problem  has arisen because the Braginski
viscosity in very weak fields leads to instabilities.
(Schekochihin \& Cowley 2004, 2006; 
 Schekochihin et. al.  2005a,b.)
The instabilities  arise because in very weak fields
the changing magnetic fields produce anisotropies
in the pressure tensor that drive fire hose and
mirror instabilities.  These are normally
suppressed in stronger fields.  How these instabilities
saturate is still under discussion.

Assuming the folded fields saturate by unfolding,
Kulsrud has attempted to build an intuitive  model for the
evolution of the protogalactic magnetic fields in Kolmogorov 
turbulence. (Kulsrud 2005) Assuming reasonable parameters
and top down formation of the protogalaxy,  he finds that the fields 
are progressively generated by larger and larger 
eddies, and in the free fall time the size of the 
most intensive fields can reach  about one thirtieth
the size of the entire protogalaxy.  Also, the field strength, 
 after compression into
the disc,   is in the microgauss range. This speculation
does not take into account the instability
problem.

Thus, the saturated fate of magnetic fields generated by turbulence, 
must wait till  numerical simulations with
the appropriate viscosity can be developed and the problems
with the nonlinear Braginski instabilities  understood.

However,  because the validity of the $ \alpha-\Omega $ 
theory is equally in doubt,  it is of interest to
consider the implications of a protogalactic field,
and  its further  evolution after the galactic disc
has formed.

\section{History of the evolution of a primordial magnetic field}

In the last two sections we consider the possibility
that the protogalaxy could generate a strong
enough magnetic  field that it might be considered
 a pregalactic  field.   
What would be the evolution  of a pregalactic  field which 
existed  prior to the formation of the Galactic disc,
and is  subject to the astrophysics of the Galactic disc?
How would it survive the criticisms of Parker discussed
in \S 10?

The answer to this question was addressed in a paper by
Howard \& Kulsrud, (1997).  A condensed  summary 
of this theory is given by  Kulsrud  in Cosmic Magnetic Fields
edited by Wielebinski and Beck (Kulsrud, 2006).  
Here an even briefer treatment of this paper is given.

\begin{figure}
\rotatebox{0}{ \scalebox{0.55}{ \includegraphics{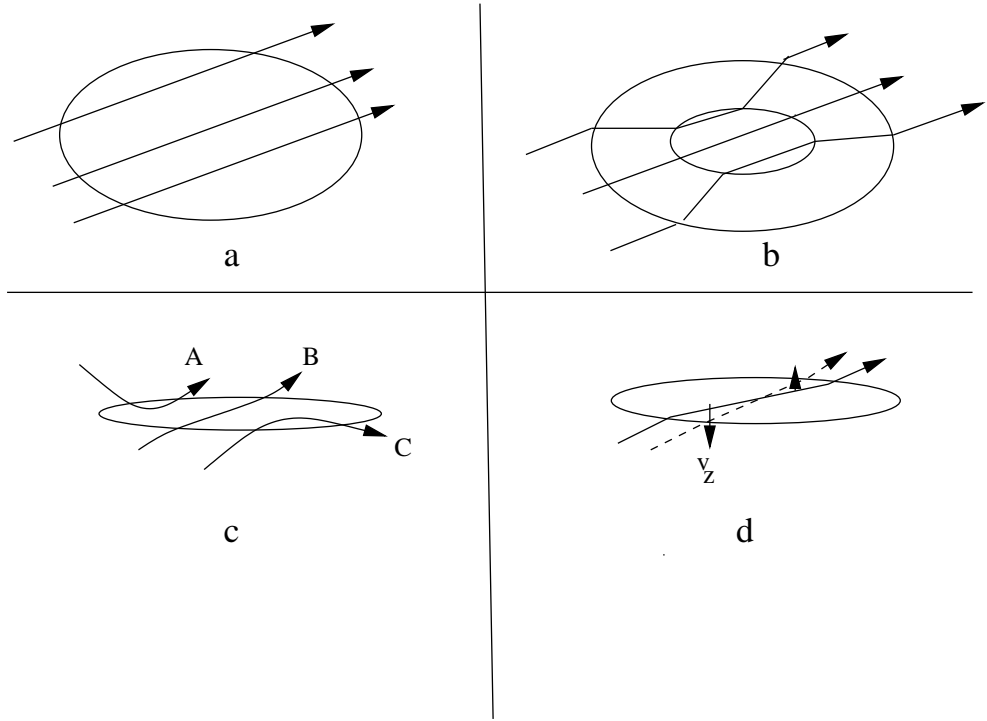} } }
\caption{The early evolution of a primordial field.
(a) The uniform field in the protogalaxy before collapse.
(b) The field lines after the protogalaxy has shrunk spherically and they
have been  drawn inward from the extragalactic region. (c)The field
lines  after the disc forms, the lines A entering and leaving from above,
lines B entering from below and leaving above and lines C entering
and leaving from below.  (d) The solid line before 
ambipolar diffusion  and the dotted line after.} 
\label{Fig4}
\end{figure}
\vspace*{1in}

First, imagine a uniform magnetic field present in the sphere
of plasma that is  to  collapse 
to form the Galactic disc. The behavior of the
field is sketched in Figure 4.  In 
panel (a) the uniform field is displayed
in the protogalactic sphere after it has separated
from  Hubble expansion. In panel (b)
the plasma collapses to a smaller virialized
radius dragging the field lines with it.  However,
it is seen that they are still connected to the
surrounding intergalactic medium.  Possible radii
for these spheres might be 100 kpc and 10 kpc.
 If the field makes a finite angle 
to the rotation axis, the horizontal  component
of the magnetic field 
would be amplified as the inverse radius squared,  or
$ 10^2 $.  Next, in panel (c), the second
sphere has collapsed to a disc which might be
100 pc thick.  In this case the horizontal 
component  would be amplified again by another
factor of 100, and the final field strength would be 
increased by $ 10^4 $ over that of the original  
primordial field.   The vertical component
would  be changed by only a factor  of $ 10^{ 2}$.  

The resulting field
lines are displayed in this third panel.  It is seen
that some lines, of type $ A $,
 enter the disc from above  and also leave
in the above  direction.  However,  some others,
 type $ B $,
enter from below and leave above. Lines of type
$ C $ both enter and leave below. Lines of type
$ B $  are
embedded in the disc and cannot escape without
violating flux freezing.  The other lines do  leave
by ambipolar diffusion.

The subsequent behavior depends on the structure of
the interstellar medium.  It is generally supposed that
the magnetic field is embedded in many clouds. 
It threads its way through these clouds and
eventually finds its way out of the disc as in
Figure 5.  The clouds  hold the magnetic
field in the disk balancing its outward force by
the gravitational field of the stars acting on the mass of the 
clouds.  (In addition, there is also a  cosmic ray pressure
which is held in by the clouds through the action of
the magnetic field and indirectly adds to  the outward pressure
exerted by the magnetic field.)  The low density material
 between these clouds
is assumed to be very light and  unable to hold the field and
cosmic ray pressure down, 
 so the field lines bow up between the clouds.

Thus, the force balance is between the outward pressure
of the magnetic field and cosmic rays acting everywhere
including the space between the clouds.  This is balanced
 by the gravitational force acting only on the clouds
so that the effect of the field and cosmic rays is concentrated
and amplified in its action on the clouds.  If the filling
factor of the clouds in the interstellar medium is
$ f $, then the force on the clouds is  intensified by
a factor of  $ 1/f $.  The effect of the magnetic field is
further increased  by the fact that the clouds
are only very  partially ionized, generally by 
 $ 10^{ -5} $ or $ 10^{ -4} $.
There is no direct magnetic force on the neutrals, but it
is transferred to them by the collisions
between the ions and neutrals.  This force only acts
on  the ions, and the field lines locked on them,
slide through the neutrals, and   through the clouds,
at some ambipolar velocity.  Because of the amplification 
factors it is possible for the ions and the field lines
embedded in them to slip entirely across a cloud.

The situation is not entirely simple since the time
to slip through a cloud is much longer than the
lifetime of the cloud.  (The slippage time is of
the order of  a billion years while the lifetime of a cloud
before it hits another cloud and is destroyed,
is in the range
of ten million years.)  However,  this complication 
is  ignored since after a cloud is destroyed
its plasma is  turned into a rarefied
plasma in which the field is still embedded by flux
freezing.  After a short time a new  cloud reforms
out of this plasma drawing the field lines back into
itself with the ionization dropping back to its
very small value again.  After the lifetime of a single
cloud the matter has worked its way outward 
from the midplane of the disc, in a non
random way.  For a given line the amount of plasma
that is below it (relative to the disc) has increased.
The amount of increase is given by the amount the 
ionized matter, holding down the field line  moves
in order to balance the magnetic and cosmic ray forces
against the ion neutral collision rate.

\begin{figure}
\rotatebox{-90}{ \scalebox{0.55}{ \includegraphics{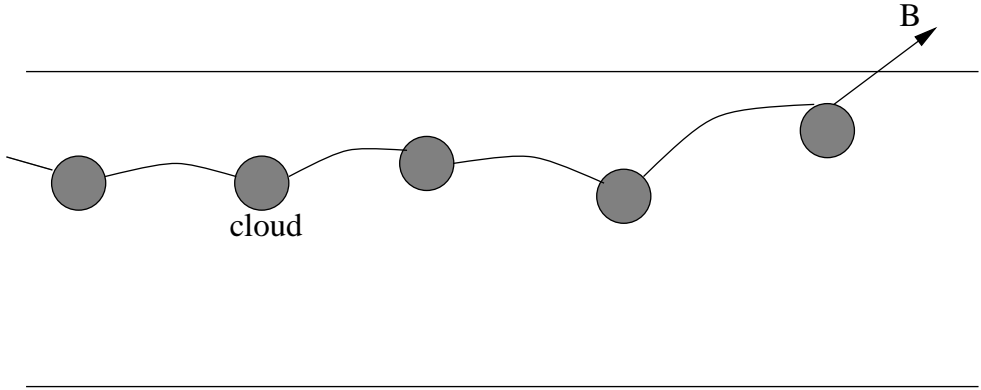} } }
\caption{A line of force penetrates many clouds and finally 
reaches a last cloud through it diffuses and leaves the disc.}
\label{Fig5}
\end{figure}
\vspace*{1in}

Finally, the field line manages to slip through the 
last cloud it threads before it leaves the disc, as
is shown in Figure 5.  Now, this motion will remove lines of type
$ A $ and $ C $,  shown in panel (c)  of figure 4,  out of the disc.
But it cannot move
field lines of type $ B $ out the disc because of their
topology.  However,  when lines of type $ B $  slip through
 the last cloud
mentioned above,  the result is that they shorten their horizontal extent
in the disc as seen in panel (d) of Figure 4.

In discussing the behavior of magnetic fields, 
the authors   ignore any interstellar turbulence 
or dynamo action.  This   enables them  us to isolate 
and simplify the
ambipolar  effects on the magnetic field and its  lines
of force.

The only motions of the interstellar medium that they  include
are the galactic  differential rotation
and the vertical ambipolar velocity.
In cylindrical coordinates these velocities are
\begin{eqnarray} \label{eq:55} 
{\bf  V} = R \Omega (R) \hat{{\bf  \theta }} 
+ V_Z \hat{{\bf  z}} , \nonumber\\ 
V_Z = -K \frac{\partial B^2 /8 \pi }{\partial Z} , 
\nonumber\\ 
K = \frac{(1+\beta/\alpha) }{\rho_i \nu_{in} f} .
\end{eqnarray} 
$ V_Z $ is  the  ambipolar velocity needed to transfer
the cosmic ray and magnetic forces to the neutrals in 
the cloud, $ \rho_i $ is the ion density, $ \nu_{in} $ 
 is the ion-neutral collision frequency and the
factor $ \beta/\alpha $ 
is the ratio of the cosmic ray pressure  to the 
 magnetic pressure, which 
is  assumed  constant with height.

To see how the fields evolve under these motions
 ignore the vertical component of the magnetic field $ B_Z $. 
Then, with these velocities, the magnetic differential 
equations 
for the field components in a frame following the galactic rotation
are (Kulsrud 1990), 
\begin{eqnarray} \label{eq:56}  
\frac{d  B_R}{d  t} =
\frac{\partial B_R}{\partial t} + V_{\theta } 
\frac{\partial B_R}{\partial \theta } =
-\frac{\partial }{\partial Z} (V_Z B_R) ,
\nonumber\\ 
\frac{d  B_{\theta}}{d  t} =
\frac{\partial B_{\theta }}{\partial t} + V_{\theta } 
\frac{\partial B_{\theta}}{\partial \theta } =
-\frac{\partial }{\partial Z} (V_Z B_{\theta} )
+R \frac{d \Omega }{d R} B_R  .
\end{eqnarray}

Let the  half thickness of the disc be $ D $.  Outside  
of the disc it is now appropriate to take these components
to vanish since they are embedded in the very tenuous
plasma that is  ionized in the clouds and has negligible  
mass. (The weakening of the horizontal components,
$ B_R $ and $ B_{\theta} $ of the  field by the $ V_Z $ motion
is due to the shortening of the field line length $ L $,  
 where the horizontal field strength is given by $  B= (L/D) B_Z$,
and  $ B_Z $ is constant.)
 
To form an idea as to the solution  assume
that the fields are parabolic in $ Z $.  The horizontal 
dependence is not given since  a single column
of plasma is followed as it is carried around  the disc
by the differential rotation.

Thus,  take 
\begin{eqnarray} \label{eq:57}
B_R(Z,t)  = B_R(0,t) ( 1 - Z^2/D^2 ) ,
\nonumber\\  
 B_{\theta} (Z,t)  = B_{\theta} (0,t) ( 1 - Z^2/D^2 )  ,
\end{eqnarray}
and solve for the time dependence of the fields at $ Z=0 $,
the coefficients of the  parabolic terms
 on the right hand side of this equation (55). 
Taking advantage of the fact that shear quickly amplifies
 $B_{\theta}$ relative to $B_r$, the solution at the midplane
$Z=0$ can be written with sufficient accuracy  as 
\begin{eqnarray} \label{eq:60} 
B_R = \frac{B_1}{\left[ 1 + (2V_{D1}/3D)\Omega^2 t^3 \right]^{1/2}  } ,
\nonumber\\ 
B_{\theta}  = \frac{-B_1 \Omega t}
{\left[ 1 + (2V_{D1}/3D)\Omega^2 t^3 \right]^{1/2}  } ,
\end{eqnarray} 
where $ B_1 $ is the initial value of $ B_R $, 
$ V_{D1} = K B_1^2/2 \pi D $ is the initial value of
$ V_D $, and   assume that initially $B_{\theta}=0$.
 $ B_{\theta} $ is plotted versus
time in Figure 6  for several  values of $ B_1 $.

\begin{figure}
\rotatebox{0}{ \scalebox{0.55}{ \includegraphics{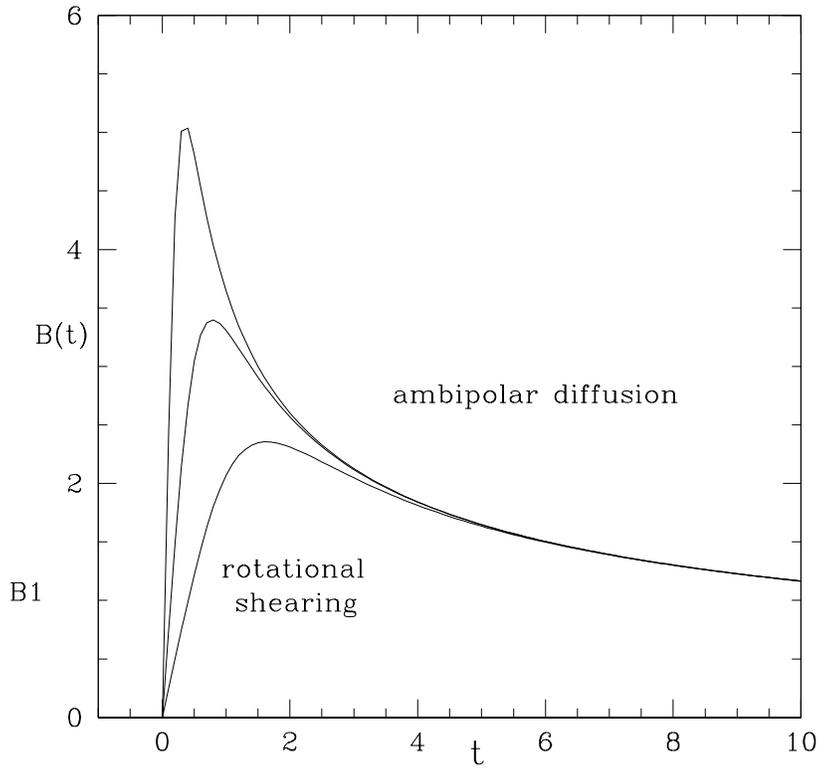} } }
\caption{The evolution of the field strength under differential 
rotation and ambipolar diffusion for initial field strengths
of $ B1=, 0.1,0.3 $ and $ 1 $ microgauss.  The ordinate
is $ B $ in microgauss and the abscissa is time in billions of
years.}
\label{Fig6}
\end{figure}
\vspace*{1in}

These equations display the behavior of the field in
the midplane of the  disc following a rotating fluid element.   
Their behavior can be described qualitatively as follows.

Consider some fluid element that initially has a
relatively weak initial  $ B_R $ component $ B_1 $. At first the
only action is the differential stretching that leads to
a linear growth of the toroidal field $ B_{\theta} $. Then,
as $ B_{\theta} $ grows stronger, the  ambipolar velocity 
starts to grow and become important.  This leads to a 
weakening of $ B_R $, but because of differential  stretching,
$ B_{\theta} $ continues to grow for a time.  Eventually,
$ B_R $ becomes too weak for  stretching to overcome
the weakening of the $ B_{\theta} $,  and 
it too starts to decrease.  Eventually, when the second
terms in the brackets in the denominator dominate,
the two components decrease as powers of $ t $.

The stronger toroidal component then  varies  as 
\begin{equation}\label{eq:61} 
B_{\theta} \approx  \pm \sqrt{ \frac{3 B_1^2 D}{2 v_{D1} }}
\frac{1}{t^{1/2} }
\end{equation}

It is easy to see that this field is just such
that the ambipolar diffusion velocity,  which is proportional
to $B_{\theta}^2 $, moves the field a distance $ D $ in the
time $ t $.  From the definition of $ v_{D1} $
one sees that at any later time, $ t $,  which is sufficiently
long, the field strength is independent of its 
initial value and only depends on the initial sign
of $ B_R $.  Thus, the entire magnetic field throughout the
disc  becomes a constant in space except for its sign.

One can take this sign into account by noting
that,  according to the   model for the primordial field, 
$ B_R = B_1 \cos \theta $ and at a  later time $ \theta
$  increases by $ \Omega(R) t $.  Because of the variation
of $ \Omega $ with $ R $ the magnetic field at the time
$ t $ and fixed $ \theta $ has a square wave behavior with
$ R $.  The spatial period is $ \Delta R =
(d \Omega /d R)^{-1} $.  [See equation (25).] This should be roughly
constant with $ R $ near the sun and is approximately
100 parsecs.  This is the wound  up field 
 discussed in \S 10, in the context of Parker's
discussion of the winding up. 

  Such a field cannot agree with the pulsar rotation measures
because most of the Faraday  rotation in the intervening
interstellar medium would cancel out.  However,  
assume that  the initial magnetic field is not  uniform, but
had a gradient with a scale comparable to the protogalaxy.
Then after collapse $ B_R $ 
 is not sinusoidal.  $ B_{\theta} $ gets its sign
from the initial sign of $ B_R $ in the same piece of plasma.
Thus,  the resultant
radial variation of the final field  $ B_{\theta} $ is 
square wave with the radial extent of one sign being
different from the radial extent of the other sign.

Now,  for  such a field the Faraday rotations
do  not average out.  Further, a field with such
 variations would change on too small a scale to be
resolved by the  analyses of  rotation measures
available at the present time.
This is just the field   introduced in \S 10 to counter
the  objection to the winding up of the pregalactic  field.

It must be remembered  that in this model  all dynamo actions 
and dynamics are  ignored. Including these
would    lead to a  different field structure.

What should be taken seriously is that the suggested
pregalactic  field could give the dynamo a proper start
to overcome the flux loss problem.  The initial
pregalactic  field could be relatively weak, and then
be developed to   sufficient strength by  the standard
$ \alpha--\Omega $ dynamo to satisfy the necessary
vacuum conditions as discussed in \S 9.

 The purpose of the work in Howard \& Kulsrud is not
to directly prove that the magnetic field is primordial 
but to examine the consequences of its being so. It
potentially could give a strong field but would
not give the Grand Design shape that generally
emerges from and $ \alpha--\Omega $ theory.
However, it cannot be so easily ruled out on these grounds
because as shown in \S 10  the perturbations produced
by compression of the field in the spiral arms
can also give a similar pattern to the field lines.

\section{Extragalactic magnetic fields }

Most of this review is  concerned with the magnetic
field of our Galaxy and its possible origin. 
The problems encountered in explaining the origin of the
 Galactic field are not universal --  galaxies with higher star
formation rates or smaller  gravitational fields 
could  expel flux -- but these are not representative of all galaxies.

A much more difficult problem is the origin  
of the magnetic fields  observed in the intracluster
medium  (ICM) of clusters of galaxies (Carilli and Taylor 2002,
 Ensslin et al. 2005).
These fields are coherent on scales of several kpc and in the 
cores of clusters appear to be several microgauss in strength,
comparable to that of galactic fields. It can be
inferred from the enrichment of the ICM with heavy elements, 
and the paucity of gas rich galaxies in clusters, that
interstellar gas is stripped from galaxies and mixed into the ICM. 
However, the gas density of the ICM is 2 - 3 orders of
magnitude lower than the mean density in galaxies, so the field
 in the ICM must be amplified by some mechanism to bring it back up
to microgauss  strength. The large intrinsic scales and lack of
 defined rotation of clusters make the operation of a mean field dynamo
problematic.
Various other  means to produce
the cluster field are discussed in Carilli \& Taylor (2002),
Schekochihin et al (2005a,b), and  Ensslin et al. (2005),
but none intrinsic to the cluster  seem  promising at this time.

 However, a   proposal that seems promising is
   based on 
 the observation that a  large magnetic flux 
  fills each of the giant radio jets whose volume
is enormous. If when such jets disband, their fields
could be dispersed uniformly throughout intergalactic 
space, then there would be enough flux to supply
significant pregalactic fields to every galaxy.
 (Kronberg, Duften \& Colgate 2001).

  This was first pointed out
by Daly \& Loeb (1990).  This result is based on
the assumption that 
radio galaxies
are rather short lived, and  new ones emerge  every few hundred
million years or so, each new one producing more flux.

When the radio jets terminate their life their plasma
expands into the intergalactic medium 
taking their  flux with it.
If the intergalactic medium is turbulent, this
flux  then merges with the surrounding
intergalactic plasma mixing its flux  with it.  
This  intergalactic  plasma  subsequently collapses 
 to  produce  galaxies which are thus already  magnetized
when they form.  
   That the total flux in radio jets is 
adequate  to produce  the observed flux
in clusters and  galaxies,  is  a striking 
numerical coincidence.
How this happens,  the expansion and mixing of the flux,
is not yet understood.  However,  given the problems with other
theories for  the origin of cosmic fields,
this origin  must be  taken seriously and developed
further. 

 Much  work has already been  done
on this problem by Colgate \& Li (2000)  and others.
They 
have developed a preliminary theory to explain the origin  
of the magnetic fields of the jets. Their theory is 
 based on ideas developed by  plasma physicists
to explain some of the interesting fusion experiments.
The principle mechanism which is employed 
is  Taylor relaxation (Taylor 1974, 1986), in which, 
by rapid reconnection, a spherical plasma relaxes to 
a particularly  stable configuration by converting
toroidal flux to poloidal flux and vice versa. 
Determining  whether
this is possible on the large megaparsec scale of radio
jets seems to depend on increasing our understanding
of the fast magnetic reconnection process. Study of
this is  a field now under intense
development. For reviews see Priest \& Forbes (2000),
 Biskamp (2000), Kronberg et al. 2001, Furlanetto \& Loeb 2001, and
Gopal-Krishna \& Witta 2003).

Presumably, the fields in the jets originate in accretion disks.
 Many of the processes already discussed in this article -- the Biermann
battery and , various types of dynamo -- should be able to generate
 and amplify magnetic fields in disks much more rapidly
than in galaxies, particularly if the disks are hot and 
collisionless (Krolik \& Zweibel 2006). Once these  fields 
are ejected into the jets,  the problem remains to amplify them,
 and generate some semblance of a coherent field.

\section{Summary and conclusions}

The origin  of cosmic magnetic fields breaks up 
into three stages: (1) The generation of very
weak seed fields from  an absolutely zero magnetic field.
(2) The amplification of these seed fields to
dynamically interesting fields.  (3) The final 
alteration and further amplification to the
fields we observe today in galaxies and clusters
of galaxies.

The first stage is generally thought to
occur in the primordial era, before and during the time
when galaxies and
their discs form.  There are a variety of mechanisms
for their production, the Biermann battery mechanism 
being the most popular.   Another mechanism is
the generation of fields in stars (by the Biermann battery) and the
later expulsion of their flux into the interstellar medium.
In short, there seems
to be no difficulty with finding mechanisms to 
produce the first stage seed fields,  and not much debate about
which is the most important.  It is just generally
accepted that they are present as needed.

The main debate centers on the second stage, as to
when do dynamos start to amplify seed fields.
Does this occur pregalactically, before galactic discs
are formed, or does it occur over the life time
of galaxies?  The bulk of opinion during  the past decades
favors the latter choice.  It has been generally accepted that the
$\alpha--\Omega $  dynamo takes these seed fields and amplifies
them up to the currently observed fields.  
There are several difficulties with this theory. 
These difficulties are primarily associated with the flux
freezing condition
that says that the total flux cannot change from zero
to a finite value in fixed amount of plasma.
This condition is satisfied in standard disc-dynamo theory
by imposing boundary conditions that imply
that during any  amplification of the field by
 a finite amount, a finite amount of flux is 
 removed from the disc to infinity, leaving behind
an increased amount of flux of the opposite sign.
Thus flux is conserved {\it in toto } but not in
the disc region alone. 

The boundary conditions in the $ \alpha--\Omega $  theory
 that enforce this condition are the 
vacuum boundary conditions.  These conditions
are formally derived by assuming that  turbulent diffusion 
outside of the disc is much larger than that in the disc.
This outside turbulent diffusion is based on 
observations  of velocities and correlation lengths made
in the halo.  But these measurements are made
when the halo has very little matter in it.  If these measurements 
were made at a time when tubes of force carrying
matter with them were diffusing out of the disc,
the gravitational force on the matter would strongly
modify these velocities.  In fact, for very weak fields,
the physics of the removal
is difficult to justify,   by models which treat
the interstellar medium as horizontally  homogeneous.
The conclusion that we can draw is that the $ \alpha--\Omega $ 
theory that treats the the interstellar medium as 
horizontally homogeneous is astrophysically incorrect
and this is likely to be true even in the case of strong fields.

  It is conceivable that a more sophisticated treatment that takes 
into account the non homogeneity of the interstellar medium 
might be justifiable. In this case, 
 when the magnetic field flux leaves,
the field lines and matter would form arcs in which the matter slides down
the tubes of force releasing the remainder of the tube to 
more freely diffuse away.  
But this is clearly not possible when the field is very
weak, so weak that its forces do not affect the motion of the
matter, which would be unaware of the direction to slide down.

  Thus, the standard theory is probably  not able to amplify these
weak fields, and some other mechanism is necessary to
provide this intermediate step in the origin  of
cosmic fields.  That is to say, a pregalactic  mechanism 
might be necessary to  provide fields of reasonable strength
for the {\it  non homogeneous}
$ \alpha--\Omega $  disc dynamo to act on.  In this sense
with a pregalactic origin  to get started such an  $ \alpha--\Omega $ 
dynamo could  play a crucial role in the magnetic field origin.

Even if the flux removal problem is solved, 
the $\alpha--\Omega$ dynamo is not necessarily 
viable. There remains a gap between the
calculations, both analytical and numerical, 
which show tremendous growth of fields at the
 resistive scale, and the mean field
$\alpha--\Omega$ dynamo models, which involve the 
transport of magnetic flux on the much larger supernova
 remnant scales. It 
is supposed  that the small scale fields can saturate,
 leading to dominance of the larger magnetic scales, but 
this remains to be
demonstrated, and may require proper numerical 
implementation of Braginski viscosity.

Two types of  observations tend to substantiate the necessity 
for this pregalactic mechanism.  Beryllium
and boron have been detected in the atmospheres of very old stars
These elements are believed to be produced
only by spallation  of low energy cosmic rays, and thus imply
the early presence of these cosmic rays\footnote{In \S 2 we mentioned that boron can also be synthesized in supernovae,
leaving the implications of its abundance ambiguous}.  Further,
to produce the amount of observed light elements
requires cosmic ray intensities at this very early  time that
are comparable to the present intensities.  This in
turn implies the presence of magnetic fields  
at this early time, although, unless most of the interstellar medium at this early time was
confined to dense clouds, these fields could have been well below equipartition with the
cosmic rays. 

A second  observational result is the detection of
Faraday rotation by magnetic fields inferred to have microgauss
strength, in high--red shift--galaxies in an early forming 
stage.   Such galaxies are too young for the standard
theory to bring seed fields up to such  strengths.

The case for pregalactic   mechanisms for magnetic field generation
is mainly a negative one, based on the nonviability
of the standard non primordial 
mechanisms to accomplish the second origin stage.
 The physics of forming galaxies and their  discs
is not well understood, and one has to rely on numerical 
simulations to draw conclusions concerning it.
However, these simulations {\it do} lead to plausible
theories for production of pregalactic  magnetic
fields by  the turbulence seen in the simulations,
provided that the problem of preferred amplification at
 the small scales can be overcome.

  Lastly,  the magnetic flux observed 
 in giant radio jets could be dispersed into the 
intergalactic  plasma magnetizing  it coherently 
before galaxies form.  This  provides
a very direct  pregalactic  origin.  Unfortunately,
  no detailed theory as to how or whether  this can
happen, has yet  developed.  Also no
observational evidence of it has yet emerged.

  In summary, the situation is as follows. The
standard theory of the generation of magnetic fields
through galactic disc dynamos suffers from the
difficulty of flux expulsion in the case of our
Galaxy and no consistent theory
that resolves this problem has yet been developed.
A further difficulty is whether the galactic dynamo 
works on very weak fields since most such dynamo theories invoke
magnetic fields of considerable strength to function.

Assuming these difficulties cannot be overcome it
is necessary that there be an initially strong pregalactic
field present before the disc forms.  Such a field can
provide the galactic dynamo with a strong enough
field to function if the flux expulsion can be handled.

The flux expulsion might be solved if there are enough
 superbubbles containing much more than the average
number of supernovae that actually blow mass 
entirely out of the galaxy carrying the flux with it.
The statistics of superbubbles are still unknown
and no theory has yet been developed to show that they
can accomplish this task.  If they can they will produce
an inhomogeneous disc field that is very discontinuous
whose lines of force consist of many finite segments 
with  ends  connected to the intergalactic medium.
These are the fields that arise if one relies on
 a galactic wind to be the  solution. Whether the
 field arising from such an inhomogeneous
theory   is compatible  with the present field
and the astrophysics  
of the interstellar medium associated with it 
is certainly an open question.

Because of these serious problems with the disc dynamo
it is necessary to ask whether the pregalactic sources
for the disc field are not the better solution  to the
origin problem at least for our Galaxy.  Because 
the effort devoted to pregalactic theories is so 
far inadequate to prove that this   is a viable origin  
for pregalactic fields,  one
cannot easily discard this possibility.

To conclude, we propose the following tentative scenario for
 the origin  and growth of galactic magnetic fields {\it is
the most likely one:}
\begin{enumerate}
\item Fields in the 10$^{-18}$ - 10$^{-20}$G range 
originate by a battery process in cosmological shocks or 
ionization fronts.

\item The battery generated fields are slightly amplified by
 compression and significantly amplified by
turbulence as protogalaxies and similar substructure forms.
 The resulting fields are in the microgauss  range,  and
are coherent over scales of order 1 kpc. This requires that
 amplification be due to the action of the largest
scale eddies, and that the small scale folded fields produced
 by the small scale eddies quickly saturate.

\item Once a magnetized galactic disk forms, its magnetic field 
is sustained against gas addition and 
subtraction processes by a supernova driven $\alpha--\Omega$ dynamo.
 Supernova energy and magnetic and cosmic
ray buoyancy efficiently expel magnetic flux of the wrong sign,
 allowing the dynamo to work. The field is
strong enough that amplification at the small scales is no 
longer a problem.
\end{enumerate}

It appears that to really resolve these questions concerning
the value of a pregalactic field preceding the 
for mation of a disc,  {\it  more}  observations are desperately needed.
 Hopefully,
these can be expected in the near future as telescopes
and detectors improve. (See Gaensler et al. 2004, Gaensler,
2006).   They will be able to 
 find many more examples of early forming extra galactic systems
(damped Lyman alpha systems) and   determine whether the majority 
already contain  strong magnetic fields.

We are happy to acknowledge useful discussions with Katia Ferri\`{e}re, 
Rainer Beck,  Jin-Lin Han, Artie Wolfe and Yansong Wang, and comments by the referees.
 The work was supported by
 NSF grants PHY 0215581, AST 0507367, and 
by the Center for Magnetic Self-Organization (CMSO).

\section{Publications}

\begin{itemize} 

\item
Alfv\'{e}n, H. and Falth\"{a}mmer 1963
{\it Cosmical Electrodynamics: Fundamental Processes} 
Clarendon Press, Oxford

\item
 Backus G 1958
{\it Class of Self-sustaining Dissipative Spherical
Dynamos}
Annals of Physics {\bf  4}, 372

\item
 Batchelor, G.K. 1950
{\it On the spontaneous Magnetic Field in a Conducting
Fluid in Turbulence Motion}
Proc. Roy. Soc. London {\bf  A201} ,405

\item 
Beck, R.	2007
{\it  	Magnetic Field Structure from Synchrotron Polarization}
EAS {\bf  23}, 19

\item 
 Beck R, Brandenburg A, Moss D, Shukurov, A., Sokoloff, D., 1996
{\it Galactic magnetism: Recent developments and perspectives }
ARA\&A  {\bf  34}: 155.

\item 
 Biermann, L. 1950,
 {\it \"{U}ber den Ursprung 
der  Magnetfelder auf Sternen und im interstellaren Raum}
Z.Naturforsch {\bf  5a} 65

Biskamp, D., 2000,
{\it Magnetic Reconnection in Plasmas}
{\it Cambridge Monographs in Plasma Physics }
Cambridge University Press, Cambridge

\item 
Blackman, E.G. Field, G.B. 1999
{\it Resolution of an ambiguity in dynamo
theory and its consequences for back-reaction studies} 
ApJ. {\bf  521}, 597

\item
Blackman, E.G. Field, G.B. 2000
{\it Constraints on the magnetic of $ \alpha $ in dynamo theory}
ApJ {\bf  524}, 984

\item
Boldyrev, S. \& Cattaneo, F. 2004
{\it Magnetic field generation in Kolmogorov turbulence}
Phys Rev Lett {\bf 92}, 144501

\item 
 Braginski, 1965
{\it Transport Processes in a Plasma }
in {\it Reviews of Plasma Physics} Vol I, p 105
ed. M.A. Leontovitch, trans. H. Lashinsky,
Consultants Bureau, New York

\item 
 Brandenburg, A.; Moss, D.; Shukurov, A.,	1995
{\it 	Galactic Fountains as Magnetic Pumps}
MNRAS. {\bf  276} ,651

\item 
 Brandenburg, A.; Subramanian, K., 2005	
{\it 	Astrophysical magnetic fields and nonlinear dynamo theory}
Phys. Reports {\bf  417}, 1

\item
Brown, J.C. \& Taylor, A.R. 2001
{\it The structure of the magnetic field in the outer galaxy 
from rotation measure observations through the disk}
ApJ {\bf 563}, 31

\item 
 Carilli, C. L.; Taylor, G. B.	2002
{\it 	Cluster Magnetic Fields}
ARA\&A..{\bf  40}, 319

\item 
 Chandrasekhar, S.	1949
{\it 	Turbulence - a Physical Theory of Astrophysical Interest.}
ApJ...{\bf  110}, 329

\item 
 Colgate, S. A.; Li, H.	2000
{\it 	The Magnetic Fields of the Universe and Their Origin} in
{\it 	Highly Energetic Physical Processes and Mechanisms for
 Emission from Astrophysical Plasmas,
 Proceedings of IAU Symposium 195} , 255

\item 
 Cowling, T.G. 1934
{\it Magnetic Fields of Sunspots}
MNRAS {\bf  94 }, 39

\item 
 Cowling, T.G. 1953
{\it Solar Electrodynamics}  in
{\it The Sun} edited by G. P. Kuiper
 University of Chicago Press, Chicago, 532

\item 
 Dobler W, Poezd AD, Shukurov A 1996
{\it Galactic dynamos have movable boundaries }
A \& A   {\bf  312} , 663

\item 
 Daly, R. A.; Loeb, A  1990
{\it 	A possible origin of galactic magnetic fields}
ApJ...{\bf  364}, 451

\item 
 Diamond, P.H., Gruzinov, A.V. 1997
{\it Suppression of Cross-Field Turbulent Transport
of Passive Scalar concentration in Two 
Dimensional Magnetohydrodynamics}
Phys. Rev. Lett. {\bf  78}, 3306

\item 
 Duncan, D. K.; Primas, F.; Rebull, L. M.;
 Boesgaard, A. M.; Deliyannis, C.; Hobbs, L. M.;
 King, J. R.; Ryan, S.  1998	
{\it 	The Evolution of Galactic Boron and
 a New Site for the Production of the Light Elements }
ASPC..{\bf  143} 354

\item
 Elsasser, W. M. 1946 {\it Inductive effects in Terrestrial 
Magnetism}  Phys. Rev. {\bf  69}, 106

\item 
 Elsasser, W. M. 1950 {\it The Earth's Interior and Geomagnetism}
 Rev. Mod. Phys. {\bf  22}, 1

\item
Ensslin, T., Vogt, C., \& Pfrommer, C. 2005
{\it Magnetic fields in clusters of galaxies},
in {\it The Magnetized Plasma in Galaxy Evolution},
 eds. K. Chyzy, K. Otmianowska-Mazur, M. Soida, \& R.
Dettmar, Jagellonian University, Krakow, p. 231

\item
 Fermi, E., 1949
{\it On the Origin  of Cosmic Radiation}
Phys. Rev. {\bf  75}, 1169

\item
Ferri\`{e}re, K. 1992a
{\it 	Effect of an ensemble of explosions on the Galactic dynamo.
 I - General formulation}
 ApJ {\bf  389} ,286

\item 
 Ferri\`{e}re, K. 1992b
{\it Effect of the explosion of supernovae and 
superbubbles on the Galactic dynamo}
ApJ.{\bf  389},  3891992ApJ...391..188F

\item 
 Ferri\`{e}re, K. 1993a
{\it The full alpha-tensor due to supernova 
explosions and superbubbles in the Galactic disk}
ApJ {\bf  404}, 162

\item 
 Ferri\`{e}re, K. 1993b
{\it 	Magnetic diffusion due to supernova 
explosions and superbubbles in the Galactic disk}
ApJ {\bf  409}, 248

\item
 Ferri\`{e}re, K. 1996
{\it Alpha-tensor and diffusivity tensor due to 
supernovae and superbubbles in the Galactic disk near the Sun.}
 A\&A, {\bf  310}, 438

\item 
 Ferri\`{e}re, K. 1998
{\it Alpha-tensor and diffusivity tensor 
due to supernovae and superbubbles in the Galactic disk}
A\&A, {\bf 335}, 488

\item 
 Ferri\`{e}re, K. and Schmitt, D  2000
{\it 	Numerical models of the galactic dynamo 
driven by supernovae and superbubbles}
A\&A, {\bf  358}, 125

\item 
 Field, G.  1995
{\it Is the Dynamo Theory of the Galactic 
Magnetic Field Wrong?}
in {\it The Physics of the Interstellar Medium and 
the Intergalactic  Medium} ed. A.Ferrara, C.F. McKee,
C. Heiles, P.R. Shapiro, Astron. Soc. Pac. Conf. Ser.{\bf  80},1

\item 
 Furlanetto, S. R.; Loeb, A. 2001	
{\it 	Intergalactic Magnetic Fields from Quasar Outflows}
ApJ.{\bf  55}, 619

\item 
 Gaensler, B. M., Beck, R., and Feretti, L.  2004
{\it The origin and evolution of cosmic magnetism
(About the kilometer square array and its use to
find primordial fields)}
 New Astronomy Reviews 48 1008-12

\item 
 Gaensler, B.M. 2006
{\it The Square Kilometre Array: a new probe of cosmic magnetism }
Astron Nach. {\bf  327},   387

\item 
 Garcia-Lopez, R J.; Lambert, D L.; Edvardsson, B; Gustafsson, B; 
Kiselman, D; Rebolo, R	1998
{\it 	Boron in Very Metal Poor Stars}
ApJ.{\bf  500} 241

\item 
 García-López, R J.	1999
{\it 	Consistent Analysis of Oxygen, Lithium,
 Beryllium, and Boron Abundances in Metal-Poor Stars}
ASP Conference Series,{\bf  171} ,77

\item
Gilmore, G; Wilkinson, M; Kleyna, J; Koch, A; Wyn Evans, N.;
 Wyse, R. F. G.; Grebel, E. K.,  2006	
{\it  	Observed Properties of Dark Matter: dynamical studies of dSph galaxies},
astro-ph/0608528

\item
Glatzmaier, G.A. \& Roberts, P.H. 1995
{\it A three-dimensional self-consistent computer simulation of a geomagnetic field reversal}
Nature, {\bf 377}, 203

\item
Gopal-Krishna, Witta, P.J., 2003
{\it Radio Galaxies and the Magnetization of the 
IGM} in
{\it Radio Astronomy at the Fringe} ASP 300th Conference 
eds. J.A. Zensus, M.H. Cohen, E. Rosner, R

\item 
Gnedin, N.Y.; Ferrara, A. \& Zweibel, E., 2000	
{\it 	Generation of the Primordial Magnetic Fields 
during Cosmological Reionization}
ApJ {\bf  539}, 505

\item
Grasso, D. \& Rubenstein, H.R. 1995, Astropart. Phys., {\bf 3}, 95

\item 
 Gruzinov, A.V., Diamond,P.H. 1994
{\it Self-Consistent Theory of Mean Field Electrodynamics}
Phys. Rev. Lett. {\bf  72}, 1651

\item 
 Gruzinov, A.V., Diamond,P.H. 1995
{\it Self-consistent mean field electrodynamics
of true dynamos}  
Phys. Plasmas {\bf  }, 1941

\item
 Han, J.L. 2001
{\it Magnetic Fields in our Galaxy: How Much Do We Know?},
Ap\&SS {\bf  278}, 181

\item 
 Han J.L. 2002
{\it Magnetic fields in our Galaxy: How much do we know? 
(II) Halo fields and the global field structure},
AIP Conference Proceedings, {\bf  609},  96

\item
 Han, J.L.,2003
{\it The Global Structure  Magnetic Fields in our Galaxy},
Supplement Issue, Acta Astronomica Sinica, {\bf 44} 148

\item 
 Han, J.L.  2006
{\it 	Magnetic fields in our Galaxy: How much do we know? 
III. Progress in the last decade},
astro-ph0603512

\item 
 Han, J.L., Manchester, R.N., Berkhuijsen, E.M., Beck, R., 1997
{\it Antisymmetric rotation measures in our Galaxy:
 Evidence for an A0 dynamo },
A \& A,  {\bf   322}, 98

\item 
 Han, J. L.; Manchester, R. N.; Qiao, G. J.	1999
{\it Pulsar rotation measures and the magnetic structure of our Galaxy},
MNRAS {\bf  306}, 371

\item
 Han, J.L.; Manchester, R.N.; Lyne, A.G.; Qiao, G.J.;
 van Straten, W. 	2006
{\it 	Pulsar Rotation Measures and the Large-Scale
 Structure of the Galactic Magnetic Field},
ApJ {\bf  642}, 868

\item
 Han, J.L., Wielebinski, R., 2002
{\it Milestones in the observations of cosmic magnetic fields }
Chinese J.F Astron. and Ast {\bf  2}293

\item 
Hanasz, M.; Kowal, G.; Otmianowska-Mazur, K.; Lesch, H., 2004	
{\it  	Amplification of Galactic Magnetic Fields by the
 Cosmic-Ray-Driven Dynamo}
ApJ {\bf  605L}, 33

\item 
 Harrison, E.R. 1970
{\it Generation of Magnetic Fields in the Radiation Era},
MNRAS {\bf  147}, 279

\item Heiles, C. 1990 {\it Clustered supernovae versus 
the gaseous disk and halo},
ApJ {\bf  354}, 483

\item
Heiles, C. 1996
{\it The local direction \& curvature of the Galactic magnetic field derived from starlight polarization},
ApJ {\bf 462}, 316

\item 
Heiles C, 1998
{\it Zeeman splitting opportunities and techniques at Arecibo },
Ast.  Lett. and Comm. {\bf  37}, 85 

\item
Heitsch, F., Zweibel, E.G., Slyz, A., \& Devriendt, J.E.G. 2004
{\it Turbulent Ambipolar Diffusion: Numerical Studies in Two Dimensions},
ApJ {\bf 603}, 165

\item 
 Howard, A. M.; Kulsrud, R. M., 1997	
{\it The Evolution of a Primordial Galactic Magnetic Field},
ApJ {\bf 483}, 648

\item 
Hoyle, F. 1958
{\it the steady state theory} 
in {\it La structure et L'evolution de L'univers:
conseil de Physique Solvay Bruxelles}, 53

\item 
 Kang, H.; Cen, R.; Ostriker, J. P.; Ryu, D.  1994	
{\it 	Hot gas in the cold dark matter scenario: 
X-ray clusters from a high-resolution numerical simulation},
ApJ {\bf  428}, 1.

 \item 
 Kazantsev, A.P., 1968
{\it Enhancement of a Magnetic Field by a Conducting Fluid},
Sov. Phys.  Sov. Phys. JETP {\bf  26}, 1031

\item 
	Kleeorin, N.; Moss, D.; Rogachevskii, I.; Sokoloff, D.	2000
{\it  Helicity balance and steady-state strength
 of the dynamo generated galactic magnetic field},
A \& A {\bf  361L}  5

\item
	Kleeorin, N.; Moss, D.; Rogachevskii, I.; Sokoloff, D. 2002	
{\it  	The role of magnetic helicity transport in nonlinear galactic dynamos},
A \&A {\bf  387}, 453
	 
\item 	Kleeorin, N.; Moss, D.; Rogachevskii, I.; Sokoloff, D.	2003
{\it  	Nonlinear magnetic diffusion and 
magnetic helicity transport in galactic dynamos},
A \& A {\bf  400}, 9

\item 
 Kraichnan, R.H., Nagarajan, S., 1967
{\it Growth of Turbulent Magnetic Fields},
Phys.  Fluids {\bf  19}, 859

\item 
 Krause, F., R\"{a}dler, K.-H. 1980
{\it Mean Field Magnetohydrodynamics and
Dynamo Theory}, Pergamon Press, Oxford

\item
Krolik, J.H. \& Zweibel, E.G. 2006
{\it The weak field limit of the magnetorotational instability},
ApJ {\bf 644}, 651

\item 
 Kronberg, P.P. 1994
{\it Extragalactic Magnetic Fields },
Rep. Prog. Phys. {\bf 57}, 325

\item 
 Kronberg, P.P., Dufton, Q.W., Li, H., \& Colgate, S.A. 2001
{\it Magnetic Energy of the Intergalactic 
Medium from Galactic Black Holes},
ApJ {\bf  560}, 178

\item 
 Kulkarni, S. R.; Fich, M.	1985
{\it 	The fraction of high velocity dispersion H I in the Galaxy}
ApJ {\bf  289}, 792

\item 
 Kulsrud, R.M. 1964
{\it  General Stability Theory in Plasma Physics } 
Lecture Notes from Course XXV, {\it Advanced Plasma Theory},
International School of Physics, Enrico Fermi,
Varenna,Italy, 1962 ed. M.N. Rosenbluth,Academic Press, New York

\item 
 Kulsrud, R.M. 1990
{\it The present state of a primordial galactic field},
in {\it  Galactic and intergalactic magnetic fields},
 Proceedings of the 140th Symposium of IAU, Heidelberg,
 1989. , Kluwer Academic Publishers, Dordrecht, Netherlands, 527

\item 
 Kulsrud, R.M. 1999
{\it A Critical Review of Galactic Dynamos},
ARA\&A..{\bf  37} 37

\item
 Kulsrud, R.M. 2000
{\it The Origin  of Magnetic Fields},
Proceedings of the International School of Physics,
Enrico Fermi, Course CXLII, 
{\it Plasmas in the Universe} ed. B. Coppi, A. Ferrari,
IOS Press, Amsterdam, 107

\item 
 Kulsrud, R.M. 2005	
{\it 	Plasma physics for astrophysics}
{\it Princeton series in astrophysics}
 Princeton University Press,  Princeton

\item 
Kulsrud, R.M. 2006
{\it The Origin  of Galactic Magnetic Fields},
in {\it Cosmic Magnetic Fields}, eds. R. Wielebinski, 
R. Beck, Springer, Berlin, Heidelberg, 69

\item 
 Kulsrud, R. M.; Anderson, S. W.	1992
	{\it The spectrum of random magnetic fields
 in the mean field dynamo theory of the Galactic magnetic field},
ApJ..{\bf  396}, 606

\item 
 Kulsrud, R. M.; Cen, R.; Ostriker, J. P.; Ryu, D.	1997
	{\it The Protogalactic Origin for Cosmic Magnetic Fields},
ApJ. {\bf  480}, 481

\item
 Kulsrud, R; Pearce, W. P.,	1969
{\it 	The Effect of Wave-Particle Interactions 
on the Propagation of Cosmic Rays},
ApJ. {\bf  156},  445

\item 
 Larmor, J. 1919
{\it  How could a rotating body Such as the Sun become
a Magnet?},  
{\it Rep. 87th Meeting Brit. Assoc. Adv. Sci, 1919}
John Murray, London

\item 
 Li, H; Lapenta, G; Finn, J M.; Li, S; Colgate, S A.	2006a
{\it 	Modeling the Large-Scale Structures of 
Astrophysical Jets in the Magnetically Dominated Limit},
ApJ {\bf  643}, 92

\item 
Li, H.; Nakamura, M.; Li, S.	2006b
{\it 	3D MHD Simulations of Large-Scale Structures of Magnetic Jets},
American Physical Society Meeting April 2006, abstract D1.092

\item 
Lin, C. C.; Yuan, C.; Shu, Frank H., 1969	
	{\it On the Spiral Structure of Disk Galaxies.
 III. Comparison with Observations},
ApJ. {\bf  155} 721L

\item 
 Lingenfelter, J.C. , Higdon, J. C. , Ramaty, R. 2000
{\it Cosmic Ray Acceleration in Superbubbles and the 
Composition of Cosmic Rays, }
astro-ph 0004166

\item 
 MacLow, M. and  McCray, R. 1988
	{\it Superbubbles in disk galaxies }
ApJ.{\bf  324}, 776

\item 
 Malyshkin, L. M.; Kulsrud, R. M., 2002
{\it 	Magnetized Turbulent Dynamo in Protogalaxies}
ApJ...{\bf  571}, 619

\item 
Manchester, R. N., 1974	
	{\it Structure of the Local Galactic Magnetic Field},
 ApJ.{\bf  188}, 637

\item
 Maron, J, Cowley, S., McWilliams, 2004
{\it The Nonlinear Cascade},
ApJ {\bf  603}, 569

\item
McCray, R. \& Kafatos, M. 1987
{\it Supershells \& Propagating Star Formation},
ApJ {\bf 317}, 190

\item 
 McCray, R.; Snow, T. P., Jr.	1979,
{\it 	The violent interstellar medium},
ARA\&A {\bf  17}, 213

\item  
 Meneguzzi, M.; Audouze, J.; Reeves, H.	1971
{\it 	The production of the elements Li, Be, B by 
galactic cosmic rays in space and its relation 
with stellar observations},
Astron. \& Astrophys. {\bf  15} , 337

\item 
 Mishustin, I.M. , Ruzmaikin, A.A. 1972
{\it Occurrence of ``Priming" Magnetic Fields During the Formation
of Protogalaxies},
Sov. Physics {\it Sov. Phys. JETP} {\bf  34} 233

\item
 Moffatt, H. K. 1978
{\it Magnetic Field Generation in Electrically 
Conducting Fluids},
Cambridge University Press, England

\item 
 Moss D, Shukurov A, Sokoloff D, 1998
{\it Boundary effects and propagating magnetic fronts in disc dynamos },
Geoph. Ast. Fluid Dynamics {\bf  89},  285

\item 
 Newcomb, W.A. 1958
{\it Motion of Magnetic Field Lines},
Ann. Phys.  {\bf  4}, 372

\item 
Ogden, D.E., Glatzmaier, G.A., \& Coe, R.S. 2006
{\it The effects of different parameter regimes in geodynamo simulations},
GAFD, {\bf 100}, 107

\item
 Oren, A. L., \& Wolfe, A.M.,  1995  
{\it  A Faraday-rotation search for magnetic fields in quasar damped
Ly-$ \alpha  $   absorption systems }
ApJ {\bf  445}, 624

\item 
 Parker, Eugene N. 1955	
{\it 	Hydromagnetic Dynamo Models},
ApJ. {\bf  122}, 293

\item 
 Parker, E. N.	1970a
{\it 	The Origin of Solar Magnetic Fields},
ARA\&A {\bf  8}, 1

\item
 Parker, E. N.	1970b
{\it 	The Origin of Magnetic Fields},
ApJ {\bf  160}, 383

\item 
 Parker, E. N.	1970c
{\it 	The Generation of Magnetic Fields in
 Astrophysical Bodies. I. The Dynamo Equations},
ApJ {\bf  162}, 665

\item
 Parker, E. N.	1971a
{\it 	The Generation of Magnetic Fields in 
Astrophysical Bodies. II. The Galactic Field},
ApJ   {\bf  163} 255

\item 
 Parker, E. N.	1971b
{\it 	The Generation of Magnetic Fields in 
Astrophysical Bodies. III. Turbulent Diffusion of 
Fields and Efficient Dynamos},
ApJ {\bf  163}, 279

\item 
 Parker, E. N.	1971c
{\it 	The Generation of Magnetic Fields in 
Astrophysical Bodies. V. Behavior at Large Dynamo Numbers},
ApJ {\bf  165}, 139

\item 
 Parker, E. N.  1971d
{\it 	The Generation of Magnetic Fields in
 Astrophysical Bodies.VI. Periodic Modes of the Galactic Field},
ApJ {\bf  166}, 295

\item
 Parker, E. N.	1971e
{\it 	The Generation of Magnetic Fields in
 Astrophysical Bodies. VII. Dynamical Considerations},
ApJ {\bf  168}, 239

\item 
 Parker, E. N.	1973a
{\it 	The Generation and Dissipation of
 Solar and Galactic Magnetic Fields},
A\&SS {\bf  22},  279

\item 
Parker, E. N.,	1973b
	{\it  Extragalactic Cosmic Rays and the Galactic Magnetic Field},
Ap\&SS,{\bf  24}, 279

\item 
  Parker, E. N.	1979
{\it
Cosmical magnetic fields: Their origin and their activity},
Oxford, Clarendon Press; New York, Oxford University Press, 1979

\item 
 Parker, E. N.	1992
{\it 	Fast dynamos, cosmic rays, and the Galactic magnetic field},
ApJ {\bf  401}, 137

\item
 Perry J.J. 1994
{\it Magnetic Fields at High Redshift} 
in {\it Cosmical Magnetic Fields: Contributed Papers
in Honor of Professor L. Mestel}, Clarendon Press, Oxford, 144

\item 
 Poezd, A.; Shukurov, A.; Sokoloff, D.	1993
{\it 	Global Magnetic Patterns in the
 Milky-Way and the Andromeda Nebula},
   MNRAS, {\bf  264}, 285

\item 
Priest,
 E., Forbes, T. 2000
{\it Magnetic Reconnection },
Cambridge University Press, Cambridge

\item
 Primas, F.; Duncan, D. K.; Peterson, R. C.; Thorburn, J. A.	1999
{\it A new set of HST boron observations. 
I. Testing light elements stellar depletion},
	 A\&A, {\bf  343}, 545

\item 
 Prochaska, J. X.; Howk, J. C.; Wolfe, A. M. 2003
{\it 	The elemental abundance pattern in a galaxy at z = 2.626},
Nature, {\bf   423}, 57

\item 
 Pudritz, R. E.; Silk, J.,  1989	
The origin of magnetic fields and primordial stars in protogalaxies
ApJ {\bf  342},  650

\item 
 Rafikov, R. R.; Kulsrud, R. M. 2000	
{\it 	Magnetic flux expulsion in powerful superbubble 
explosions and the $ \alpha-\Omega $  dynamo},
MNRAS {\bf  314}, 839

\item
Ramaty, R., Kozlovsky, B., Lingenfelter, R.E., \& Reeves, H. 1997
{\it Light elements and cosmic rays in the early Galaxy},
ApJ {\bf 488}, 730

\item 
 Ramaty, R.; Lingenfelter, R. E.; Kozlovsky, B. 2000
{\it 	LiBeB Evolution: Three Models} in
{\it 	The Light Elements and their Evolution,
 Proceedings of IAU Symposium 198}, 
 Edited by L. da Silva, R. de Medeiros,\& M. Spite, 51

\item 
 Ramaty, R; Scully, S T.; Lingenfelter, R E.; Kozlovsky, B, 2000	
{\it 	Light-Element Evolution and Cosmic-Ray Energetics},
ApJ {\bf  534} 747

\item 
 Ramaty, R; Tatischeff, V; Thibaud, J. P.; Kozlovsky, B;
 Mandzhav,  N, 2000	
{\it 	6LI from Solar Flares},
ApJ.{\bf  534L} 207

\item  
 Rand, R. J.; Kulkarni, S. R. 1989 
{\it 
The local Galactic magnetic field}
 ApJ {\bf 343}, 760

\item 
 Rees, M. 1987
{\it the Origin and Cosmogonic Implications of
Seed Fields},
Q. J. Roy. Ast. Soc. {\bf  28}, 197

\item 
 Rees, M., 1994
{\it Orig of the Seed Magnetic Field for a Galactic
Dynamo}, in {\it Cosmical Magnetism} ed. D.  Lynden-Bell
Kluwer Academic Publishers,  Netherlands, 155

\item
 Rees, M. 2005
{\it Magnetic Fields in the Early Universe},
in {Cosmic Magnetic Fields} eds. R. Wielebinski, 
R. Beck, Springer, Berlin, Heidelberg, 1

\item
 Rees, M. 2006
{\it Origin   of cosmic magnetic fields},
Astron. Nachr. {\bf  327}, 395

\item Rees, M. J.\& Setti, G., 1968	
	{\it Model for the Evolution of Extended Radio Sources},
Nature, {\bf 219},  127. 

\item
 Reeves, H 1994
{\it On the Origin  of the Light Elements (Z<6) },
Rev Mod Phys {\bf   66}, 193

\item 
 Reeves, H. 2007
{\it The Origin  of Early $ 6\mbox{Li} $ and the Reionization of the
Universe}

\item
Reeves, H. 2005
{\it Element stratification in Stars, 40 years of atomic diffusion},
eds G. Alecian, O. Richard and S. Vauclair
EAS Publication Series {\bf  9} astro-ph 0509380

\item
Roberts, W.W. \& Yuan, C. 1970
{\it Application of the density wave theory to the spiral structure of the Milky Way system. III. Magnetic field: hydromagnetic shock
formation},
ApJ {\bf 161}, 887

\item
Rosner, R
 \& Deluca, E 1988
{\it On the Galactic Dynamo}
in  {\it The Center of the Galaxy}, IAU Symposium
{\bf  136}, ed M. Morris, 319

\item 
 Ruzmaikin, A. A.; Sokoloff, D.D.; Shukurov, A. M., 1988	
{\it Magnetic fields of galaxies},
Kluwer Academic Publishers , Dordrecht

\item
 Ryu, D.; Ostriker, J. P.; Kang, H.; Cen, R.  1993	
{\it 	A cosmological hydrodynamic code based on the
 total variation diminishing scheme},
ApJ. {\bf  414}, 1

\item 
 Ryu, D., Ostriker, J. P., Kang, H.,  Cen, R. 1994,
	{\it Hot gas in the cold dark matter scenario: X-ray 
clusters from a high-resolution numerical simulation},
 ApJ. {\bf  428}, 1

\item 
  Sagdeev, R.Z., Galeev, A.A. 1969
{\it Nonlinear Plasma Theory},
edited by T.M.  O'Neill and D.L. Book
W.A. Benjamin, New York

\item 
 Schekochihin, A. A.; Cowley, S. C.	2005
{\it 	Turbulence and Magnetic Fields in Astrophysical Plasmas},
astro-ph/0505686

\item 
 Schekochihin, A. A.; Cowley, S. C.; 
Maron, J. L.; McWilliams, J. C.	2004
{\it 	Self-Similar Turbulent Dynamo},
Phys. Rev. Lett.,  {\bf  92}, 4501

\item 
 Schekochihin, A.; Cowley, S.; Kulsrud, R.; 
Hammett, G.; Sharma, P.	,2005a
{\it 	Magnetized plasma turbulence in clusters of galaxies},
in {the Magnetized Plasma in galaxy Evolution} Krakow Poland
eds K.T. Chyzy. K. Otmianowska,  M,  Soida, R.-J. Dettmar, 86

\item 
 Schekochihin, A. A.; Cowley, S. C.; Kulsrud, R. M.; Hammett,
 G. W.; Sharma, P.	2005b
{\it 	Plasma Instabilities and Magnetic Field Growth in
 Clusters of Galaxies},
ApJ.,{\bf  629}, 139	

\item 
 Schekochihin, A. A.; Cowley, S. C.	2006
{\it Fast growth of magnetic fields in galaxy clusters:
a self-accelerating dynamo},
Astron. Nachr.  {\bf  327}, 599

\item
  Shukurov, A. 2004
{\it Introduction to galactic dynamos},
astro-ph 0411739

\item
Smith, M.C. \& 22 coauthors 2007
{\it The RAVE survey: constraining the local Galactic escape speed},
MNRAS {\bf 379}, 755

\item
 Steenbeck, M., Krause, F., and R\"{a}dler, Ke-H 1966
{\it Berechnung der  mitteren Lorentz-Feldstarke
$ {\bf  v} \times {\bf B} $ fur ein elektrisch leitendes Medium
in turbulenter, durch Coriolis-Kr\"{a}fte beeinflusster
Bewegung}, Z. Naturforsch {\bf  21a} 369.

\item 
Spergel, D. N.; Bean, R.; Doré, O.; Nolta, M. R.; Bennett, C. L.;
 Dunkley, J.; Hinshaw, G.; Jarosik, N.; Komatsu, E \& Page, L.; 
and 12 coauthors, 2007
{\it Three-Year Wilkinson Microwave Anisotropy Probe (WMAP) 
Observations: Implications for Cosmology},
ApJS {\bf  170}, 335

\item 
Spitzer, L.  1962 
{\it Physics of fully Ionized gases}
Interscience publishers

\item 
 Spitzer, L 1978
{\it 	Physical processes in the interstellar medium}
	New York Wiley-Interscience, 1978. 333

\item 
Subramanian K.; Barrow J D, 2002
{\it Small-scale cosmic microwave background polarization anisotropies
 due to tangled primordial magnetic fields},   	
MNRAS.{\bf  335}, 57	

\item 
Taylor, J.B. 1974
{\it Relaxation of toroidal plasma and generation of reversed
magnetic fields}
Phys. Rev. Lett.  {\bf  33} 1139

\item 
Taylor, J.B. 1986
{\it Relaxation and magnetic reconnection in plasmas},
Rev Modern Phys. {\bf  58}, 741 

\item
Tegmark, M., Silk, J., Rees, M.J., Blanchard, A., Abel, T., \& Palla, F. 1997
{\it How small were the first cosmological objects?}
ApJ, {\bf 474}, 1

\item
 Vainshtein, S.I. \& Ruzmaikin, A.A., 1972
{\it Generation of the Large Scale Galactic Magnetic Field}
Sov. Astron. {\bf  15}, 714

\item
 Vallee JP,  2004
{\it Cosmic magnetic fields - as observed in the Universe, in galactic 
dynamos, and in the Milky Way}, 
New Astron. Rev. {\bf  48} 763

\item 
 Watson, A. \& Perry, J.	1991
{\it 	QSO absorption lines and rotation measures},
MNRAS {\bf  248} 58

\item 
Weaver, R.; McCray, R.; Castor, J.; Shapiro, P.; Moore, R., 1977	
{\it 	Interstellar bubbles. II - Structure and evolution},
ApJ {\bf  218}. 377

\item
Weisberg, J.M., Cordes, J.M., Kuan, B., Devine, K.E., 
Green, J.T., \& Backer, D.C., 2004
{\it Arecibo 430 MHz pulsar polarimetry: Faraday rotation measures \& morphological classifications},
ApJS {\bf 150}, 317

\item 
 Welter, G. L.; Perry, J. J.; Kronberg, P. P. 1984
{\it 	The rotation measure distribution of QSOs
 and of intervening clouds - Magnetic fields and column densities},
ApJ, {\bf  279}, 19

\item 
 Widrow L.M.  2002
{\it Origin of galactic and extragalactic magnetic fields },
Rev. Mod Phys {\bf  74} 775

\item 
 Wolfe, A M , Lanzetta, K.M., Oren, A.L. 1992
{\it Magnetic Fields in Damped $ \mbox{Ly}\alpha $ Systems},
ApJ {\bf  388}, 17

\item
Woltjer, L, 1969
{\it Remarks on the galactic magnetic field }
in {\it  the Galactic system} edited by Hugo van Woerden,
Academic Press, p. 1967

\item
Zweibel, E.G. 1996
{\it Polarimetry and the theory of the Galactic magnetic field},
in {\it Polarimetry of the Interstellar Medium}, eds. W. G. Roberge \& D.C.B. Whittet, A.S.P. Conference Series Volume 97, p. 486

\item
 Zweibel,  E.G. 2003
{\it Cosmic-Ray History and its Implications
 for Galactic Magnetic Fields}
ApJ. {\bf  587}, 625

\item 
Zweibel,E.G., Heiles, C., 1997
{\it 	Magnetic fields in galaxies and beyond}
Nature, {\bf  385}, 131

\end{itemize} 

\end{document}